\documentclass[prd,reprint,onecolumn,superscriptaddress,tightenlines,nofootinbib,eqsecnum]{revtex4-2}

\usepackage{amsmath}
\usepackage{cancel}
\usepackage{amsfonts}
\usepackage{amssymb}
\usepackage{bm}
\usepackage[colorlinks]{hyperref}
\usepackage{mathrsfs}
\usepackage{graphicx}
\usepackage{multirow}
\usepackage{empheq}
\usepackage{ulem}
\usepackage{tensor}
\usepackage{tabularx}
\usepackage[usenames]{color}
\definecolor{darkgreen}{rgb}{0,0.6,0}
\hypersetup{urlcolor=darkgreen, citecolor=darkgreen}
\usepackage{cleveref}
\usepackage{stmaryrd}
\newcolumntype{Y}{>{\centering\arraybackslash}X}

\allowdisplaybreaks

\DeclareSymbolFontAlphabet{\mathrsfs}{rsfs}
\DeclareMathAlphabet{\mathcal}{OMS}{cmsy}{m}{n}

\usepackage{parskip}
\definecolor{darkgreen}{rgb}{0,0.5,0}
\definecolor{mygreen}{rgb}{0,0.8,0}

\hypersetup{
    unicode=false,          
    pdftoolbar=true,        
    pdfmenubar=true,        
    pdffitwindow=false,     
    pdfstartview={FitH},    
    pdftitle={My title},    
    pdfauthor={Author},     
    pdfsubject={Subject},   
    pdfcreator={Creator},   
    pdfproducer={Producer}, 
    pdfkeywords={keyword1} {key2} {key3}, 
    pdfnewwindow=true,      
    colorlinks=true,       
    linkcolor=red,          
    citecolor=cyan,        
    filecolor=magenta,      
    urlcolor=darkgreen,           
    linktocpage=true}

\newcommand{\be}{\begin{equation}}
\newcommand{\ee}{\end{equation}}

\newcommand{\mutp}{\widetilde{\mu}^{(2)}_{+}}
\newcommand{\mutm}{\widetilde{\mu}^{(2)}_{-}}
\newcommand{\sigmatp}{\widetilde{\sigma}^{(2)}_{+}}
\newcommand{\sigmatm}{\widetilde{\sigma}^{(2)}_{-}}
\newcommand{\zetatp}{\widetilde{\mu}^{(3)}_{+}}

\newcommand\calO{{\mathcal{O}}}
\newcommand{\dd}{\mathrm{d}}

\newcommand{\nn}{\nonumber}
\newcommand{\fid}{\dot{\phi}}
\newcommand{\rd}{\dot{r}}
\newcommand{\Et}{\widetilde{E}}
\newcommand{\Lt}{\widetilde{L}}
\newcommand{\tmass}{M}

\newcommand{\etidal}{\epsilon_\text{tidal}}
\newcommand{\et}{e_t}

\usepackage{ulem}
\allowdisplaybreaks

\usepackage{etoolbox}
\makeatletter
\patchcmd{\frontmatter@abstract@produce}
  {\vskip200\p@\@plus1fil
   \penalty-200\relax
   \vskip-200\p@\@plus-1fil}
\makeatother

\begin{document}

\title{Adiabatic tides in compact binaries on quasi-elliptic orbits: \\ Dynamics at the second-and-a-half relative post-Newtonian order}

\author{Quentin \textsc{Henry}}
\email{quentin.henry@uib.es}
\affiliation{Departament de Física, Universitat de les Illes Balears, IAC3 – IEEC, Crta. Valldemossa km 7.5, E-07122 Palma, Spain}

\author{Anna \textsc{Heffernan}}
\email{anna.heffernan@uib.eu}
\affiliation{Departament de Física, Universitat de les Illes Balears, IAC3 – IEEC, Crta. Valldemossa km 7.5, E-07122 Palma, Spain}

\date{\today}

\begin{abstract}
GW200105 is the first gravitational wave detection to show signs of eccentricity, it also is a neutron star - blackhole binary. This raises the need for waveforms that incorporate tidal effects on quasi-elliptic orbits. We tackle the problem of finite size effects within the post-Newtonian framework, including the mass-type quadrupole and octupole, as well as the current-type quadrupole deformations in the adiabatic approximation. The computations are performed at the second-and-a-half relative post-Newtonian order. We first derive the quasi-Keplerian parametrization of the conservative motion; we then express the radial separation and phase with their time derivatives in terms of the orbital frequency, the time eccentricity and the eccentric anomaly. To obtain these as functions of time, we invert the generalized Kepler equation while also discussing the convergence of eccentricity expanded results. We provide those results to the fourteenth order in eccentricity. Finally, we exploit the already known radiation reaction term of the acceleration in order to derive the secular and oscillatory evolutions of the orbital elements. The companion paper contains the derivation of the radiated fluxes and the amplitude modes of the strain. All relevant results are provided in an ancillary file.
\end{abstract}

\pacs{04.25.Nx, 04.25.dg, 04.30.-w, 97.80.-d, 97.60.Jd, 95.30.Sf}

\maketitle

\section{Introduction}\label{sec:intro}

Binary neutron star systems (BNSs) were thrown into the limelight with the gravitational wave (GW) detection, GW170817~\cite{LIGOScientific:2017vwq}, which to-date is still the first and only multi-messenger detection with an electromagnetic (EM) counterpart~\cite{LIGOScientific:2017ync}. Receiving both EM and GW information from a single event not only enabled investigations into both nuclear and neutron star (NS) physics~\cite{LIGOScientific:2017pwl, LIGOScientific:2017zic} but also provided constraints for beyond-Einstein theories of relativity~\cite{Sotiriou:2017obf} and had implications in cosmology~\cite{LIGOScientific:2017adf}. Indeed, this event proved so scientifically valuable that the current network of ground-based detectors, the LIGO-Virgo-KAGRA collaboration (LVK), now releases near immediate alerts with sky localization and the probability of whether a NS is involved. If a NS is present, the chances of an EM counterpart increase exponentially, making this information invaluable for followup observations by all EM telescopes. As is true for all GW detections, one's ability to detect, localize and parameterize an event relies not only on the detector capabilities and data pipelines but also the accuracy of the waveform models.

Most detections thus far have emanated from binary blackholes (BBHs)~\cite{LIGOScientific:2025pvj}, however several have involved neutron stars~\cite{LIGOScientific:2017vwq, LIGOScientific:2021qlt, LIGOScientific:2024elc}. Of particular interest is the detection of GW200105~\cite{LIGOScientific:2021qlt}, which came from a neutron star - blackhole~(NSBH) system and has recently claimed to have an eccentric signature~\cite{Fei:2024ruj,Morras:2025xfu,Planas:2025plq,Kacanja:2025kpr,Jan:2025fps,Tiwari:2025fua}. Parameter-estimation analysis indicates that the system entered the detector band with a significant eccentricity of $\sim 0.13$. As the LVK now begins to power down for planned upgrades, which themselves will make the LVK more sensitive to NS systems (see Table~1 of \cite{LVK:2025T2400403}), knowledge of the eccentric NSBH event, GW200105, brings a renewed need for more complete eccentric waveforms for systems with matter. The expected increase in sensitivity for the upcoming fifth LVK observing run~(O5) will allow for an order of magnitude increase in detections of both BNSs~\cite{Shah:2023ozh} and NSBHs~\cite{Colombo:2023une}, with also the prospect for larger signal-to-noise ratios later in the signal. This is of particular interest for BNS and NSBH systems, where tidal effects due to the deformability of NSs are expected but are yet to be detected.

The imprint of tidal interactions on the GW signal is only seen in the late inspiral, hence its elusiveness with current detectors. The possibility of seeing such an effect grows significantly when looking at third-generation ground-based detectors, like the Einstein Telescope~\cite{ET:2025xjr} and with it a plethora of subatomic physics (see S6 of~\cite{ET:2025xjr}). For the future space-based detector LISA, modeling waveforms with tidal effects for eccentric orbits was noted as a prospective challenge in the mission redbook~\cite{LISA:2024hlh}. This arises for galactic binaries with white dwarfs (WDs) or NSs where the early inspiral can be captured by LISA, which is also where one is more likely to see eccentricity. One should note that too early and the large separation of the binary will inhibit tidal effects; although this can be somewhat countered by high eccentricity. LISA is expected to resolve $\sim 10^4$ double WD systems (DWDs) (in addition to 10's of BNS and NSBH), which, prior to mass transfer, can also be described by the action below. Although most DWDs are expected to be quasi-circular, those formed through dynamical capture in globular clusters will not necessarily be so~\cite{Kremer:2018tzm}. Meanwhile NSBHs and BNSs are expected to have even more variety in their eccentricity distribution; with dynamical formation allowing for eccentricity $>0.1$~\cite{Sedda:2020wzl} and $>0.3$~\cite{Andrews:2019plw} respectively in the LISA band.

On its initial detection, GW170817, was analysed with waveform models describing only the leading order (LO) tidal deformation, parametrized by the first Love number $k^{(2)}$, on a quasi-circular motion~\cite{LIGOScientific:2018hze}. More recent models for systems with a NS incorporate more physical effects such as spin precession, and use several modeling techniques: effective one body (EOB) models, \textit{e.g.} \texttt{TEOBResumS-Dalí}~\cite{Gamba:2023mww,Albanesi:2025txj}, \texttt{SEOBNRv5THM}~\cite{Haberland:2025luz}, phenomenological models (Phenom) as \texttt{IMRPhenomXP\_NRTidalv2}~\cite{Colleoni:2023ple}, \texttt{NRTidalv3}~\cite{Abac:2023ujg} and \texttt{IMRPhenomXPHM\_NSBH}~\cite{IMRPhenomXPHMNSBH} as well as the so-called reduced-order models~\cite{Purrer:2014fza,Lackey:2016krb,Lackey:2018zvw}. In analysing GW200105, both Phenom~\cite{Morras:2025nlp, Planas:2025feq} and EOB~\cite{Haberland:2025luz, Gamba:2023mww, Albanesi:2025txj} eccentric models have been used; among these, only \texttt{TEOBResumS-Dalí} can account for tidal effects. Both families of models are built upon post-Newtonian (PN) predictions to describe the inspiral phase of the binary, however such results are incomplete in the PN literature; tidal effects for a binary on a quasi-elliptic orbit have not been calculated previously in the PN framework. This work focuses on those missing terms. 

Over the past decades, many works have tackled the motion of eccentric binaries in general relativity within the PN framework, notably motivated by the discovery of the Hulse-Taylor pulsar~\cite{Hulse:1974eb}. Over time, more terms in the PN expansion, as well as physical effects, have been considered. The conservative motion has been solved for point-particle effects up to the 3PN order~\cite{DamourDeruelle1,
damour1985general,Damour:1988mr,Schafer:1993pkg,Memmesheimer:2004cv,Boetzel:2017zza}, and recently some results at the 4PN order~\cite{Cho:2021oai,Trestini:2025yyc}. The spins of the companions have also been taken into account~\cite{Tessmer:2010hp,Klein:2010ti,Tessmer:2012xr,Cho:2022syn,Henry:2023tka} up to the 3PN order. The radiative dynamics, at 2.5 and 3.5PN was derived for point-particles in~\cite{Damour:2004bz,Konigsdorffer:2006zt,Boetzel:2019nfw}. Furthermore, the secular dynamics of the orbital elements including the nonspinning and spinning parts were obtained at higher PN orders using the flux balance equations in~\cite{Arun:2007sg,Arun:2009mc,Arun:2007rg,Henry:2023tka}. On the other hand, finite size effects were also broadly studied within the PN or post-Minkowskian framework, see the non-exhaustive list~\cite{Vines:2010ca,Damour:2012yf,Steinhoff:2016rfi,Banihashemi:2018xfb,Abdelsalhin:2018reg,HFB19,HFB20a,HFB20b,Cheung:2020sdj,Kalin:2020lmz,Mougiakakos:2022sic,Mandal:2023hqa,Patil2024,Bernard:2023eul,Dones:2024odv,Dones:2025zbs}, but not for bound eccentric orbits. The motivation of this project is straightforward: to derive consistently at the relative second-and-a-half PN order, the dynamics and the GW strain taking into account adiabatic tidal interactions within a binary system of compact objects on quasi-elliptic orbits. Tackling this problem is not only interesting from a theoretical point of view, as we solve for the most general motion that a binary system of spinless compact objects in the presence of adiabatic tides at 2.5PN beyond Newtonian gravity, but it also has the practical aim to be directly implemented in waveform models. 

This project relies on a series of previous works~\cite{HFB19,HFB20a,HFB20b,Dones:2024odv}, in which the observables were derived for tidal effects and reduced at a late stage to quasi-circular orbits. The procedure to derive all relevant quantities for an eccentric system is well understood: i) Solve the conservative equations of motion (EOM), this is done using a systematic method to derive a Keplerian-like parametrization, called quasi-Keplerian parametrization (QKP); ii) Solve the radiative dynamics, that is we include the influence of the GW radiation on the motion by adding it to the conservative dynamics; iii) Compute the radiated energy and angular momentum, and deduce the secular evolution of the orbital elements to higher order; iv) Calculate the waveform amplitude. This work is divided into two companion papers. In the present article, referred to as Paper I, we compute the conservative and radiative dynamics while in Paper~II~\cite{paperII}, we derive the radiated fluxes and amplitude of the GW, decomposed in spin-weighted spherical harmonics.

The matter action describing adiabatic tidal effects can be written through effective field theory as a skeletonized action defined on the worldlines of each body $A=(1,2)$. Here we consider the effective matter action
\begin{equation}\label{eq:Stidal}
S_\text{matter} = \sum_{A} \int \dd \tau_{A} \left[ -m_A c^2 + \frac{\mu_{A}^{(2)}}{4} G^{A}_{\mu\nu}G_{A}^{\mu\nu} + \frac{\sigma_{A}^{(2)}}{6c^{2}}H^{A}_{\mu\nu} H_{A}^{\mu\nu} + \frac{\mu_{A}^{(3)}}{12} G^{A}_{\lambda\mu\nu} G_{A}^{\lambda\mu\nu} \right]\,.
\end{equation}
The first term is the point-particle action, while the remaining three describe the tidal interactions. The tensors $G$ and~$H$ constitute the mass- and current-type decomposition of (derivatives of) the Weyl tensor; we refer to, \textit{e.g.}~\cite{Damour:1990pi,Damour:1991yw,Bini:2012gu} for more details regarding the construction of this action. At the 2PN order, three contributions appear: the mass-type quadrupole, the current-type quadrupole and the mass-type octupole. Each interaction is parametrized by a set of constants $\mu^{(\ell)}$ and $\sigma^{(\ell)}$, called tidal polarizabilities, encoding the internal structure of the bodies. They are defined as
\begin{equation}\label{eq:muAlsigmaAl}
\mu_A^{(\ell)} = \frac{2}{G (2\ell-1)!!} k_A^{(\ell)}R_A^{2\ell +1}\,, \qquad \sigma_A^{(\ell)} = \frac{\ell-1}{4 G(\ell+2)(2\ell-1)!!} j_A^{(\ell)}R_A^{2\ell +1}\,,
\end{equation}
where $R_A$ is the radius and $\{k_A^{(\ell)},j_A^{(\ell)}\}$ are the $\ell^\text{th}$ dimensionless mass-type and current-type Love numbers~\cite{love1911_geodynamics} of body~$A$. For nonspinning BHs, the static Love numbers vanish~\cite{Damour:2009vw,Binnington:2009bb,Landry:2015zfa,LeTiec:2020bos} but they are non-zero for NSs. We would like to emphasize again here that this work assumes adiabatic tides, however, for realistic systems dynamical tides should play a qualitative role when the binary follows an eccentric motion. Vibration modes of NSs can be triggered as their radial separation decreases, for eccentric orbits these modes lead to interesting dynamics as they are triggered upon iterative pericentre passages. Considering adiabatic tides is nonetheless a first step towards a more complete solution of the motion that would include dynamical tidal interactions.

The paper is organized as follows: In~\Cref{sec:Newtonian}, we perform a detailed didactic derivation of the dynamics at Newtonian order, and discuss the order of magnitude of the perturbation that we are considering. 
In~\Cref{sec:QK}, we extend the computation of the conservative dynamics at next-to-next-to leading order (NNLO). To do so, we derive the QKP in terms of the conserved quantities of the system; we then express the separation, phase and their derivatives in terms of the orbital frequency, time-eccentricity and eccentric anomaly; finally we invert the generalized Kepler equation to express the results in terms of the mean anomaly through an eccentricity expansion. We also discuss the validity of eccentricity expansions.
In~\Cref{sec:rad}, we consider the radiation reaction to derive the radiative part of the dynamics, including the secular and oscillatory evolutions. Finally, in the Appendices, we derive general integrals required for the QKP and display lengthy results. All relevant results are also provided in the ancillary file~\cite{SuppMaterial1} as a \textit{Mathematica} file. Its detailed content is given in the concluding section.


\section{The leading order: a didactic approach}\label{sec:Newtonian}


\subsection{Notations}

 Throughout this paper, we will use the following notations: the total mass of the system is denoted $M = m_1 + m_2$ with $m_1 \geq m_2$, $\nu=m_1 m_2/\tmass^2$ is the symmetric mass ratio and $\delta = (m_1-m_2)/\tmass = \sqrt{1-4\nu}$ is the normalized mass difference. The dot notation refers to the time-differentiation. The separation is denoted $r$ and the phase angle $\phi$, thus the relative velocity can be written as $\bm{v} = \rd \bm{n} + r \fid \bm{\lambda}$ where $\bm{n}$ is the unit vector pointing from body 1 to body 2 and $\bm{\lambda}$ is its perpendicular such that the vectorial product $\bm{n}\times\bm{\lambda}$ matches the direction of the orbital angular momentum. We also define the convenient combinations of the tidal polarizabilities 
\begin{equation}\label{eq:polarpm}
\mu_\pm^{(\ell)} = \frac{1}{2}\left(\frac{m_{2}}{m_{1}}\,\mu_{1}^{(\ell)} \pm
  \frac{m_{1}}{m_{2}}\,\mu_{2}^{(\ell)}\right)\,,\qquad \sigma_\pm^{(\ell)} =
\frac{1}{2}\left(\frac{m_{2}}{m_{1}}\,\sigma_{1}^{(\ell)} \pm
  \frac{m_{1}}{m_{2}}\,\sigma_{2}^{(\ell)}\right)\,,
\end{equation}
as well as their normalized versions
\begin{equation}\label{eq:musigmatilde}
\widetilde{\mu}_\pm^{(\ell)} = \left(\frac{c^2}{G \tmass}\right)^{2\ell+1}
\!\!\!G\,\mu_\pm^{(\ell)}\,,\qquad \widetilde{\sigma}_\pm^{(\ell)} =
\left(\frac{c^2}{G \tmass}\right)^{2\ell+1} \!\!\!G\,\sigma_\pm^{(\ell)}\,.
\end{equation}
The tidal polarizabilities~\eqref{eq:muAlsigmaAl} determine the formal PN order at which the tidal effects appear
\begin{equation}\label{eq:epstidal}
\mu^{(2)}\sim\sigma^{(2)} 
= \calO\left(\epsilon_\text{tidal}\right)\,, \qquad \mu^{(3)}
= \calO\left(\frac{\epsilon_\text{tidal}}{c^4}\right)\,.
\end{equation}
They are considered as a perturbation of the Newtonian quantities since they appear as a 5PN orbital effect.


\subsection{The setup} \label{sec:setup}


The computation starts from the conserved energy and angular momentum, $E$ and $\bm{J}$ respectively, of the binary system. One can show that the motion is planar, so we need only consider the norm of the angular momentum.
At leading order (LO), they read
\begin{subequations}\label{eq:ELtildeLO}
\begin{align}
\Et &= \frac{\rd^2}{2}+\frac{r^2\fid^2}{2} -\frac{G M}{r} -\frac{3\,G^2M}{r^6}\mu_+^{(2)} 
\,,\\ \label{eq:LtildeLO}
\widetilde{L} &= r^2\fid 
\,,
\end{align}
\end{subequations}
where we have defined the following normalized quantities
\begin{equation}\label{eq:tildEhdef}
\Et \equiv \frac{E - Mc^2}{\tmass \nu}\,, \qquad
\widetilde{L} \equiv \frac{\vert \bm{J}\vert}{\tmass\nu}\,, \qquad h \equiv \frac{\widetilde{L}}{G \tmass}\,.
\end{equation}
Introducing the dimensionless quantities $y = G \tmass h^2/r$ and $\varepsilon = 18\mutp/(hc)^{10}$ and inverting Eqs.~\eqref{eq:ELtildeLO}, we get the system of first-order non-linear differential equations
\begin{subequations}\label{eq:syst}
\begin{align}
(y')^2 &= 2 \Et h^2 + 2y-y^2+\frac{\varepsilon}{3}y^6\,,\label{eq:yd2}\\
\frac{\dd \phi}{\dd t}&= \frac{y^2}{G\tmass h^3}\,,\label{eq:fidy}
\end{align}
\end{subequations}
where $y'= \tfrac{\dd y}{\dd\phi}$. To solve for the motion of the binary, we differentiate~\eqref{eq:yd2} with respect to $\phi$ and discard the case $y'=0$, arriving at the second-order differential equation
\begin{equation}\label{eq:ydd}
y'' + y = 1 + \varepsilon \, y^5\,.
\end{equation}
Of course, the ideal case would be to solve~\eqref{eq:ydd} exactly, however, in the absence of exact solutions, one requires a perturbative approach (assuming that $\varepsilon\ll 1$). We discuss the numerical value of $\varepsilon$ in~\Cref{subsec:valeps}. Notice that this differential equation is very similar to a particular case of the Duffing oscillator~\cite{duffing1918erzwungene}. There are various ways to solve this differential equation perturbatively, we will present two: the first is \textit{la méthode bourrine} (naive derivation), of which we will show the limitations, followed by a refined method that recovers the LO results from the QKP method employed in~\Cref{sec:QK}.


\subsection{The unperturbed Kepler equation}


In the unperturbed case, where $\varepsilon=0$, Eq.~\eqref{eq:ydd} is the differential equation of a simple harmonic oscillator of frequency 1. The solutions are given by $y_0 = 1 + A\cos(\phi - \phi_0)$, where $A$ and $\phi_0$ are constants of integration. For simplicity, we assume that the phase origin vanishes, $\phi_0 = 0$, for the remainder of this section. By injecting $y_0$ into~\eqref{eq:yd2}, we retrieve $A = \sqrt{1+2 \Et h^2} \equiv e$, which we recognize as the Keplerian eccentricity. Injecting this solution, $y_0=1+e \cos{\phi}$, into~\eqref{eq:fidy}, we find the time parameter as a function of the phase angle $\phi$
\begin{equation}
l \equiv \frac{(1-e^2)^{3/2}}{G\tmass h^3} (t-t_0) = 2\arctan\left[\sqrt{\frac{1-e}{1+e}}\tan\frac{\phi}{2} \right] -e\sqrt{1-e^2}\frac{\sin\phi}{1+e\cos\phi}\,. 
\end{equation}
By introducing the eccentric anomaly $u$, related to the phase through
\begin{equation} \label{eqn:u0}
\tan\frac{u}{2}=\sqrt{\frac{1-e}{1+e}}\tan\frac{\phi}{2}\,,
\end{equation}
the Keplerian parametrization of a bound orbit becomes
\begin{subequations}
\begin{align}
r &= a(1-e\cos u)\,,\\
l = n (t - t_0) &= u - e\sin u\,,\\
\phi &= 2\arctan\left[\sqrt{\frac{1+e}{1-e}}\tan\frac{u}{2}\right]\,,
\end{align}
\end{subequations}
where $a = -G\tmass/2\Et>0$ is the semi-major axis and $n = (-2\Et)^{3/2}/G\tmass$ is the mean motion.


\subsection{The naive perturbative solution}


The straightforward method for solving the system~\eqref{eq:ydd} is to
allow $y = y_0 + \varepsilon y_1$, where $\varepsilon \ll 1$. We then inject $y_0$ into the $\calO(\varepsilon)$ of~\eqref{eq:ydd},
\begin{equation}
y''_1 + y_1 = y_0^5 = \bigl(1+e\cos\phi\bigr)^5 = \sum_{k=0}^5\alpha_k \cos(k\phi)\,,
\end{equation}
where the coefficients $\alpha_k$ are functions of $e$\footnote{These are $\alpha_0=1+5e^2+15e^4/8$, $\alpha_1 = 5e(1 + 3e^2/2+e^4/8)$, $\alpha_2 = 5e^2(1+e^2/2)$, $\alpha_3 = 5e^3/2(1+e^2/8)$, $\alpha_4 = 5e^4/8$ and $\alpha_5 = e^5/16$.}. The solutions of this differential equation are given by
\begin{equation}
y_1(\phi) = B \cos\phi+ \alpha_0 + \frac{\alpha_1}{2} \phi\sin\phi + \sum_{k=2}^5 \frac{\alpha_k}{1-k^2} \cos(k\phi) \,.
\end{equation}
Now, plugging $y_0$ and $y_1$ into~\eqref{eq:yd2} allows us to fix $B$ as a function of $e$ and yields a final solution
\begin{equation}\label{eq:yold}
y(\phi) =  1 + e \cos\phi+ \varepsilon\left[ \alpha_0 + \frac{\Bar{\alpha}_1}{e} \cos\phi + \frac{\alpha_1}{2} \phi\sin\phi + \sum_{k=2}^5 \frac{\alpha_k}{1-k^2} \cos(k\phi) \right]\,,
\end{equation}
with $\Bar{\alpha}_1 = 1/6 + 5e^2/2 + 45e^4/16 + 5e^6/24$. Although formally correct, this solution is not realistic. The function~$\phi\sin\phi$ is unbounded while the solution of~\eqref{eq:ydd} must be bounded since it is the solution of a weakly perturbed harmonic oscillator. Such problematic terms, called secular, are an artifact of the perturbative expansion. They arise when the frequency of the harmonic oscillator matches the one of a cosine or sine in the source of the differential equation. In general, one can treat these terms by introducing some different scalings in the differential equation. Various methods exist, \textit{e.g.} multiscale, Poincaré-Lindstedt, Krylov-Bogoliobov-Mitropolsky, etc. For more details, see~\cite{mickens1996oscillations}, where the Duffing oscillator is treated, and from which we adapt the Poincaré method to the (closely-related) present case.


\subsection{Treating the secular terms}\label{subsec:secterms}


The Poincaré-Lindstedt method, developped in~\cite{Poincare1893MethodesNouvelles2}, is employed in order to treat these secular terms by not only perturbing the unknown function $y$ but also the frequency. It introduces a new variable linearly related to the previous one, $\theta = \omega \phi$. The free parameter $\omega$ adopts a power series in $\varepsilon$ of the form $\omega(\varepsilon) = \omega_0 + \sum_{k\geq 1} \omega_k \varepsilon^k$, where $\omega_0$ is chosen to recover the unperturbed differential equation, which is 1 in this case. Thus, at linear order, we have
\begin{subequations}
\begin{align}
y(\theta,\varepsilon) &= y_0(\theta) + \varepsilon y_1(\theta) + \calO(\varepsilon^2)\,,\\
\omega(\varepsilon) &= 1 + \varepsilon \omega_1 + \calO(\varepsilon^2)\,.
\end{align}
\end{subequations}
Now, the linear order of~\eqref{eq:ydd} can be recast in terms of the new variable $\theta$ as
\begin{equation}
\frac{\dd^2 y_1}{\dd\theta^2} + y_1 = 2 \omega_1( y_0 - 1 ) + y_0^5 \,,
\end{equation}
We can see that the source of this differential equation contains the constant $\omega_1$, which can be chosen to cancel the term proportional to $\cos\theta$. More precisely, when injecting $y_0(\theta)$, we get
\begin{equation}\label{eq:y1temp}
\frac{\dd^2 y_1}{\dd\theta^2} + y_1 = \alpha_0 + (2 e\, \omega_1 + \alpha_1) \cos\theta + \sum_{k=2}^5\alpha_k \cos(k\theta)\,,
\end{equation}
so by setting $\omega_1 = -\alpha_1/2e$, the term generating secular contributions vanishes. The remainder of the particular solution is then the same as in the previous case, arriving at
\begin{equation}\label{eq:ynew}
y(\theta) = 1+e \cos\theta + \varepsilon\left[ \alpha_0 + \frac{\Bar{\alpha}_1}{e} \cos\theta + \sum_{k=2}^5 \frac{\alpha_k}{1-k^2} \cos(k\theta) \right]\,,
\end{equation}
with $\theta=(1-\varepsilon\alpha_1/2e)\phi$. The Taylor expansion in $\varepsilon\rightarrow 0$ of~\eqref{eq:ynew} yields exactly~\eqref{eq:yold}. Thus, the natural choice is to keep using $\theta$ in the intermediate computations.


\subsection{Keplerian-like parametrization}


Moving on to a Keplerian-like parametrization, we suppose that there exists an angle $u$ that allows one to write the separation at $\calO(\varepsilon)$ in the form $r=a_r(1-e_r\cos u)$, where $a_r=a+\varepsilon a_1$, $e_r =e+\varepsilon e_1$ and $u=u_0(\theta)+\varepsilon u_1 (\theta)$ with $u_0(\theta)$ taking the previous definition of $u$ in~\eqref{eqn:u0} but with $e \rightarrow \tilde{e} = e(1+\varepsilon\lambda_0)$. Recalling $y=\tfrac{G \tmass h^2}{r}$ and slotting in our ansatz for $r$, the resulting expression at $O(\varepsilon)$ in $y$ takes the form $u_1(\theta) (A \sin{\theta}+B \sin{2 \theta})+C \cos{\theta} + D \cos{2 \theta} +F $, which tells us $u_1(\theta)$ will have the profile $\sum_{n=1}^{3} \lambda_n \sin{n \theta}$ to deliver the desired sum of $\cos{k \theta}$ in~\eqref{eq:ynew} (i.e. even powers of $\sin{\theta}$ are required). We now have
\begin{equation}\label{eq:ugen}
u = 2\arctan\left[\sqrt{\frac{1-\tilde{e}}{1+\tilde{e}}}\tan \frac{\theta}{2}\right] + \varepsilon \bigl[  \lambda_1 \sin\theta + \lambda_2\sin 2\theta + \lambda_3\sin 3\theta \bigr]+\calO(\varepsilon^2)\,,
\end{equation}
where the  $e \rightarrow \tilde{e} = e(1+\varepsilon\lambda_0)$ substitution in $u_0(\theta)$ contributes to a coefficient proportional to $\lambda_0 \frac{\sin\theta}{1+e \cos\theta}$, or can be seen as the free parameter allowing you to fix the coefficient of $\cos(2 \theta)$ in $y(\theta)$. Indeed, all unknown coefficients can now be fixed by comparing our resulting expression from $y=\tfrac{G \tmass h^2}{r}$ with the ansatz for $r$ to~\eqref{eq:ynew}. Ultimately, the separation expressed in terms of the eccentric anomaly $u$ is given by
\begin{equation} \label{eq:ruN}
r(u)= \frac{G\tmass h^2}{1-e^2} \left( 1 - \varepsilon\, \frac{2(1-e^4)}{3} \right)\left[ 1 - e \left( 1 + \varepsilon\, \frac{1+9e^2-5e^4-5e^6}{6e^2}  \right) \cos u \right] +\calO(\varepsilon^2)\,.
\end{equation}
For completeness we turn to the time dependency. One can rewrite~\eqref{eq:ynew} in the following convenient form $y(\theta) = 1 + e\cos\theta + \varepsilon\sum_{k=0}^5 \gamma_k (1 + e \cos(\theta))^k$ 
in order to insert it in~\eqref{eq:fidy}. After integrating with respect to $\theta$, recalling $\dd\theta = \omega \dd \phi$, this leads to
\begin{align}
\frac{(1-e^2)^{3/2}}{G \tmass h^3} (t-t_0) & = 2 \arctan\left[\sqrt{\frac{1-e}{1+e}}\tan\frac{\theta}{2} \right] -e\sqrt{1-e^2}\frac{\sin\theta}{1+e\cos\theta} + \varepsilon (1-e^2)^{3/2} \left[ \left( 1+\frac{e^2}{4} \right)\theta  \right.\nn \\
&\left.  + \sin\theta\left(\frac{5}{12}e + \frac{e^2}{24}\cos\theta - \frac{8+82e^2+15 e^4}{48e(1+e\cos\theta)} - \frac{8+194e^2+113e^4}{48e(1+e\cos\theta)^2} \right)\right]+\calO(\varepsilon^2)\,.
\end{align}
Finally, when expressed in terms of the parameter $u$ defined in~\eqref{eq:ugen}, we get
\begin{align}\label{eq:ltemp}
\frac{(1-e^2)^{3/2}}{G \tmass h^3} (t-t_0) & = u - e\left( 1+ \varepsilon\frac{1+5e^2-5 e^4-e^6}{6e^2} \right) \sin u \\
& + \varepsilon (1-e^2)^{3/2}\biggl[ \left(1+\frac{e^2}{4}\right)2\arctan\left[\sqrt{\frac{1+e}{1-e}}\tan\frac{u}{2}\right] +\frac{e\sqrt{1-e^2}}{12} \frac{\sin u}{1-e\cos u}\left( 7 + \frac{1-e^2}{1-e\cos u} \right)\biggr]\,.\nn
\end{align}
Of course, the above results are in agreement with the generalized ones~\eqref{eq:QKgen} derived in the following Section. We can see that in the intermediate computations, various eccentricity-like parameters $e_r$, $\tilde{e}$ and the coefficient in front of $\sin u$ in~\eqref{eq:ltemp}, arise. Although they all reduce to the Keplerian eccentricity at $\calO(\varepsilon^0)$, the eccentricity is not unique. This means that on top of making a choice of the form of the parametrization, we need to choose a certain definition of eccentricity in the rest of the computations. See for example~\cite{Loutrel:2018ydu}, for a list of different eccentricity parameters that exist in GW analysis. The method used in~\Cref{sec:QK} to derive the QKP imposes a unique choice of parametrization. It also automatically treats the secular contributions uniquely because it imposes the separation to be periodic.

\subsection{Numerial values of the perturbation parameter}\label{subsec:valeps}

Before moving to the QKP at higher orders, let us discuss the numerical values that $\varepsilon$ can take. We can express it by replacing the normalized angular momentum as a function of the semi-major axis and the Keplerian eccentricity through $h^2 = a (1-e^2)/G \tmass$. Recalling $\varepsilon = 18\mutp/(hc)^{10}$ from Sec.~\ref{sec:setup}, this gives
\begin{equation}\label{eq:epsilonval}
\varepsilon = 6\left(\frac{G}{ac^2(1-e^2)}\right)^5\left( m_2 m_1^4 \frac{k_1^{(2)}}{\mathcal{C}_1^5} + m_1 m_2^4 \frac{k_2^{(2)}}{\mathcal{C}_2^5} \right)\,,
\end{equation}
where $k_A^{(2)}$ and $\mathcal{C}_A = G m_A/R_A c^2$ are respectively the first Love number and the compactness of body $A$. For numerical applications regarding NSs, the typical values are $k^{(2)}\sim 0.1$ and $\mathcal{C}\sim 0.15$~\cite{Flanagan:2007ix,Rezzolla:2025pft}, although recent work~\cite{Rezzolla:2025pft} suggest that the maximum compactness of a NS is $\mathcal{C}_\text{max} = 1/3$. We can see that the maximum value of $\varepsilon$ for fixed masses, compactness and Love numbers are obtained for minimal value of the semi-major axis and maximal value of the eccentricity. Here let us consider identical bodies, and take the extreme case of $e=0.6$ and $m_{1,2} = 2M_\odot$. One finds that $\varepsilon$ is smaller than $0.1$ for a semimajor axis bigger than $\sim 30$km. This separation is the same order of magnitude than the diameters of the NSs and thus is already late inspiral. Hence for the purposes of our computations, we can safely always consider $\varepsilon\ll 1$. Finally, let us mention that $\varepsilon$ depends on the semi-major axis, or equivalently the orbital frequency. This means that in the context of GW emission, $\varepsilon$ becomes time dependent since the orbital frequency grows, making the perturbation parameter grow as the system evolves. For the event GW200105 that was mentioned in the introduction, we take the values of Table I of~\cite{Morras:2025xfu}, which leads to a value $\varepsilon \sim 10^{-8}$ at the 20Hz frequency (when it entered the detector band). 


\section{Conservative dynamics: the quasi-Keplerian parametrization to NNLO}\label{sec:QK}


\subsection{In terms of the conserved quantities}\label{subsec:QKCons}


We start from the conserved energy $E$ and angular momentum $\bm{J}$ including tidal effects at NNLO as computed in~\cite{HFB19,Dones:2024odv}, which correspond to Eqs.~\eqref{eq:ELtildeLO} including higher-order terms. These expressions, in standard harmonic coordinates, are provided for point-particles in Eqs.~(4.6) and (4.7) of~\cite{Blanchet:2002mb} and their tidal parts in Eqs.~(A2) and (A3) of~\cite{Dones:2024odv}\footnote{Thanks to the work performed in~\cite{Mandal:2024iug}, it was found that the conserved quantities in~\cite{HFB19} were not in the same gauge as the multipole moments derived in~\cite{HFB20a}, which lead to an error in the radiated flux and (2,2) mode at NNLO.}. We express them in terms of $(r,\rd,\fid)$, which leads to a coupled system of 2 equations of two variables as functions of $t$. 
We employ a well-understood systematic method to derive the QKP (nicely explained in~\cite{Memmesheimer:2004cv}). First, we iteratively invert the relations for $\Et$ and $h$ to obtain an expression for $\rd$ and $\fid$ in terms of $\Et$, $h$ and $s = G \tmass/r$
\begin{subequations}\label{eq:rdfid}
\begin{align}
\dot{r}^2 &= \sum_{k=0}^{10} p_k\bigl(\Et,h\bigr)\, s^k = \mathcal{P}(s)\,,\label{eq:rd2}\\
\dot{\phi}&= \sum_{k=2}^{10} q_k\bigl(\Et,h\bigr)\, s^k = s^2 \mathcal{Q}(s)\label{eq:phid}\,,
\end{align}
\end{subequations}
where $\mathcal{P}$ and $\mathcal{Q}$ are two polynomials of degrees 10 and 8 respectively, and whose coefficients are explicit expressions of $\Et$ and $h$. Next, we impose that the radial motion follows the ansatz
\begin{equation} \label{eq:r}
r = a_r(1-e_r \cos u)\,,
\end{equation}
with $0\leq e_r < 1$ and where $u$ is the eccentric anomaly. By doing so, we impose that $s$ goes through a maximum and a minimum as $u$ varies between 0 and $2\pi$, thus one can factorize $\mathcal{P}$ as
\begin{equation}\label{eq:Pfac}
\mathcal{P}(s) = (s_+ - s)(s - s_-) \mathcal{R}(s)\,,
\end{equation}
where $\mathcal{R}$ is also a polynomial of degree 8 which reduces to 1 at LO. The above relation defines $s_+$ and $s_-$, which, by convention\footnote{Here we define $s_- < s_+$ , hence $s_- \equiv G M / r_+$; some literature defines $s_- \equiv 1/r_-$.}, are
\begin{equation} \label{eq:spm}
s_+\equiv\frac{G\tmass}{a_r(1 - e_r)}\,, \qquad s_- \equiv \frac{G\tmass}{a_r(1 + e_r)}\,,
\end{equation}
and thus the semi-major axis $a_r$ and radial eccentricity $e_r$ are given by
\begin{equation}\label{eq:arerspsm}
a_r=\frac{G\tmass}{2}\frac{s_++ s_-}{s_+ s_-}\,, \qquad e_r = \frac{s_+-s_-}{s_++s_-}\,.
\end{equation}
The explicit results are provided in~\Cref{app:qkpCons} due to their unwieldiness. We will continue to give larger expressions in the Appendix throughout to keep the flow clean.

We next focus on the integration of Eq.~\eqref{eq:phid}. The phase variable is computed as a function of the eccentric and true anomalies through
\begin{equation}\label{eq:phimphi0}
\phi-\phi_0 = \int
\dd \phi' 
= G\tmass \int_{s}^{s_+}\dd x\, \frac{ \mathcal{Q}(x) \left[\mathcal{R}(x) \right]^{-1/2}}{\sqrt{(s_+ -x)(x-s_-)}}\,.
\end{equation}
On expanding, this integral can be simply re-written as a sum of kernel integrals of the form~\eqref{eq:Indef}. In~\Cref{app:In}, we derive a general formula for those kernel integrals, whose results are available in Eqs.~\eqref{eq:I0}, \eqref{eq:I1} and \eqref{eq:InG2}, and expressed in terms of $a_r$, $e_r$ and
\begin{equation}
\tilde{v} = 2 \arctan \left[\sqrt{\frac{1+e_r}{1-e_r}}\tan\frac{u}{2}\right]. 
\end{equation}
With those results at hand, we are able to express the phase~\eqref{eq:phimphi0} in terms of the conserved quantities and $\tilde{v}$. Next, the advance of periastron angle, defined as the angular integral over one orbit, can be computed directly from our expression for $\phi-\phi_0$ with
\begin{equation}\label{eq:Phiintegral}
\Phi 
= 2 \lim_{s\rightarrow s_-} (\phi-\phi_0)= 2(\phi-\phi_0)\vert_{\Tilde{v}=\pi}\,,
\end{equation}
where the limit is obtained by substituting $\Tilde{v}=\pi$. The periastron advance is simply defined as $K\equiv \Phi/2\pi$. Next, we define the true anomaly as
\begin{equation}
v= 2 \arctan\left[\sqrt{\frac{1+e_\phi}{1-e_\phi}} \tan \frac{u}{2} \right].
\end{equation}
After expressing $\tilde{v}$ in terms of $v$, the phase reads
\begin{equation}\label{eq:QKphiA}
\frac{\phi-\phi_0}{K} = v + g_v \sin{v} + \sum_{k=2}^{8}g_{k v} \sin(k v)\,.
\end{equation}
The last step for the phase parametrization is to fix the phase eccentricity $e_\phi$ by imposing that $g_{v} = 0$, which results in a unique relation between both eccentricities $e_r$ and $e_\phi$.

We treat Eq.~\eqref{eq:rd2} in the same way as Eq.~\eqref{eq:phid}. The time dependency of the QKP is given by the integral
\begin{equation}\label{eq:tmt0}
t-t_0 = \int
\dd t' = G\tmass \int_{s}^{s_+}\dd x \, \frac{ \left[\mathcal{R}(x) \right]^{-1/2}}{x^2\sqrt{(s_+ -x)(x-s_-)}}\,.
\end{equation} 
This integral is computed by using the general formulas of~\Cref{app:In} whose results are expressed in terms of $a_r$, $e_r$, $u$ and~$\Tilde{v}$. The radial period $P$, defined as the time integral over one orbit then reads
\begin{equation}\label{eq:Pintegral}
P 
= 2 \lim_{s\rightarrow s_-} (t-t_0) = 2(t-t_0)\vert_{u=\Tilde{v}=\pi}\,,
\end{equation}
which is computed the same way as $\Phi$. The mean motion is defined as $n = 2\pi/P$, which in turn leads to the mean anomaly $l=n(t-t_0)$.

In the end, the QKP used to parametrize the motion of the binary is given by
\begin{subequations}\label{eq:QKgen}
\begin{align}
r &= a_r(1-e_r \cos u)\,, \label{eq:QKr}\\
l = n(t-t_0) &= u - e_t \sin u + f_{v-u} (v-u) + \sum_{k=1}^{6}f_{kv} \sin(k v)\,,\label{eq:QKl}\\
\frac{\phi-\phi_0}{K} &= v+ \sum_{k=2}^{8}g_{kv} \sin(kv)\,,\label{eq:QKphi}\\
v&= 2 \arctan\left[\sqrt{\frac{1+e_\phi}{1-e_\phi}} \tan \frac{u}{2} \right]\,, \label{eq:QKv}
\end{align}
\end{subequations}
where the explicit expressions of the various coefficients in terms of the conserved quantities are displayed in Appendix~\ref{app:qkpCons} and provided in the supplementary material~\cite{SuppMaterial1}. The specific bounds of the sums in~\eqref{eq:QKl} and~\eqref{eq:QKphi} are determined by the degrees of the polynomials $\mathcal{P}$ and $\mathcal{Q}$. Of course, this parametrization is not unique, but is built in order to match with previous works, such as point-particle at 3PN (\textit{e.g.}~\cite{Memmesheimer:2004cv}), spin-aligned at 3PN~\cite{Henry:2023tka} and magnetic dipole interactions~\cite{Henry:2023guc, Henry:2023len}. Note that some other references such as~\cite{Damour:2004bz,Konigsdorffer:2006zt,Boetzel:2017zza}, include the periastron advance $K$ in the definition of $g_{kv}$ as $\phi = Kv +\sum g'_{kv}\sin(kv)$. This can lead to some errors when combining results from different articles.

\subsection{In terms of $(x,e_t,u)$}

In this Section, we aim at expressing $(r,\dot{r},\phi,\dot{\phi})$ in terms of $(x,e_t,u)$, where $x$ is the usual gauge-independant adimensonned PN parameter
\begin{equation}\label{eq:QKx}
x = \left(\frac{G M \Omega}{c^3}\right)^{2/3}\,,
\end{equation}
with $\Omega = K n$ being the orbital frequency. To do so, since we know the expressions of $x$ and $e_t$ in terms of $\Et$ and $h$, we start by iteratively inverting the system
\begin{subequations}
\begin{align}
x \,c^2 &= -2\Et + f(\Et,h)\,,\\
\et^2 &= 1 + 2\Et h^2 + g(\Et,h)\,.
\end{align}
\end{subequations}
At LO, it reads
\begin{subequations}\label{eq:Ethofxet}
\begin{align}
-\Et &= \frac{c^2 x}{2} \left[ 1 -\frac{3}{4} \frac{x^5\mutp}{(1-\et^2)^5} \left( 5(8+12\et^2+\et^4) - 4(1-\et^2)^{3/2}(4+\et^2)\right) \right] \,,\\
h &= \frac{\sqrt{1-\et^2}}{c\sqrt{x}}\left[ 1 + \frac{3\mutp x^5}{8(1-\et^2)^5} \biggl( 48 + 108 \et^2 + 13 \et^4
 -4 \sqrt{1-\et^2}\bigl(4 + 9 \et^2 + 2 \et^4 \bigr) \biggr)  \right]\,.
\end{align}
\end{subequations}
Next, we can express all the quantities in Eqs.~\eqref{eq:QKgen} in terms of $(x,e_t,u)$. The time derivatives of the separation and phase are derived using chain rules of the type $\tfrac{\dd}{\dd t}= \tfrac{\dd u}{\dd t}\tfrac{\dd}{\dd u}$, and $\tfrac{\dd u}{\dd t} = \frac{n}{\dd l/\dd u}$. Hence, we have
\begin{align}
r = a_r(1-e_r \cos u)\qquad ;& \qquad \rd = \frac{n\, a_r e_r}{\dd l/\dd u}\sin u\,,\\
\phi-\phi_0 = K\left[ v+\sum_{k=2}^8  g_{kv} \sin(kv)\right] \,;& \qquad \fid = \frac{\Omega}{\dd l/\dd u}\frac{\dd v}{\dd u}\left[1+\sum_{k=2}^8 k \, g_{kv} \cos(kv)\right]\,.
\end{align}
All constant coefficient (such as $a_r$, $e_r$, $g_{kv}$, $f_{kv}$...) appearing in these expressions have been derived in~\Cref{subsec:QKCons} in terms of $\Et$ and $h$. We express them in terms of $(x,\et)$ using Eqs.~\eqref{eq:Ethofxet} and replace $v$ as a function of $u$ in order to derive the expressions of $(r,\rd,\phi,\fid)$ uniquely in terms of $(x,\et,u)$. We also obtained the value of $K$ in terms of $(x,\et)$ and remark that its value is in agreement with Eq.~(4.5) of~\cite{HFB20b} in the limit $\et\rightarrow 0$. Let us provide here a the handy expression of the derivative of the mean anomaly
\begin{equation}\label{eq:dldu}
\frac{\dd l}{\dd u} = 1-e_t\cos u + f_{v-u}\left(\frac{\dd v}{\dd u}-1\right) + \frac{\dd v}{\dd u} \sum_{k=1}^6 k \,f_{kv} \,T_k \bigl(\cos v\bigr)\,,
\end{equation}
where $T_k$ is the $k^\text{th}$ Chebyshev polynomials of the first kind, $\tfrac{\dd v}{\dd u} = \tfrac{\sqrt{1-e_\phi^2}}{1-e_\phi\cos u}$ and $\cos v = \tfrac{\cos u -e_\phi}{1-e_\phi\cos u}$.

When combining all the above expressions, we are able to get the expressions of $(r,\rd,\phi,\fid)$ in terms of $(x,\et,u)$. They are given at LO by
\begin{subequations}\label{eq:QKPxeu}
\begin{align}
r (x,\et,u) =& \, \frac{G\tmass}{x \, c^2} \biggl\{ 1-\et \cos u + \frac{3}{4}\frac{x^5 \mutp}{(1-\et^2)^5}\Bigl[3(8+20\et^2+7\et^4) -4(1-\et^2)^{3/2}(4+\et^2)\nn \\
& \qquad  - \et \left( 5(8+12\et^2+\et^4) +2(1-\et^2)^{3/2}(4+\et^2) \right) \cos u \Bigr]\biggr\}+ \calO\left(\frac{1}{c^2},\frac{\etidal}{c^2} \right)\,,\\
\rd (x,\et,u) =& \, \frac{c \, \sqrt{x}\,\et\sin u}{1-\et \cos u} -3  \frac{\et \,c \, x^{11/2}\mutp\sin u}{(1-\et^2)^{5}}\biggl\{ \frac{(1-\et^2)^4}{(1-\et\cos u)^5} + \frac{2(1-\et^2)^3}{(1-\et\cos u)^4}  + \frac{(1-\et^2)^2(3+\et^2)}{(1-\et\cos u)^3}\nn \\
& - \frac{3(1-\et^2)^{3/2}(4+\et^2)}{2(1-\et\cos u)^2} + \frac{5 (8+12\et^2+\et^4)-4(1-\et^2)^{3/2}(4+\et^2)}{8(1-\et\cos u)} \biggr\}+ \calO\left(\frac{1}{c^2},\frac{\etidal}{c^2} \right)\,,\\
\phi (x,\et,u) =& \, 2\arctan\left[\sqrt{\frac{1+\et}{1-\et}}\tan \frac{u}{2} \right]\left( 1 + \frac{45(8+12\et^2+\et^4)}{8(1-\et^2)^5}\mutp x^5 \right) \nn \\
& + \frac{3 \et \mutp x^5\sin u}{4(1-\et^2)^{9/2}} \Biggl\{\frac{(1-\et^2)^3}{(1-\et\cos u)^4} + \frac{5(1-\et^2)^2}{(1-\et\cos u)^3}  + \frac{34-27\et^2-7\et^4}{2(1-\et\cos u)^2}  \nn \\
& \qquad \qquad \qquad \qquad \qquad + \frac{146+105\et^2+12\sqrt{1-\et^2}(4+\et^2)}{2(1-\et\cos u)}\Biggr\}+ \calO\left(\frac{1}{c^2},\frac{\etidal}{c^2} \right)\,,\\
\fid (x,\et,u) =& \, \frac{x^{3/2}c^3}{G\tmass}\Biggl\{ \frac{\sqrt{1-\et^2}}{(1-\et \cos u)^2} + \frac{\mutp x^5}{(1-\et^2)^5 } \biggl[ \frac{9(1-\et^2)^2(4+\et^2)+24\sqrt{1-\et^2}(1-\et^4)}{(1-\et\cos u)^3} \nn \\
& \qquad \qquad - \frac{36(1-\et^2)(4+\et^2)+3\sqrt{1-\et^2}(112+132\et^2+7\et^4)}{8(1-\et\cos u)^2}\biggr] \Biggr\} + \calO\left(\frac{1}{c^2},\frac{\etidal}{c^2} \right)\,.
\end{align}
\end{subequations}
The full NNLO expressions are available in the supplementary file~\cite{SuppMaterial1}. As a check of these results, we have replaced them in the full NNLO expression of $\Et(r,\rd,\phi,\fid)$ which gave an expression of $\Et(x,\et)$ (independent of $u$) and then replaced $x(\Et,h)$ and $\et(\Et,h)$ which gave back exactly $\Et$, same for $h$. 

\subsection{Inverting $l(u)$}\label{sec:lofu}

In this section, we want to invert~\eqref{eq:QKl} in order to express $(r,\rd,\phi,\fid)$ in terms of $(x,\et,l)$. This problem has been tackled and very well explained in~\cite{Boetzel:2017zza}. However, the formulas provided in that paper are not valid in the presence of tidal effects. Thus, we generalize here the inversion. The goal is to write $u-l$ as a Fourier series of the form
\begin{equation}\label{eq:ulgen}
u - l = \sum_{k=1}^\infty A_k \sin(kl)\,.
\end{equation}
To do so, we first express $f_{v-u} (v-u) + \sum_{k=1}^6 f_{kv} \sin(k v)$ as a series $\sum_{j=1}^\infty \alpha_j \sin(ju)$ and compute the coefficients~$\alpha_j$, their expression are given in Eq.~\eqref{eq:alphajval}. The second step is to link the coefficients $A_k$ to $\alpha_j$, their expressions are given in Eq.~\eqref{eq:Akvalue}. In this second step, the computation differs from~\cite{Boetzel:2017zza}.

\subsubsection{Computing $\alpha_j$}

As mentionned above, we start from~\eqref{eq:QKl} which we generalize here for any bound $N$ of the sum
\begin{equation}
l = u - e_t \sin u + f_{v-u} (v-u) + \sum_{k=1}^N f_{kv} \sin(k v)\label{eq:lgenkv}
\end{equation}
As detailed in Appendix B of~\cite{Boetzel:2017zza}, one can express $v-u$ and $\sin(kv)$ as series of $\sin(nu)$. We have checked these expressions that read
\begin{equation}
v-u = 2 \sum_{n=1}^\infty \frac{\beta_\phi^n}{n} \sin(nu)\,, \qquad \text{and} \qquad \sin(k v) = \sum_{n=1}^\infty \mathcal{E}^k_n \sin(nu)\,,
\end{equation}
where $\beta_\phi = \tfrac{1-\sqrt{1-e_\phi^2}}{e_\phi}$ and the value of $\mathcal{E}^i_j$ given in Eqs.~(B18b) and~(B18c) of~\cite{Boetzel:2017zza}, can be alternatively expressed under the form
\begin{equation}
\forall (i,j)\in\mathbb{N}^2, \quad \mathcal{E}^i_j = i\sum_{n=0}^j \frac{(-)^{n+i-j}(i+n-1)!}{(i-j+n)!(j-n)!n!}\beta_\phi^{2n+i-j}\,.
\end{equation}
When injecting these expressions in~\eqref{eq:lgenkv}, which we can rewrite as
\begin{equation}\label{eq:lgenju}
l = u - \et \sin u +\sum_{j=1}^\infty \alpha_j \sin(ju)\,,
\end{equation}
where the $\alpha$ coefficients are given by
\begin{equation}\label{eq:alphajval}
\alpha_j = 2\frac{\beta_\phi^j}{j}f_{v-u} + \sum_{i=1}^N f_{iv} \mathcal{E}^i_j \,.
\end{equation}
%

\subsubsection{Computing $A_k$}

Now, we generalize the expressions of the Fourier coefficients $A_k$ in terms of the coefficients $\alpha_j$. Indeed, in the case of point-particle, it is sufficient to truncate the intermediate results at $\calO(\alpha_j)$ because they are 2PN point particle quantities, but with the present QKP, we need to perform the computation up to $\calO(\alpha_j^2)$ because the tidal effects enter the $f_{kv}$ at LO. We start by writing the Fourier series of the function $f(l)= u(l)-l$. This function is continuous on $[-\pi,\pi]$ and $2\pi$-periodic, so in virtue of the Dirichlet theorem, its value for a fixed $l\in[-\pi,\pi]$ is equal to its Fourier series. Furthermore, it is an odd function, leading to the coefficients of $\cos(kl)$ in the series to vanish. Thus, we get
\begin{equation}\label{eq:uofl}
u-l = \sum_{k=1}^\infty A_k \sin(kl) \,,
\end{equation}
which defines the Fourier coefficients $A_k$ as
\begin{equation}
A_k = \frac{2}{\pi} \int_0^\pi \dd l \bigl(u(l)-l\bigr)\sin(kl) =\frac{2}{k\pi}\int_0^\pi \dd u \cos\bigl(k \,l(u)\bigr)\,.
\end{equation}
The second equality is obtained by performing an integration by parts and using $u(l=0)=0$ and $u(l=\pi) = \pi$. Then we inject $l(u)$, Eq.~\eqref{eq:lgenju}, in the cosine to obtain
\begin{equation}
A_k = \frac{2}{k\pi}\int_0^\pi \dd u \left[ \cos\Bigl( ku-ke_t\sin(u)\Bigr)\cos\left(k\sum_{j=1}^\infty \alpha_j \sin(ju) \right)-\sin\Bigl( ku-ke_t\sin(u)\Bigr)\sin\left(k\sum_{j=1}^\infty \alpha_j \sin(ju) \right)\right]\,.
\end{equation}
The cosine and sine of the series are expanded to $\calO\bigl(\alpha_j^2\bigr)$ as\footnote{We used the Cauchy product
\begin{equation*}
\left( \sum_{i=1}^\infty a_i \right) .  \left( \sum_{j=1}^\infty b_j \right) = \sum_{k=1}^\infty c_k\,, \qquad \text{with} \qquad c_k = \sum_{\ell=1}^k a_\ell b_{k-\ell+1}\,.
\end{equation*}}
\begin{subequations}
\begin{align}
\cos\left(k\sum_{j=1}^\infty \alpha_j \sin(ju) \right) & = 1-\frac{k^2}{2}\left( \sum_{j=1}^\infty \alpha_j \sin(ju) \right)^2 + \calO\bigl(\alpha_j^4\bigr)\nn\\
& = 1-\frac{k^2}{2} \sum_{j=1}^\infty \sum_{i=1}^j \frac{\alpha_i \alpha_{j-i+1}}{2} \Bigl[ \cos\bigl( (2i-j-1)u \bigr) - \cos\bigl((j+1) u\bigr)\Bigr]+\calO\bigl(\alpha_j^4\bigr)\,,\\
\sin\left(k\sum_{j=1}^\infty \alpha_j \sin(ju) \right) &= k \sum_{j=1}^\infty \alpha_j \sin(ju)+ \calO\bigl(\alpha_j^3\bigr)\,.
\end{align}
\end{subequations}
Finally, by injecting these expansions in $A_k$ and using the Bessel functions of the first kind defined as
\begin{equation}
J_n(x) = \frac{1}{\pi}\int_0^\pi \dd\tau \cos(n\tau -x\sin\tau)\,,
\end{equation}
we find that $A_k$ reads at quadratic order in $\alpha_j$
\begin{align}\label{eq:Akvalue}
A_k =& \, \frac{2}{k} J_k(k e_t) + \sum_{j=1}^\infty\alpha_j\Bigl(J_{k+j}(k e_t) - J_{k-j}(k e_t)\Bigr)\nn\\
& + \frac{k}{4} \sum_{j=1}^\infty\sum_{i=1}^j \alpha_i \alpha_{j-i+1}\Bigl(J_{k+j+1}(k e_t) + J_{k-j-1}(k e_t) - J_{k+2i-j-1}(k e_t) - J_{k-2i+j+1}(k e_t)  \Bigr) + \calO\bigl(\alpha_j^3\bigr)\,.
\end{align}
We now have everything at hand to derive $u(l)$ and consequently $(r,\rd,\phi,\fid)$ as functions of $(x,\et,l)$ up to a given eccentricity order. In principle, one should rewrite this Fourier series as a power series in $\et$ (see discussion below), however it is not required in the present case since $J_n(k\et) = \calO\bigl(\et^n\bigr)$. Thus, one can neglect the terms above a certain value of the index summation $j_\text{max}$ and perform an eccentricity expansion. 

\subsubsection{Comments on the convergence of this series}

We would like to point-out a subtle but important feature of this inversion. To this end, we will restrain ourselves to the Newtonian case in the absence of tidal effects, \textit{i.e.} $\alpha_j=0$. We obviously recover the well-known result
\begin{equation}\label{eq:invKepler}
u - l = \sum_{k=1}^\infty \frac{2}{k} J_k(ke)\sin(kl)\,.
\end{equation}
This Fourier series is convergent on the interval $[-\pi,\pi]$ for any $e<1$. However, it is not convenient for waveform modeling purposes because it an infinite sum of non-elementary functions. Thus, in practice, we perform an eccentricity expansion of~\eqref{eq:invKepler}, which is equivalent to express this Fourier series as a power series in $e$ and ignoring terms above a certain power of $e$. To do so, we can use the expression of the Bessel function of the first kind as a hypergeometric function. Hence, the Fourier series~\eqref{eq:invKepler} can be recast (after rearrangement of the summation indices) as the following power series in $e$
\begin{equation}\label{eq:ulJ}
u - l = \sum_{n=1}^\infty \frac{e^n}{n!} \sum_{k=0}^{\lfloor n/2\rfloor}(-)^k\binom{n}{k}\left(\frac{n}{2}-k\right)^{n-1} \sin\bigl((n -2k) l\bigr)\,.
\end{equation}
Equivalently, one can use the Lagrange inversion theorem~\cite{lagrange1770} (which has been independently discovered by Bürmann~\cite{lagrange_legendre1799})
, as done by Laplace~\cite{Laplace1825}, 
resulting in
\begin{equation}\label{eq:ulLap}
u = l + \sum_{n=1}^\infty \frac{e^n}{n!} \frac{\dd^{n-1}}{\dd l^{n-1}}\Bigl( \sin^{n} l  \Bigr)\,.
\end{equation}
Both power series~\eqref{eq:ulJ} and~\eqref{eq:ulLap} are equal. The crucial point here is that they are not convergent for any $e<1$ on the interval $[-\pi,\pi]$. As shown in~\cite{Laplace1825,tisserand1889traité}, the maximum eccentricity $e_\text{max}$ at which this power series is convergent, also called the Laplace limit (easily obtained using the Lagrange theorem), is given by $e_\text{max} = \tfrac{\rho_0}{\cosh \rho_0}$ where $\rho_0$ is the only real positive root of the equation $\cosh\rho_0 - \rho_0 \sinh \rho_0 = 0$. It takes the numerical value of $e_\text{max} \simeq 0.6627434$. Beyond this value of the Keplerian eccentricity, the series~\eqref{eq:ulJ} is not convergent for all $l\in[-\pi,\pi]$ anymore, making $u(l)$ discontinuous.

Although this result has been known for a long time (at least since 1825), we thought it would be good to recall this property. Indeed, the practical computations that we perform uses the eccentricity-truncated expression of~\eqref{eq:uofl} (especially at the level of the GW amplitude modes), which is equivalent to taking the power series~\eqref{eq:ulJ} and ignoring terms beyond a certain power of $e$. Therefore, both eccentricity-expanded Newtonian and PN results cannot be taken above $e_\text{max}$. Of course, the closer the value of the eccentricity is to $e_\text{max}$, the slower the series converges. For a fixed value of the eccentricity, the truncation order depends on the precision level one wants to achieve in the inversion. In the present work, one should compute as well the radius of convergence of the power series in $\et$ of the complete result~\eqref{eq:uofl}, however we have not tried to tackle this problem.

\subsubsection{Results}

For a better accuracy of Phenom waveform models, explained in Paper~II~\cite{paperII}, the overall goal of this project is to provide the observables up to $\calO(\et^{12})$. To have a consistent eccentricity expanded waveform amplitude, one needs to expand the dynamics up to the next eccentricity order due to a division by $\et$ in some intermediate steps. Thus, $u(l)$ has to be expanded up to $\calO(\et^{14})$. To this end, we inserted~\eqref{eq:Akvalue} in~\eqref{eq:uofl} and performed PN and eccentricity expansions in order to obtain $u(x,e_t,l)$ at NNLO and up to $\calO(\et^{14})$. We display here its expression at LO to $\calO(\et^3)$
\begin{align}
u(l) =& \, l + \et\sin(l)\left( 1-\frac{\et^2}{8} -\mutp x^5\left( 30 +\frac{747}{8}\et^2 \right) \right) + \frac{\et^2}{2}\sin(2l)\left( 1 - \frac{165}{2} \mutp x^5 \right) \nn \\
&  + \frac{\et^3}{8}\sin(3l)\left( 3 - 447 \mutp x^5 \right) + \calO\left(\frac{1}{c^2},\frac{\etidal}{c^2}\right) + \calO\bigl(\et^4\bigr)\,.
\end{align}
To check the full result, we injected it in Eq.~\eqref{eq:QKl}, expressed it in terms of $(x,e_t,u)$, and verified that we recover $l + \calO(1/c^6,\etidal/c^6) +\calO(\et^{16})$. In the end, we replace $u(l)$ in Eqs.~\eqref{eq:QKPxeu} and perform PN and eccentricity expansions which lead to the expressions of $(r,\rd,\phi,\fid)$ as functions of $(x,\et,l)$ at NNLO and $\calO\bigl(\et^{14}\bigr)$. Their LO expressions read
\begin{subequations}\label{eq:QKPxel}
\begin{align}
r (x,\et,l) &= \frac{G\tmass}{x \,c^2}\biggl[ 1 + \frac{\et^2}{2} + \mutp x^5\left( 6+93\et^2\right) -\et \cos(l) \left( 1 - \frac{3}{8}\et^2 + \mutp x^5\left( 36 + \frac{1857}{8}\et^2 \right) \right)\nn \\
& \qquad \qquad -\frac{\et^2}{2} \cos(2l) \left( 1 + 6 \mutp x^5 \right) - \frac{\et^3}{8} \cos(3l) \left( 3 - 117 \mutp x^5 \right) \biggr]+ \calO\left(\frac{1}{c^2},\frac{\etidal}{c^2}\right) + \calO\bigl(\et^4\bigr)\,,\\
\rd (x,\et,l) &=c\sqrt{x}\biggl[ \et \sin(l) \left( 1 - \frac{3}{8}\et^2 + \mutp x^5\left( -9 - \frac{87}{2}\et^2 \right) \right) + \et^2 \sin(2l) \left(1 - 39 \mutp x^5 \right)\nn \\
& \qquad \qquad  + \frac{\et^3}{8} \sin(3l) \left( 9 - 756 \mutp x^5  \right) \biggr]+ \calO\left(\frac{1}{c^2},\frac{\etidal}{c^2}\right) + \calO\bigl(\et^4\bigr) \,,\\
\fid (x,\et,l) &=\frac{x^{3/2}c^3}{G\tmass}\biggl[ 1 +\et \cos(l) \left( 2 - \frac{\et^2}{4} + \mutp x^5\left( 60 + \frac{1215}{4}\et^2 \right) \right) + \frac{\et^2}{2} \cos(2l) \left( 5 + 180 \mutp x^5  \right) \nn \\
& \qquad \qquad  + \frac{\et^3}{4} \cos(3l) \left( 13 + 405 \mutp x^5 \right) \biggr]+ \calO\left(\frac{1}{c^2},\frac{\etidal}{c^2}\right) + \calO\bigl(\et^4\bigr)\,.
\end{align}
\end{subequations}
The phase $\phi$ is treated separately as it can be decomposed into a  rapidly oscillating part $W(l)$ and a slowly growing part $\lambda = K l$, \textit{i.e.}
\begin{equation}\label{eq:phixetl}
\phi (x,\et,l) -\phi_0 = \lambda + W(l) \,,
\end{equation}
where $W$ expanded in eccentricity reads
\begin{align}
W(l) =& \, \et\sin(l)\left( 2 - \frac{\et^2}{4} + \mutp x^5\left( 150 +\frac{1755}{2}\et^2 \right) \right)  + \frac{\et^2}{4} \sin(2l)\left( 5 + 405 \mutp x^5 \right)\nn \\
& + \frac{\et^3}{12} \sin(3l)\left( 13 + 990 \mutp x^5\right) + \calO\left(\frac{1}{c^2},\frac{\etidal}{c^2}\right) + \calO\bigl(\et^4\bigr)\,.
\end{align}
The full results are provided in the ancillary file~\cite{SuppMaterial1}.

The problem of the conservative dynamics is now solved and we have everything at hand to compute the full waveform at NNLO. However, we aim at computing the observables at relative 2.5PN. There is no half-PN contributions of the conservative problem (at this order) but we need to include radiation effects to the dynamics in order to be consistent at the relative second-and-a-half PN order. We tackle this problem in the following Section.


\section{Radiation reaction dynamics}\label{sec:rad}


In the absence of energy and angular momentum loss, a binary system on a quasi-elliptic orbit evolves over two time scales: the orbital period and the periastron precession period. When considering GW radiation, the additional time scale of the reaction to the radiation.
In~\cite{Dones:2024odv}, we derived the 2.5PN beyond LO term in the acceleration for an arbitrary motion of the isolated binary. This contribution is called the radiation reaction term as it contains the information of the GW emission, resulting in the time-dependency of the orbital elements and in particular the shrinking of the separation. The relative acceleration (neglecting NLO terms) takes the form
\begin{equation}\label{eq:aRR}
\bm{a} = -\frac{G\tmass}{r^2}\left( 1 + 18 \frac{G \,\mu^{(2)}_+}{r^5} \right)\bm{n} + \frac{\bm{a}_\text{RR}}{c^5}\,.
\end{equation}
These contributions to the EOM are the ones that will dictate the time evolution of the orbital elements that we wish to calculate.


\subsection{The method}\label{subsec:method}


One method to take into account the radiation reaction is to use the variation of the constants on these conserved quantities and allow a time dependency on them. These will be seen as small perturbations of the conservative dynamics, as they are subdominant within the PN series (2.5PN beyond LO). More precisely, we need to impose this variation on four constants and make use of a time-scale separation. This method has been employed for point-particles at 2.5 and 3.5PN~\cite{Damour:2004bz,Konigsdorffer:2006zt,Boetzel:2019nfw} and we adopt the same notations. One needs to make a choice of the time-varying variables; in~\cite{Damour:2004bz,Konigsdorffer:2006zt}, the authors use the conserved energy and angular momentum or the mean motion $n$ and time eccentricity $\et$. Here we adopt the choice of~\cite{Boetzel:2019nfw} by choosing the four variables $\{x,\et,l,\lambda\}$, which are assumed to be independent and time-dependent. In particular, the two angles $l$ and $\lambda$ are now given by
\begin{subequations} \label{eq:llambda}
\begin{align}
l(t) &= \int_{t_0}^t n(t')\,\dd t' + c_l(t)\,,\\
\lambda(t) &= \int_{t_0}^t K(t')\,n(t')\, \dd t' + c_\lambda(t)\,,
\end{align}
\end{subequations}
where $c_l$ and $c_\lambda$ are two time-dependent constants of integration. We recall that the angles $l$ and $\lambda$ evolve over two separate time-scales as $l$ corresponds to the rapidly oscillating orbital angle and $\lambda$ the slowly secularly evolving part of the phase. The goal is to determine the time evolution of $\{x,\et,c_l,c_\lambda\}$. The first step consists in splitting their secular and oscillatory contributions as
\begin{equation}
c_\alpha(l) = \Bar{c}_\alpha(l) + \tilde{c}_\alpha(l)\,,
\end{equation}
where $c_\alpha \in \{x,\et,c_l,c_\lambda\}$. Furthermore, we impose that $\tilde{c}_\alpha$'s average over one orbit vanishes. Next, as discussed in~\Cref{subsec:application}, the differential equations followed by $c_\alpha$ are found by enforcing the now time-varying `constants' to verify the EOM~\eqref{eq:aRR}. They take the symbolic form
\begin{equation}
\frac{\dd c_\alpha}{\dd t} = F_\alpha(c_a,l)\,,
\end{equation}
where it is important to keep in mind that the functions $F_\alpha$ are at the radiation reaction PN order, \textit{i.e.} 2.5PN beyond~LO. This means that the differential equations over $l$ read
\begin{equation}
\frac{\dd c_\alpha}{\dd l} = \frac{F_\alpha(c_a,l)}{n(c_a)+F_l(c_a,l)} = \frac{F_\alpha(c_a,l)}{n(c_a)} + \calO\left(\frac{1}{c^{10}},\frac{\etidal}{c^{10}}\right)  \equiv G_\alpha(c_a,l)\,.
\end{equation}
Finally, we split the contributions of the secular and oscillating parts of $c_\alpha$, which are
\begin{subequations}
\begin{align}
\frac{\dd \bar{c}_\alpha}{\dd l} &= \bar{G}_\alpha(\bar{c}_a) \equiv \langle G_\alpha \rangle\,,\\
\frac{\dd \tilde{c}_\alpha}{\dd l} &= \tilde{G}_\alpha(\bar{c}_a,l) = G_\alpha(\bar{c}_a,l) - \bar{G}_\alpha(\bar{c}_a)\label{eq:dctildealpha}\,,
\end{align}
\end{subequations}
where $\langle . \rangle$ refers to the orbit average operation defined in~\eqref{eq:orbavgdef}. The subtraction in~\eqref{eq:dctildealpha} comes from the fact that $\tilde{c}_\alpha$ has been defined to have a 0 orbit average, while $\tilde{c}_a$ does not appear as it comes in at higher order. Once the integrand is known, it is trivial to integrate since $(\bar{x},\bar{e}_t)$ are constant at this order. Thus, it amounts to simply find the primitive:
\begin{equation}\label{eq:ctildealphagen}
\tilde{c}_\alpha = \int \dd l \, \tilde{G}_\alpha(\bar{c}_a,l)\,,
\end{equation}
which is uniquely defined as we supposed a vanishing orbit average. Now, we apply this general method to the QKP~\eqref{eq:QKgen}.


\subsection{Application to the problem}\label{subsec:application}


In this section, we adapt the above method to the tidal effects problem. Let us denote $S$ and $W$ the functionals of $(x,\et,l)$ of the separation and the phase. As nicely outlined in~\cite{Damour:2004bz}, we consider the conservative 2PN QKP equations~\eqref{eq:QKgen} as the non-perturbed state, and account for the radiation reaction as a perturbation on these. They adopt the following functional form
\begin{subequations}  \label{eq:QKPfunct}
\begin{align}
r = S(x,\et,l) \quad &; \quad \rd = n\frac{\partial S}{\partial l}(x,\et,l)\,,\\
\phi = \lambda + W(x,\et,l) \quad &; \quad \fid = K n + n\frac{\partial W}{\partial l}(x,\et,l)\,.
\end{align}
\end{subequations}
We now denote the functionals of the eccentric and true anomalies with capital letters $U$ and $V$ to emphasize that they depend on $(x,\et,l)$. The functionals $S$ and $W$, expressed using~\eqref{eq:QKr} and~\eqref{eq:QKphi} (together with~\eqref{eq:phixetl}), read
\begin{align}
S(x,\et,l) &= a_r(x,\et) \Bigl[ 1- e_r(x,\et)\cos\bigl( U(x,\et,l) \bigr) \Bigr]\label{eq:Sdef}\,,\\
W(x,\et,l) &= K(x,\et)\left[ V\Bigl( e_\phi(x,\et) ,U(x,\et,l)\Bigr) -l + \sum_{k=2}^8 g_{kv}(x,\et)\sin\left(k V\Bigl( e_\phi(x,\et) ,U(x,\et,l)\Bigr)\right) \right]\label{eq:Wdef}\,.
\end{align}
Next, we account for the radiation reaction by keeping the same structure of the QKP equations as in the non-perturbed format~\eqref{eq:QKPfunct} but allow $x$ and $e_t$ to be functions of time (seen as a perturbation). 
We also now have time-dependent `constants' of integration from $\dot{l}$ and $\dot{\lambda}$ as illustrated in~\eqref{eq:llambda}. This allows us to produce the four time-differential equations~\cite{Damour:2004bz}, 
\begin{subequations}\label{eq:xeclclambdasource}
\begin{align}
\frac{\dd x}{\dd t} &= \frac{\partial x(\bm{x},\bm{v})}{\partial v^i} a_\text{RR}^i\,,\label{eq:dxdtRR}\\
\frac{\dd \et}{\dd t} &= \frac{\partial \et(\bm{x},\bm{v})}{\partial v^i} a_\text{RR}^i\label{eq:detdtRR}\,,\\
\frac{\dd c_l}{\dd t} &= - \left(\frac{\partial S}{\partial l} \right)^{-1} \left( \frac{\partial S}{\partial x}\frac{\dd x}{\dd t} +\frac{\partial S}{\partial \et}\frac{\dd \et}{\dd t} \right) \label{eq:dcldtRR} \,,\\
\frac{\dd c_\lambda}{\dd t} &= -\frac{\partial W}{\partial x}\frac{\dd x}{\dd t} -\frac{\partial W}{\partial \et}\frac{\dd \et}{\dd t} -\frac{\partial W}{\partial l}\frac{\dd c_l}{\dd t} \label{eq:dclambdadtRR}\,,
\end{align}
\end{subequations}
where $x$ is the adimensional PN parameter defined in~\eqref{eq:QKx} and $\bm{x} = r\bm{n}$ is the separation vector. We have assumed $(x,\et,l,\lambda)$ to be independent, so that $\partial l/\partial x = \partial l/\partial \et = 0$. The second two equations are derived by using chain rules on $r$ and $\phi$ of~\eqref{eq:QKPfunct} and equating to their time derivatives. 

The right-hand side of the first two equations, ~\eqref{eq:dxdtRR} and~\eqref{eq:detdtRR}, are calculated by taking the expression of $x$ and $\et$ in terms of $(\Et,h)$, replacing the conserved quantities expressed by their values in terms of the relative separation and velocities, then differentiating with respect to $v^i$ and finally contracting with the radiation reaction part of the acceleration. The right-hand side of the second two equations,~\eqref{eq:dcldtRR} and~\eqref{eq:dclambdadtRR}, are calculated by using chain rules and expressing all of them in terms of the derivatives of $U$; these are obtained by explicitly calculating both $\partial l/\partial x$ and $\partial l/\partial \et$ and equating to zero, obtaining
\begin{subequations}
\begin{align}
\frac{\partial U}{\partial l} &= \left[ 1 - \et\cos u+f_{v-u}\left(\frac{\partial v}{\partial u}-1\right) +\frac{\partial v}{\partial u}\sum_{k=1}^6k f_{kv}\cos (kv) \right]^{-1}\,,\label{eq:dudl}\\
\frac{\partial U}{\partial x} &= - \frac{\partial U}{\partial l}\left[ (v-u)\frac{\partial f_{v-u}}{\partial x} + \frac{\partial v}{\partial e_\phi}\frac{\partial e_\phi}{\partial x}\left( f_{v-u} +\sum_{k=1}^6 kf_{kv}\cos(kv) \right)+\sum_{k=1}^6\frac{\partial f_{kv}}{\partial x}\sin(kv)\right]\,,\\
\frac{\partial U}{\partial \et} &= \frac{\partial U}{\partial l}\left[ \sin u -(v-u)\frac{\partial f_{v-u}}{\partial \et} - \frac{\partial v}{\partial e_\phi}\frac{\partial e_\phi}{\partial \et}\left( f_{v-u} +\sum_{k=1}^6 kf_{kv}\cos(kv) \right)-\sum_{k=1}^6\frac{\partial f_{kv}}{\partial \et}\sin(kv)\right]\,.
\end{align}
\end{subequations}
In the point-particle case up to NLO, they reduce to $\partial U/\partial l = (1-\et\cos u)^{-1}$, $\partial U/\partial \et = \sin u/(1-\et\cos u)$ and $\partial U/\partial x = 0$, which is significantly simpler than the tidal LO. Indeed, in the adopted parametrization, $f_{v-u}$, $f_{v}$ and $f_{2v}$ contribute to the LO, which complicates the later computations. Furthermore, in the definition of $W$ appears~$V$ which is a functional of $e_\phi $ and $U$, thus we need the following relations as well for $A =\{x,\et,l \}$
\begin{equation}
\frac{\partial V}{\partial A} = \frac{\partial v}{\partial e_\phi}\frac{\partial e_\phi}{\partial A} + \frac{\partial v}{\partial u} \frac{\partial U}{\partial A}\,,
\end{equation}
with
\begin{equation}
\frac{\partial v}{\partial e_\phi} = \frac{\sin u}{\sqrt{1-e_\phi^2}(1-e_\phi \cos u)}\,, \qquad \text{and}\qquad 
\frac{\partial v}{\partial u} = \frac{\sqrt{1-e_\phi^2}}{1-e_\phi\cos u}\,.
\end{equation}

All of the above expressions combined lead to differential equations for $c_\alpha=(x,\et)$ of the general form
\begin{equation}\label{eq:dcalphadtnovmu}
\frac{\dd c_\alpha}{\dd t} = \frac{c^3 x^4}{G\tmass} \sum_k \frac{\lambda_k + \mu_k \, x^5}{(1-\et \cos u)^k} \,,
\end{equation}
where $\lambda_k$ and $\mu_k$ are functions of the time eccentricity, the masses and $\mu_k$ also contain the tidal polarizabilities $\widetilde{\mu}^{(2)}_\pm$ (the power of $x$ in the prefactor of $\dd x/\dd t$ is 5). However, for $c_\alpha = (c_l,c_\lambda)$, the equations contain an additional type of term
\begin{equation}\label{eq:dcalphadtvmu}
\frac{\dd c_\alpha}{\dd t} = \frac{c^3 x^4}{\et G\tmass} \sum_k \left[ \frac{\sin u}{(1-\et \cos u)^k} \bigl(\lambda_k + \mu_k \, x^5\bigr)+   \mu'_k x^5 \frac{v-u}{(1-\et \cos u)^k} \right]\,.
\end{equation}
As one can see, the additional term proportional to $v-u$ is a tidal contribution. It does not appear in the point-particle case and it significantly complicates the computations since the primitives of those terms are not expressible in terms of elementary functions.

\subsection{The secular part}

As explained in~\Cref{subsec:method}, the secular part of Eqs.~\eqref{eq:xeclclambdasource} are obtained by computing the orbit average of their sources, where the average of a given $P$-periodic function $F$ is given by the following integral
\begin{equation}\label{eq:orbavgdef}
\langle F\rangle = \frac{1}{P} \int_0^P \dd t \, F \,.
\end{equation}
In the present case, $F$ is always at least of order $\calO(1/c^{5},\etidal/c^5)$. Thus, since $c_l$ is also of this order, recalling~\eqref{eq:llambda}, one can write $F \, \dd t = F (\dd l - \dd c_l)/n = F \, \dd l/\bar{n} + \calO(c^{-5})$. This leads to
\begin{equation}
\langle F\rangle = \int_0^{2\pi}\frac{\dd l}{2\pi} F(l) = \int_0^{2\pi}\frac{\dd u}{2\pi} \frac{\partial l}{\partial u} F(u)\,.
\end{equation}
Contrary to the point-particle case where $\partial l/\partial u = 1-\et\cos u$, $\partial l/\partial u$ at LO is replaced by~\eqref{eq:dldu} (or equivalently by the inverse of~\eqref{eq:dudl}). In practice, the orbit average integrals that we encounter in this computation are all of the form
\begin{subequations}
\begin{align}
\int_0^{2\pi}\frac{\dd u}{2\pi} \frac{\cos{(ku)}}{(1-\et\cos{u})^n}&=  \frac{(n+k-1)!}{(n-1)!} \beta_t^k \sum_{\ell=0}^{n-1}\frac{1}{2^\ell \ell! (k+\ell)!}\frac{(n+\ell-1)!}{ (n-\ell-1)!}\frac{(1-\sqrt{1-\et^2})^{\ell}}{(1-\et^2)^{(n+\ell)/2}}\,,\label{eq:Ikn}\\
\int_0^{2\pi}\frac{\dd u}{2\pi} \frac{\sin{(ku)}}{(1-\et\cos{u})^n}&= 0 \,,\\
\int_0^{2\pi}\frac{\dd u}{2\pi} \frac{\bigl( v - u \bigr)\cos{(ku)}}{(1-\et\cos{u})^n} &= 0\label{eq:Iknvmu}\,.
\end{align}
\end{subequations}
with $\beta_t = \tfrac{1-\sqrt{1-\et^2}}{\et}$. These formulas are valid for $(k,n) \in \mathbb{N}^2$ except for~\eqref{eq:Ikn} where its value is $\delta_{k0}$ for $n=0$. Note that~\eqref{eq:Ikn} is the closed form of Eq.~(3.8a) of~\cite{Henry:2023tka} and a proof of that expression (together with a generalization containing logarithms) is provided in Appendix~A of Paper II~\cite{paperII}. The last integral~\eqref{eq:Iknvmu} is easily obtained using the odd behaviour of the following function\footnote{The proof of this formula is provided in Appendix A of~\cite{Konigsdorffer:2006zt}.}
\begin{equation}
v-u = 2\arctan\left( \frac{\beta_\phi \sin u}{1-\beta_\phi \cos u} \right)\,,
\end{equation}
where we recall the definition $\beta_\phi = \tfrac{1-\sqrt{1-e_\phi^2}}{e_\phi}$. 

Finally, we apply the orbit-averaging operation to the differential equations verified by the secular part of $(x,\et,c_l,c_\lambda,n,k)$. We find that, just like the point-particle case, $\dd\bar{c}_l/\dd t = \dd\bar{c}_\lambda/\dd t = 0$. Next, the equations satisfied by $\bar{x}$ and $\bar{e}_t$ read
\begin{subequations}\label{eq:xetdotSec}
\begin{align}
\frac{\dd\Bar{x}}{\dd t} =& \, \frac{64 \, x^5\nu c^3}{5G\tmass(1-\et^2)^{7/2}}\left(1+\frac{73}{24} \et^2+\frac{37}{96}\et^4\right)\nn\\
& + \frac{384\, x^{10} c^3}{5G\tmass(1-\et^2)^{17/2}} \Biggl\{\Bigl(\mutp+ \delta\,\mutm\Bigr)\left[ 1 + \frac{211}{8}\et^2+\frac{3369}{32}\et^4+\frac{6275}{64}\et^6+\frac{10355}{512}\et^8 + \frac{225}{512}\et^{10}\right] \nn \\
&  \qquad \qquad \qquad \qquad \qquad + \nu\,\mutp \left[27 + \frac{549}{4}\et^2+\frac{7303}{24}\et^4+\frac{57727}{384}\et^6+\frac{39199}{3072}\et^8 \right. \nn\\
& \qquad \qquad \qquad \qquad \qquad \qquad \left. + \sqrt{1-\et^2}\left( -5 + \frac{1237}{24}\et^2+\frac{3869}{64}\et^4+\frac{1813}{192}\et^6 -\frac{29}{128}\et^8\right) \right] \Biggr\} 
\,,\\
\frac{\dd\Bar{e}_t}{\dd t} =& \, - \frac{304 \, x^4\nu c^3 \et}{15G\tmass(1-\et^2)^{5/2}}\left(1+\frac{121}{304} \et^2\right)\nn \\
& - \frac{2448\, x^9  c^3\et}{5G\tmass(1-\et^2)^{15/2}} \Biggl\{\Bigl(\mutp+ \delta\,\mutm\Bigr) \left[ 1 + \frac{487}{68}\et^2+\frac{1245}{136}\et^4+\frac{2545}{1088}\et^6+\frac{65}{1088}\et^8 \right] \nn \\
&  \qquad \qquad \qquad \qquad \qquad + \nu\, \mutp \left[\frac{479}{102} + \frac{27611}{1224}\et^2+\frac{96463}{4896}\et^4+\frac{100463}{39168}\et^6+\frac{5}{272}\et^8 \right. \nn\\
& \qquad \qquad \qquad \qquad \qquad \qquad \left. + \sqrt{1-\et^2}\left( \frac{550}{153} + \frac{7391}{1224}\et^2+\frac{4057}{4896}\et^4 -\frac{13}{288}\et^6 \right) \right] \Biggr\} 
\,.
\end{align}
\end{subequations}
The variables $x$ and $\et$ in the right-hand side of these expressions should be understood as their secular part $\bar{x}$ and $\bar{e}_t$. We recover the LO expressions of $\langle \dot{x} \rangle$ and $\langle \dot{e}_t \rangle$ derived in Paper II~\cite{paperII} in which we use the energy and angular momentum flux balance equations. These expressions are available to 2.5PN beyond LO including $\langle \dot{k} \rangle$,~$\langle \dot{a}_r \rangle$ and $\langle \dot{n} \rangle$ as well. The system~\eqref{eq:xetdotSec} of two coupled differential equations can be solved analytically using an eccentricity expansion, see \textit{e.g.}~\cite{Sridhar:2024zms}, however its resolution is far from being trivial and is left for future works.

\subsection{The oscillatory part}

We now turn to the oscillatory parts of $c_\alpha = (x,\et,c_l,c_\lambda)$; starting from~\eqref{eq:dctildealpha}, as the sources~\eqref{eq:dcalphadtnovmu} and \eqref{eq:dcalphadtvmu} are written in terms of $u$, one can integrate over this variable through
\begin{equation} \label{eq:ctilde}
\tilde{c}_\alpha = \int \dd u \frac{\partial l}{\partial u} \, \tilde{G}_\alpha(\bar{c}_a,u)\,.
\end{equation}
For $c_\alpha = (x,\et)$, these can be easily computed using the reader's favorite book of integrals. However, the presence of $v-u$ in~\eqref{eq:dcalphadtvmu} makes it more complicated to integrate for $c_\alpha = (c_l,c_\lambda)$. Note that Refs.~\cite{Damour:2004bz,Konigsdorffer:2006zt,Boetzel:2019nfw}, which tackle the point-particle case, also encounter such integrals at the later stage of $\tilde{l}$ and $\tilde{\lambda}$ (see~\eqref{eq:llambdatildegen} below) but leave them in a symbolic form. Just as Ref.~\cite{Boetzel:2019nfw}, we replace $u$ by $u(l)$~\eqref{eq:uofl}  and perform an eccentricity expansion up to $\calO(\et^{14})$ before integrating. Note that we need to make sure that $\tilde{c}_\alpha$ is 0 under the orbit average operation. As explained in~\cite{Boetzel:2019nfw}, one needs to manually add a term to $\tilde{c}_l$ and $\tilde{c}_\lambda$ to ensure this condition. 
Once $\tilde{c}_l$ and $\tilde{c}_\lambda$ are known, we turn to the integration of $\tilde{l}$ and $\tilde{\lambda}$ using
\begin{subequations}\label{eq:llambdatildegen}
\begin{align}
\tilde{l}(\bar{l}) &= \int \frac{\tilde{n}}{\bar{n}}\dd l + \tilde{c}_l(\bar{l})\,,\\
\tilde{\lambda}(\bar{l}) &= \int \left[\bar{K}\frac{\tilde{n}}{\bar{n}} + \tilde{k}\right]\dd l + \tilde{c}_\lambda(\bar{l})\,,
\end{align}
\end{subequations}
where $K = 1+k$. The time-varying quantities $n$ and $K$ are computed exactly in the same way as $x$ and $\et$. Finally, the oscillatory parts of $(x,\et,l,\lambda)$ read
\begin{subequations}
\small
\begin{align}
\tilde{x} &=  \nu x^{7/2}\Biggl\{ \et \sin(l)\left[ 80 +\frac{4538}{15}\et^2 + x^5\left(\frac{\mutp+\delta \mutm}{\nu}\left(834 +\frac{346899}{20}\et^2\right) + \mutp\left(13872+\frac{901252}{5}\et^2\right)\right) \right] \nn\\
& \qquad + \et^2 \sin(2l)\left[ \frac{1436}{15}  + x^5\left(\frac{3012}{5}\frac{\mutp+\delta \mutm}{\nu} +  \frac{73324}{5}\mutp\right) \right] \nn\\
& \qquad + \et^3 \sin(3l)\left[ \frac{6022}{45}  + x^5\left(-\frac{52113}{20}\frac{\mutp+\delta \mutm}{\nu} +  \frac{209468}{15}\mutp\right) \right] \Biggr\} + \calO\bigl(\et^4\bigr)\,,\\
\tilde{e}_t &=  -\nu x^{5/2}\Biggl\{ \sin(l)\left[ \frac{64}{5} +\frac{1138}{15}\et^2 + x^5\left(\frac{\mutp+\delta \mutm}{\nu}\left(\frac{384}{5} +\frac{75789}{20}\et^2\right) + \mutp\left(\frac{7488}{5}+\frac{168442}{5}\et^2\right)\right) \right] \nn\\
& \qquad + \et \sin(2l)\left[ \frac{352}{15} +\frac{842}{15}\et^2 + x^5\left(\frac{\mutp+\delta \mutm}{\nu}\left(\frac{669}{5} +\frac{24164}{5}\et^2\right) + \mutp\left(\frac{16108}{5}+\frac{208598}{5}\et^2\right)\right) \right] \nn\\
& \qquad + \et^2 \sin(3l)\left[ \frac{358}{9}  + x^5\left(-\frac{987}{4}\frac{\mutp+\delta \mutm}{\nu} +  \frac{15358}{3}\mutp\right) \right] \nn\\
& \qquad + \et^3 \sin(4l)\left[ \frac{1289}{20}  + x^5\left(-\frac{13999}{5}\frac{\mutp+\delta \mutm}{\nu} +  \frac{22369}{4}\mutp\right) \right] \Biggr\} + \calO\bigl(\et^4\bigr)\,,\\
\tilde{l} &=  -\frac{\nu x^{5/2}}{\et}\Biggl\{ \cos(l)\left[ \frac{64}{5} +\frac{1654}{15}\et^2 + x^5\left(\frac{\mutp+\delta \mutm}{\nu}\left(\frac{384}{5} +\frac{52407}{20}\et^2\right) + \mutp\left(\frac{7488}{5}+\frac{140606}{5}\et^2\right)\right) \right] \nn\\
& \qquad + \et \cos(2l)\left[ \frac{352}{15} +\frac{694}{15}\et^2 + x^5\left(\frac{\mutp+\delta \mutm}{\nu}\left(\frac{669}{5} +\frac{26098}{5}\et^2\right) + \mutp\left(\frac{16108}{5}+\frac{206926}{5}\et^2\right)\right) \right] \nn\\
& \qquad + \et^2 \cos(3l)\left[ \frac{358}{9} + x^5\left(-\frac{987}{4}\frac{\mutp+\delta \mutm}{\nu} +  \frac{15358}{3}\mutp\right) \right] \nn\\
& \qquad + \et^3 \cos(4l)\left[ \frac{1289}{20} + x^5\left(-\frac{13999}{5}\frac{\mutp+\delta \mutm}{\nu} +  \frac{22369}{4}\mutp\right) \right] + \calO\bigl(\et^4\bigr) \Biggr\} \,,\\
\tilde{\lambda} &=  -\nu x^{5/2}\Biggl\{ \et \cos(l)\left[ \frac{296}{3} +\frac{2188}{5}\et^2 + x^5\left(\frac{\mutp+\delta \mutm}{\nu}\left(1887 +\frac{600057}{20}\et^2\right) + \mutp\left(16000+\frac{1236281}{5}\et^2\right)\right) \right] \nn\\
& \qquad + \et^2 \cos(2l)\left[ \frac{199}{5} + x^5\left(\frac{31293}{10}\frac{\mutp+\delta \mutm}{\nu} + \frac{2678}{5}\mutp\right) \right] \nn\\
& \qquad + \et^3 \cos(3l)\left[ \frac{176}{9}  + x^5\left(\frac{31285}{4}\frac{\mutp+\delta \mutm}{\nu} -5857\mutp\right) \right] \Biggr\} + \calO\bigl(\et^4\bigr)\,.
\end{align}
\end{subequations}
As in Eqs.~\eqref{eq:xetdotSec}, one should understand $x$, $\et$ and $l$ as their secular part $\bar{x}$, $\bar{e}_t$ and $\bar{l}$. Notice the presence of a division by $\et$ in $\tilde{l}$; this is due to the fact that $\tilde{c}_l$ has a division by $\et$. For this reason, the ancillary file contains the expressions of $(\tilde{x},\tilde{e}_t)$ to $\calO(\et^{14})$ and $(\tilde{l},\tilde{\lambda})$ to $\calO(\et^{13})$. In verifying these results, the point-particle part is in agreement with~\cite{Boetzel:2019nfw} up to $\calO(\et^5)$\footnote{We find disagreement with the sixth order of that Ref. due to the division by the eccentricity, which requires one to push the expansion at the next non-vanishing order (their expressions terminate at this sixth order).}. This completes the computation of the full dynamics at relative 2.5PN, required to compute the waveform at the same PN order, which we provide in Paper II~\cite{paperII}. 

\section{Conclusion}

We solved the problem of the motion of compact binaries in general relativity including adiabatic tides to the second-and-a-half post-Newtonian order beyond leading order in standard harmonic coordinates. We employed a quasi-Keplerian parametrization consistent with previous results for point particles and spins. These results include the conservative motion at next-to-next-to-leading order, as well as the radiation reaction at leading order. The tidal mass quadrupole, current quadrupole and mass octupole interactions are taken into account. More precisely, we started from the conserved energy and angular momentum, which allowed us to derive the QKP in terms of the conserved quantities, then expressed the separation, phase and their derivatives in terms of the orbital frequency, the time eccentricity and the eccentric anomaly. We inverted the generalized Kepler equation to obtain expressions in terms of the mean anomaly at the cost of an eccentricity expansion. Finally, we included the effect of GW radiation on the dynamics, which needs to be added to the conservative motion. All relevant results of the present paper are provided in the supplementary \textit{Mathematica} file~\cite{SuppMaterial1} which contains: 
\begin{itemize}
\item $(\Et,\Lt)$ in terms of $(r,\rd,\fid)$,
\item the QKP in terms of the conserved quantities at NNLO,
\item $(\Et,h,K,n,a_r)$ in terms of $(x,\et)$ at NNLO,
\item $(r,\rd,\phi,\fid)$ in terms of $(x,\et,u)$ at NNLO,
\item $(u,r,\rd,\fid,W)$ in terms of $(x,\et,l)$ at $\calO \bigl( \et^{14} \bigr)$ and NNLO,
\item $(\tilde{x},\tilde{e}_t)$ at $\calO \bigl( \et^{14} \bigr)$ and $(\tilde{\lambda},\tilde{l})$ at $\calO \bigl( \et^{13} \bigr)$ at leading post-adiabatic order.
\end{itemize}

The results of the present paper are used in Paper~II which provides the radiated energy and angular momentum fluxes, the secular evolution of the orbital elements, and the amplitude modes of the GW all consistent to the relative 2.5PN order. These results are both theoretical and practical as they aim at being readily included in semi-analytical waveform models such as Phenom and EOB. Detections of NSBH or BNS systems on eccentric orbits within the LVK detectors band, as well as Einstein Telescope and LISA could require such results for precise data analysis, fundamental physics and cosmology. They also constitute a first step towards a better PN description of the dynamical tides and a more complete modelling of compact binaries.

\section*{Acknowledgments}

We are grateful to Tim Dietrich, Guillaume Faye, François Larrouturou and Antoni Ramos-Buades for useful discussions. AH is supported by grant PD-034-2023 co-financed by the Govern Balear and the European Social Fund Plus (ESF+) 2021-2027. This work was supported by the Universitat de les Illes Balears (UIB); the Spanish Agencia Estatal de Investigación grants PID2022-138626NB-I00, RED2024-153978-E, RED2024-153735-E, funded by MICIU/AEI/10.13039/501100011033 and the ERDF/EU; and the Comunitat Autònoma de les Illes Balears through the Conselleria d'Educació i Universitats with funds from the European Union - NextGenerationEU/PRTR-C17.I1 (SINCO2022/6719) and from the European Union - European Regional Development Fund (ERDF) (SINCO2022/18146).

\appendix

\section{Kernel integrals for the QKP}\label{app:In}

In order to derive the QKP, and more precisely Eqs.~\eqref{eq:phimphi0} and~\eqref{eq:tmt0}, one needs to evaluate the following integrals for $n\in \mathbb{N}$,
\begin{equation}\label{eq:Indef}
I_n = \int_s^{s_+} \frac{x^{n-2}}{\sqrt{(s_+ - x)(x-s_-)}} \dd x\,,
\end{equation}
where $s_- \leq s \leq s_+$. In the present case, the degrees of the polynomials $\mathcal{P}$ and $\mathcal{Q}$, which are respectively 10 and 8, forces us to go up to $n=10$. In this section we derive the general formula for $I_n$. We can rewrite $I_n$, by performing the change of variables $y=\tfrac{s_+ -x}{s_+ - s}$, as
\begin{equation}
I_n = \sqrt{b} \, s_+^{n-2}\int_0^1 \dd y \, y^{-1/2}(1-ay)^{-(2-n)}(1-by)^{-1/2}\,,
\end{equation}
with $a=\tfrac{s_+ - s}{s_+}$ and $b = \tfrac{s_+ - s}{s_+ - s_-}$. It can be integrated using Eq. (3.211), p.318 of~\cite{gradshteyn2007}\footnote{
The Appell series of two variables $F_1$ is defined as
\begin{equation*}
F_1 (\alpha,\beta,\beta';\gamma;x,y) = \sum_{i,j=0}^\infty \frac{(\alpha)_{i+j} (\beta)_i (\beta')_j} {(\gamma)_{i+j} \,i! \,j!} \,x^i y^j\,,
\end{equation*}
where $(x)_k = \tfrac{\Gamma(x+k)}{\Gamma(x)}$ is the Pochhammer symbol.}
\begin{equation}\label{eq:Ingennosimp}
I_{n} = 2 \sqrt{b} \,s_+^{n-2} F_1\left( \frac{1}{2}, 2-n,\frac{1}{2};\frac{3}{2};a,b \right)\,.
\end{equation}
In order to simplify~\eqref{eq:Ingennosimp}, let us first notice that $(x)_{i+j} = (x+i)_j (x)_i$. Then, the Appell series containing a negative integer argument turns into the following finite sum\footnote{The hypergeometric function ${}_2 F_1$ is defined as
\begin{equation*}
{}_2 F_1(\alpha,\beta; \gamma;z) = \sum_{k=0}^\infty \frac{(\alpha)_k(\beta)_k}{(\gamma)_k}\frac{z^k}{k!}\,.
\end{equation*}}
\begin{equation}\label{eq:F1to2F1}
\forall m\in \mathbb{N}, \quad F_1 (\alpha,-m,\beta';\gamma;x,y) = \sum_{i=0}^m \binom{m}{i}\frac{(\alpha)_i}{(\gamma)_i}(-x)^i \, {}_2 F_1\left( \beta',\alpha+ i ;\gamma + i;y \right)\,,
\end{equation}
because of the following property of the Pochhammer symbol of a negative integer
\begin{equation}\label{eq:negPochh}
(-m)_i = \begin{cases}
    (-)^i\frac{m!}{(m-i)!} & \text{if $i \leq m$} \\
    0 & \text{otherwise}
  \end{cases}\,.
\end{equation}
Note that formula~\eqref{eq:F1to2F1} is also valid for $m=0$. At this stage, we need to differentiate several cases: $n\in \{0,1\}$ and $n\geq 2$. Let us first focus on $I_0$, it reads
\begin{align}
I_0 &= \frac{2 \sqrt{b}}{s_+^2} F_1\left( \frac{1}{2}, 2,\frac{1}{2};\frac{3}{2};a,b \right)\,,\\
&=\frac{2}{s_+^2}\sqrt{\frac{b}{1-a}}F_1\left( \frac{1}{2}, -1,\frac{1}{2};\frac{3}{2};\frac{a}{a-1},\frac{b-a}{1-a} \right)\,,\label{eq:I0line2}
\end{align}
where~\eqref{eq:I0line2} is obtained using line two of Eq. (9.183.1), p.1020 of~\cite{gradshteyn2007}. We can thus apply~\eqref{eq:F1to2F1} for $m=1$ and, as $a$ and $b$ can be expressed in terms of $a_r$, $e_r$ and $u$ using Eqs.~\eqref{eq:r} and~\eqref{eq:spm}, we find (after some simplification)
\begin{equation}\label{eq:I0}
I_0 = \left(\frac{a_r}{G\tmass}\right)^2\sqrt{1-e_r^2} (u - e_r \sin u)\,.
\end{equation}
For the case $n=1$, we get
\begin{align}
I_1 &= \frac{2 \sqrt{b}}{s_+}  F_1\left( \frac{1}{2}, 1,\frac{1}{2};\frac{3}{2};a,b \right)\,,\\
&=\frac{2}{s_+} \sqrt{\frac{b}{1-a}}F_1\left( \frac{1}{2}, 0,\frac{1}{2};\frac{3}{2};\frac{a}{a-1},\frac{b-a}{1-a} \right)\,,
\end{align}
which leads to
\begin{equation}\label{eq:I1}
I_1 = \frac{a_r}{G\tmass}\sqrt{1-e_r^2}\, u\,.
\end{equation}
Now for the case $n\geq 2$, we can directly apply~\eqref{eq:F1to2F1} to~\eqref{eq:Ingennosimp}. We obtain the intermediate result
\begin{equation}
I_{n} = 2 \sqrt{b} \,s_+^{n-2} \sum_{i=0}^{n-2}\binom{n-2}{i}\frac{(-a)^i}{2i+1}{}_2 F_1\left( \frac{1}{2}, i+\frac{1}{2};i+\frac{3}{2};b \right)\,,
\end{equation}
which can be rewritten in terms of $a_r$, $e_r$ and $\Tilde{v}=2 \arctan\sqrt{\tfrac{s_+ - s}{s-s_-}} = 2 \arctan\sqrt{\tfrac{1+e_r}{1-e_r}}\tan\tfrac{u}{2}$ as
\begin{equation}
I_{n} = \frac{(G\tmass)^{n-2}}{a_r^{n-2}(1-e_r^2)^{n-2}}\sum_{i=0}^{n-2}\binom{n-2}{i}\binom{i-\frac{1}{2}}{i}(-2 e_r)^i(1+e_r)^{n-2-i}\left[\Tilde{v} - \sum_{j=0}^{i-1} \frac{j!}{(2j+1)!!}\sin\Tilde{v}(1-\cos\Tilde{v})^j \right]\,.
\end{equation}
Finally, we can express the latter sum as a sum of $\sin(k \Tilde{v})$ in order to match the nice expressions (D.11)-(D.18) of~\cite{Tessmer:2012xr} modulo the $(G\tmass)^{n-2}$ which comes from the fact that in this paper we defined $s=G\tmass/r$ and not $s=1/r$. We finally find the general formula, $\forall n\geq 2$,
\begin{align} \label{eq:InG2}
I_{n} = \frac{(G\tmass)^{n-2}}{a_r^{n-2}(1-e_r^2)^{n-2}} \sum_{i=0}^{n-2}&\binom{n-2}{i}\binom{i-\frac{1}{2}}{i}(-2 e_r)^i(1+e_r)^{n-2-i}\nn \\
& \times\left[\Tilde{v} - \sum_{j=0}^{i-1}\sum_{k=0}^{j} \frac{(-)^{j-k}(j!)^2}{k!(2j-k+1)!}\biggl[\sin\bigl((j-k+1)\Tilde{v}\bigr)+\sin\bigl((j-k)\Tilde{v}\bigr)\biggr] \right]\,.
\end{align}
We have checked that this formula reduces to the values of $I_n$ directly integrated with \textit{Mathematica} up to $n=10$, which correspond to the ones that are required in this project. We recall that to compute the period $P$ and the angle of periastron advance $\Phi$, one simply needs to take the limit $u=\tilde{v}\rightarrow \pi$ in these results, which kills the sum over $j$.

\section{Explicit QKP in terms of conserved quantities}\label{app:qkpCons}

Here we present the expressions of the coefficients appearing in the QKP~\eqref{eq:QKgen} in terms of the conserved quantities $\Et$ and $h$. First, the semi-major axis~\eqref{eq:QKr}
\begin{align}
\small \frac{a_r}{G \tmass} =& \, -\frac{1}{2\Et} -\frac{7-\nu}{4c^2} - \frac{8(4-7\nu)+\Et h^2(1+\nu^2)}{8c^4h^2} - \frac{24(1+\Et h^2)\mutp}{c^{10}h^8}\nn\\
& + \frac{1}{c^{12}h^{10}}\biggl[ - 12 \mutp\left(56-12\nu + \Et h^2(79-28\nu) +6\Et^2h^4(3-2\nu) \right)  \nn \\
& \qquad \qquad + 60 \delta\, \mutm \Et h^2(1+\Et h^2)  - 192 \sigmatp \left( 4+8 \Et h^2 +3 \Et^2 h^4 \right) \biggr]\nn \\
& +\frac{1}{c^{14}h^{12}}\biggl[ -\frac{2}{7}\mutp \left( 4(12654 - 7241 \nu + 420 \nu^2) +24 \Et h^2 (3637-2997\nu+273\nu^2) \right. \nn \\
& \left. \qquad \qquad \qquad \qquad +6\Et^2h^4 (5564-7404\nu+1155\nu^2) +7 \Et^3 h^6(189-722\nu+246\nu^2) \right) \nn \\
& \qquad \qquad - \frac{6}{7}\delta\,\mutm \left(744 -140\nu -696 \Et h^2 +10 \Et^2 h^4(-290+63\nu) +7\Et^3 h^6(-149+50\nu) \right) \nn \\
&  \qquad \qquad + 32 \sigmatp \left( 48(-15+4\nu) + 4 \Et h^2(-397+150\nu) + 8 \Et^2 h^4(-109+63\nu) + 3 \Et^3 h^6(-31+32\nu) \right) \nn \\
& \qquad \qquad + 32 \delta\, \sigmatm \left(-48 -100\Et h^2-32 \Et^2 h^4 + 9 \Et^3h^6 \right) -120\zetatp \left( 4+8 \Et h^2 + 3 \Et^2 h^4 \right) \biggr]\,.
\end{align}
Next, the three eccentricity parameters
\begin{subequations}
\small
\begin{align}
e_r =& \, \sqrt{1+2\Et h^2} - \frac{\Et}{2 c^2\sqrt{1+2\Et h^2}}\left( 12-2\nu + \Et h^2(15-5\nu) \right) \nn\\
& +\frac{\Et}{ 8c^4h^2(1+2\Et h^2)^{3/2}}\left( 32(-4+7\nu)+8 \Et h^2(-35+99\nu) +4\Et^2h^4(50+148\nu+\nu^2) +\Et^3h^6(415-210\nu+7\nu^2)\right)\nn\\
& - \frac{24\Et \mutp}{c^{10}h^8\sqrt{1+2\Et h^2}}\left( 4+10\Et h^2 + 5\Et^2 h^4\right)\nn\\
& - \frac{\Et}{c^{12}h^{10}(1+2\Et h^2)^{3/2}}\biggl[ 12\mutp \left( 224 - 48\nu +8\Et h^2(131 - 34\nu) + 2\Et^2h^4(783- 266\nu) + \Et^3 h^6(752- 401\nu)  \right. \nn\\
& \left.  +\Et^4 h^8(31-83\nu) \right) - 60 \delta\,\mutm \Et h^2(1+2\Et h^2)(4+10 \Et h^2 +5 \Et^2 h^4) +384 \sigmatp(1+2\Et h^2)\left(8+ 28\Et h^2 + 28 \Et^2 h^4 + 7 \Et^3 h^6\right) \biggr]\nn\\
& - \frac{\Et}{c^{14}h^{12}(1+2\Et h^2)^{5/2}}\Biggl\{ \frac{\mutp}{7} \left[ 32(12654-7241\nu + 420\nu^2) + 16 \Et h^2 (174522-111625\nu+7644\nu^2) \right. \nn\\
& \left. + 24 \Et^2 h^4 (299041 -221650\nu +18389\nu^2) + 28 \Et^3 h^6 (295437-270526\nu+28335\nu^2) + 14 \Et^4 h^8 (278682-363658\nu +51777\nu^2)  \right. \nn\\
& \left.  + 42 \Et^5 h^{10} (9285-31039\nu +7090\nu^2) - 7 \Et^6 h^{12}(5019+6850\nu-5049\nu^2) \right] \nn\\
& + \frac{6}{7}\delta\,\mutm (1+2\Et h^2) \left[2976 -560\nu + 8 \Et h^2 (954 - 245\nu) - 140 \Et^2 h^4 (81-5\nu) - 56 \Et^3 h^6 (807 - 160\nu) \right. \nn \\
& \left.\qquad \qquad \qquad \qquad - 7 \Et^4 h^8 (4899 - 1360\nu) - 7 \Et^5 h^{10} (729 - 325\nu) \right] \nn\\
& + 64\sigmatp (1+2\Et h^2) \left[ 96(15 - 4\nu) + 416\Et h^2(19- 6\nu) +4 \Et^2 h^4 (3785 - 1494\nu) + 6 \Et^3 h^6 ( 1916 - 1053\nu)\right. \nn \\
& \qquad \qquad \qquad \qquad \left. + 3 \Et^4 h^8 (843 - 902\nu) - \Et^5 h^{10} (163+297\nu) \right] \nn\\
& + 64\delta\,\sigmatm (1+2\Et h^2)^2 \left( 96 + 344\Et h^2 + 340\Et^2 h^4 + 52\Et^3 h^6 - 23\Et^4 h^8\right) \nn \\
& + 240 \zetatp (1+2\Et h^2)^2 \left( 8+28\Et h^2 + 28\Et^2 h^4 + 7 \Et^3 h^6 \right) \Biggr\} \,,\\
e_\phi =& \, \sqrt{1+2\Et h^2} - \frac{\Et}{2 c^2\sqrt{1+2\Et h^2}}\left( 12 + \Et h^2(15-\nu) \right) \nn \\
& - \frac{\Et}{16 c^4h^2(1+2\Et h^2)^{3/2}}\left[ 416 -91\nu -15\nu^2 +2 \Et h^2(600-125\nu-33\nu^2) +12\Et^2 h^4(20-5\nu-7\nu^2)-2\Et^3h^6 (415-94\nu+11\nu^2) \right] \nn\\
& - \frac{3\Et \mutp}{c^{10}h^8\sqrt{1+2\Et h^2}}\left( 93 +244 \Et h^2 +124\Et^2 h^4\right) \nn \\
& - \frac{\Et}{4c^{12}h^{10}(1+2\Et h^2)^{3/2}}\left[ 3 \mutp\left( 3(3839-651\nu) +2\Et h^2(27968 -5769\nu ) + 10\Et^2 h^4(8807 -2375\nu) \right. \right. \nn \\
& \left.\left.  +24\Et^3 h^6(1937 -802\nu) +8\Et^4 h^8(471-571\nu) \right)  + 15\delta\,\mutm\left(1+2\Et h^2 \right) \left(297+948\Et h^2 +772\Et^2 h^4 +112 \Et^3 h^6 \right) \right. \nn \\
& \left. + 96 \sigmatp \left( 1+2\Et h^2 \right)\left( 375 +1306\Et h^2 +1268\Et^2 h^4 +296\Et^3 h^6 \right) \right] \nn \\ 
& - \frac{\Et}{1792c^{14}h^{12}(1+2\Et h^2)^{5/2}} \Biggl\{ \mutp\left[ 9( 40887016 - 14415209\nu + 926121\nu^2) + 4\Et h^2 (657662112 -255147281\nu + 19587309\nu^2) \nn \right.\\
& \left. + 4\Et^2 h^4 (1778575392 -778237451\nu + 73435803\nu^2) + 32\Et^3 h^6 (278130768 - 142631977\nu + 17311581\nu^2) \nn \right.\\
& \left. + 16\Et^4 h^8 (310962510 - 199484659\nu + 33771717\nu^2) + 64\Et^5 h^{10} (15152022 - 13622281\nu + 3843399\nu^2)\nn \right.\\
& \left. + 448\Et^6 h^{12} (92958 - 86215\nu + 80769\nu^2 )\right] \nn\\
& + 6 \delta\,\mutm \left(1+2\Et h^2\right) \left[ 15182828 - 2077425\nu + 70 \Et h^2 (1210924 - 194551\nu) + 8\Et^2 h^4 (20847558 - 4073335\nu) \nn \right.\\
& \left. + 48\Et^3 h^6 (2803616 - 705215\nu)  + 80 \Et^4 h^8(476480 -171157\nu) + 224 \Et^5 h^{10}(8404 - 5575\nu) \right] \nn\\
& + 448 \sigmatp \left(1+2\Et h^2\right) \left[ 1140203 - 270510\nu +2\Et h^2 (3145943 - 884622\nu) + 8\Et^2 h^4 (1513337 - 532788\nu) \nn \right.\\
& \left. + 48\Et^3 h^6 (192751 - 94648\nu)  +48 \Et^4 h^8(44401 -41246\nu) - 32\Et^5 h^{10}(2153 + 7290\nu) \right] \nn\\
& + 448 \delta\,\sigmatm \left(1+2\Et h^2\right)^2 \left( 130619 + 524424 \Et h^2 + 653224 \Et^2 h^4 + 260512 \Et^3 h^6 + 17968 \Et^4 h^8 \right) \nn\\
& + 53760 \zetatp\left(1+2\Et h^2\right)^2\left( 237 + 842 \Et h^2 +844\Et^2 h^4 + 208 \Et^3 h^6\right)\Biggr\}\,,\\
\et =& \,  \sqrt{1+2\Et h^2} + \frac{\Et}{2 c^2\sqrt{1+2\Et h^2}}\left( 4-4\nu + \Et h^2(17-7\nu) \right) \nn\\
& -\frac{\Et}{8c^4h^2(1+2\Et h^2)^{3/2}} \left[ 64-112\nu +8\Et h^2(15-50\nu-3\nu^2) - 4\Et^2 h^4 (90 +73\nu +22\nu^2) \right. \nn \\
& \left. \qquad \qquad \qquad \qquad -\Et^3 h^6(607-138\nu+79\nu^2) -24\sqrt{-2\Et h^2}\left(1+2\Et h^2\right)^2(5-2\nu) \right] \nn \\
& -\frac{3\Et \mutp}{c^{10}h^8\sqrt{1+2\Et h^2}}\left[ 16-15(-2\Et h^2)^{1/2} -16 (-2\Et h^2) +18 (-2\Et h^2)^{3/2} +2(-2\Et h^2)^2 -3(-2\Et h^2)^{5/2} \right] \nn\\
&  -\frac{3\Et}{8 c^{12}h^{10}(1+2\Et h^2)^{3/2}} \Biggl\{ \mutp \biggl[ 256(14-3\nu) -70 (-2\Et h^2)^{1/2}(68-15\nu) - 32 (-2\Et h^2)(256-69\nu) \nn \\
& +25 (-2\Et h^2)^{3/2}(487-123\nu) +16  (-2\Et h^2)^2(375-134\nu) -2 (-2\Et h^2)^{5/2}(5191-1542\nu) -4 (-2\Et h^2)^{3}(370 -189\nu)  \nn \\
&  + 21 (-2\Et h^2)^{7/2}(151-55\nu) +2 (-2\Et h^2)^{4}(35-27\nu) -12 (-2\Et h^2)^{9/2}(17-8\nu) \biggr] \nn\\
& - 10 \delta\,\mutm \biggl[  35 (-2\Et h^2)^{1/2} - 16 (-2\Et h^2) - 85(-2\Et h^2)^{3/2} + 32 (-2\Et h^2)^2 + 65(-2\Et h^2)^{5/2}\nn \\ 
& -18 (-2\Et h^2)^{3} -15 (-2\Et h^2)^{7/2} +2 (-2\Et h^2)^{4} \biggr] \nn \\
& + 16 \sigmatp \biggl[256 -315 (-2\Et h^2)^{1/2} - 640 (-2\Et h^2) + 840 (-2\Et h^2)^{3/2} + 528 (-2\Et h^2)^2 - 750 (-2\Et h^2)^{5/2} \nn \\ 
& -152 (-2\Et h^2)^{3} +240  (-2\Et h^2)^{7/2} +8 (-2\Et h^2)^{4} -15(-2\Et h^2)^{9/2} \biggr]\Biggr\} + \calO\left(\frac{\etidal}{c^4} \right)\,.
\end{align}
\end{subequations}
We have not written the 2PN tidal term in $\et$ due to its heavy length. Now, the quantities entering the time-evolution equation~\eqref{eq:QKl}
\begin{subequations}
\small
\begin{align}
G \tmass n =& \, (-2\Et)^{3/2} + (-2\Et)^{5/2} \frac{(-15+\nu)}{8c^2} + \frac{\Et^3}{16c^4h}\left(192(5-2\nu) -\sqrt{-2\Et h^2}(555+30 \nu+11\nu^2)\right) \nn\,, \\
& + \frac{36\mutp\Et^3(5+2\Et h^2)}{c^{10}h^7} \nn\\
& + \frac{\Et^3}{c^{12}h^{9}}\Biggl\{ 3 \mutp \left(35(68-15\nu) + 30 \Et h^2(92-29\nu) + 12 \Et^2h^4(33-13\nu) \right) + 75 \delta\,\mutm  \left(7+6\Et h^2 \right) \nn \\
&  \qquad \qquad + 360 \sigmatp \left(21 +28\Et h^2 +4\Et^2h^4\right)\Biggr\}\nn\\
& + \frac{\Et^3}{c^{14}h^{11}}\Biggl\{ \frac{3\mutp}{28} \left[63(28608-12577\nu+980\nu^2)+70\Et h^2\left(5(8661-4715\nu+567\nu^2)+144\sqrt{-2\Et h^2}(5-2\nu) \right) \right. \nn \\
&  \qquad \qquad \left. + 12 \Et^2h^4 \left( 5(21648-13665\nu+2527\nu^2) +336\sqrt{-2\Et h^2}(5-2\nu)\right) + 168\Et^3h^6(597-366\nu+97\nu^2) \right] \nn \\
&  \qquad \qquad + \frac{3\delta\,\mutm}{28} \left( 189(1528-245\nu) +70\Et h^2(6702-1435\nu) + 60\Et^2h^4(2433-665\nu) \right) \nn \\
&  \qquad \qquad + 6 \sigmatp\left( 231(199-52\nu) +210\Et h^2(415-141\nu) + 140 \Et^2h^4(287-120\nu) + 40\Et^3h^6(71-31\nu) \right) \nn \\
&  \qquad \qquad + 714 \delta\,\sigmatp  \left( 33 +60 \Et h^2 +20\Et^2h^4\right) + 225 \zetatp \left(21+28\Et h^2 +4\Et^2 h^4 \right) \Biggr\}\,,\\
\frac{f_{v-u}}{(-2\Et)^{3/2}} =& \, \frac{15-6\nu}{2c^4 h} + \frac{9\mutp}{2c^{10}h^7}\left(5+2\Et h^2\right) \nn\\
&  + \frac{1}{8 c^{12}h^9}\left[\mutp\left(7140 -1575\nu +45 \Et h^2(169- 57\nu) +18\Et^2 h^4(51-25\nu)\right) \right. \nn\\
& \left. \qquad \qquad \qquad + 75\delta\,\mutm\left(7+6\Et h^2 \right) + 360\sigmatp\left( 21+28\Et h^2 +4 \Et^2 h^4 \right)\right] \nn \\
& + \frac{1}{4c^{14}h^{11}}\Biggl\{ \frac{9}{112} \mutp \biggl[ 42(28608 -12577\nu +980\nu^2) + 70 \Et h^2( 26490 -15033\nu + 1855\nu^2) \nn \\
& + 5 \Et^2 h^4(133809- 95194\nu + 19341\nu^2 ) +182 \Et^3 h^6(207-162\nu+55\nu^2) - 672 (-2\Et h^2)^{3/2}\left( 5+ 2 \Et h^2 \right)(5-2\nu)\biggr] \nn \\
& + \frac{9}{56} \delta\,\mutm \left(63(1528-245\nu) + 945\Et h^2(159-35\nu) +10\Et^2 h^4(4341-1295\nu)\right) \nn \\
& + \sigmatp\left( 693(199-52\nu) +315\Et h^2(785-279\nu) + 420 \Et^2 h^4(242-117\nu) + 60\Et^3 h^6(97-59\nu) \right) \nn \\
& + 357\delta\,\sigmatm\left(33+60\Et h^2 +20 \Et^2 h^4 \right) + \frac{225}{2}\zetatp\left( 21+28\Et h^2 +4 \Et^2 h^4 \right)\Biggr\}\,,\\
\frac{f_v}{(-2\Et)^{3/2}} =& \, \frac{\sqrt{1+2\Et h^2}}{8c^4 h}(15-\nu)\nu + \frac{12\sqrt{1+2\Et h^2}\mutp}{c^{10} h^7} \nn\\
& + \frac{3}{2\sqrt{1+2\Et h^2}c^{12} h^9}  \left[ \mutp \left( 421 -80\nu + 18\Et h^2(59-13\nu) +4\Et^2 h^4(119-36\nu)\right) \right. \nn\\
& \left. + 10\delta\,\mutm \left( 1+2\Et h^2\right)\left( 7+4 \Et h^2\right) + 8\sigmatp \left( 1+2\Et h^2\right) \left( 61 + 42\Et h^2\right) \right] \nn\\
& + \frac{(1+2\Et h^2)^{-3/2}}{896c^{14} h^{11}}\Biggl\{ \mutp \biggl[16937376 - 5344775\nu +467481\nu^2 + 8 \Et h^2( 10564026 - 3602663\nu + 365589\nu^2) \nn\\
& + 24\Et^2 h^4(5853210-2217535\nu+269283\nu^2) + 224 \Et^3 h^6(383910 -169037\nu +25779\nu^2)  \nn \\
& + 112 \Et^4 h^8(134592 -70745\nu +14415\nu^2)  - 504 (-2\Et h^2)^{3/2}\left( 1 + 2\Et h^2\right)^2 \left(160+11 \nu - 5\nu^2 + 2\Et h^2(15-\nu)\nu\right)\biggr] \nn\\
& + 12 \delta\,\mutm \left(1 + 2\Et h^2\right) \left[431784 - 70805\nu +2 \Et h^2(678296 - 134645 \nu) + 980 \Et^2 h^4(1136 - 295\nu) +56 \Et^3 h^6(3644 - 1305\nu) \right]\nn\\
& + 896\sigmatp (1+2\Et h^2) \left[ 28014-6519\nu + \Et h^2(87059 - 23985\nu) + 2 \Et^2 h^4(35543 - 12315\nu) + 96 \Et^3 h^6(143 -62\nu)  \right] \nn\\
& + 1792\delta\,\sigmatm\left(1 + 2\Et h^2\right)^2\left(1632 +2041 \Et h^2+ 354 \Et^2h^4 \right) + 6720\zetatp \left(1 + 2\Et h^2\right)^2\left(61 + 42\Et h^2\right)\Biggr\} \,,\\
\frac{f_{2v}}{(-2\Et)^{3/2}} =& \, \frac{3(1+2\Et h^2)\mutp}{4c^{10} h^7} + \frac{3}{16 c^{12} h^{9}} \left[ \mutp\left( 428-52\nu + \Et h^2(937-125\nu) + 2\Et^2h^4(99-19\nu) \right) \right. \nn\\
& \left. + 10\delta\,\mutm \left( 1 + 2\Et h^2\right)\left( 14 + 3 \Et h^2\right)+32 \sigmatp \left( 1 + 2\Et h^2\right)\left( 17 + 4 \Et h^2\right) \right] \nn\\
& + \frac{1}{896c^{14} h^{11}}\Biggl\{ \mutp \biggl[2895684 - 474413\nu +25788\nu^2 + 2\Et h^2( 3750240 - 722891\nu + 42084\nu^2) \nn\\
& + 3\Et^2 h^4(1393585-354110\nu+20517\nu^2) + 42 \Et^3 h^6(8411 -2970\nu +159\nu^2) - 1008 (-2\Et h^2)^{3/2}(1+2\Et h^2)(5-2\nu)\biggr] \nn\\
& + 3 \delta\,\mutm \left[481206 - 89285\nu +2 \Et h^2(650894 - 145915 \nu) + 20 \Et^2 h^4(38601 - 11606\nu) +56 \Et^3 h^6(1081 - 450\nu) \right]\nn\\
& + 56\sigmatp \left[ 15(4597-776\nu) + 2 \Et h^2(89245 - 17796\nu) + 4 \Et^2 h^4(24509 - 6036\nu) + 8 \Et^3 h^6(979 -240\nu)  \right] \nn\\
& + 56\delta\,\sigmatm\left(1 + 2\Et h^2\right)\left(12129+9428\Et h^2+580\Et^2h^4 \right) + 3360\zetatp \left(1 + 2\Et h^2\right)\left(17 + 4\Et h^2\right)\Biggr\} \,,\\
\frac{f_{3v}}{(-2\Et)^{3/2}} =& \, \frac{(1+2\Et h^2)^{3/2}}{2 c^{12}h^{9}}\left( 19 \mutp + 10 \delta\,\mutm + 24 \sigmatp\right) \nn\\
& + \frac{\sqrt{1+2\Et h^2}}{448 c^{14}h^{11}} \Biggl\{\mutp \left( 269064 +25190\nu-9303\nu^2 + 4\Et h^2(155490 +10376\nu-6321\nu^2) + 28\Et^2 h^4(8040 - 406\nu -477\nu^2) \right) \nn \\
& +3\delta\,\mutm\left( 64208 -14455\nu + 36\Et h^2(4208-1085\nu) + 28 \Et^2 h^4(2008-685\nu) \right) \nn \\
& + 224 \sigmatp\left( 5(254-5\nu) +2 \Et h^2(1431 -37\nu) +4\Et^2 h^4(242-3\nu)\right)\nn \\
& + 896 \delta\,\sigmatm \left( 92+ 221\Et h^2 + 74\Et^2 h^4 \right) +3360 \zetatp \left( 1+ 2\Et h^2 \right) \Biggr\}\,,\\ 
\frac{f_{4v}}{(-2\Et)^{3/2}} =& \, \frac{(1+2\Et h^2)^{2}}{32 c^{12}h^{9}}\left( 3 \mutp (8+\nu) + 15 \delta\,\mutm + 24 \sigmatp\right) \nn\\
& + \frac{(1+2\Et h^2)}{32 c^{14}h^{11}} \Biggl\{\frac{3}{28} \mutp \left( 31336 +7923\nu-2856\nu^2 +\Et h^2(66298 +15531\nu-6419\nu^2) + 14\Et^2 h^4(806 +23\nu-97\nu^2) \right) \nn \\
& +\frac{3}{28}\delta\,\mutm\left( 27136 -7455\nu + \Et h^2(57751-16765\nu) + 14 \Et^2 h^4(677-245\nu) \right) \nn \\
& + \sigmatp\left( 2346 +744\nu +2 \Et h^2(2447+801\nu) +4\Et^2 h^4(209+69\nu)\right)\nn \\
& + 2 \delta\,\sigmatm \left( 1+ 2\Et h^2 \right) \left( 579+80\Et h^2 \right) +15 \zetatp \left( 1+ 2\Et h^2 \right) \Biggr\}\,,\\
\frac{f_{5v}}{(-2\Et)^{3/2}} =& \, \frac{(1+2\Et h^2)^{5/2}}{640c^{14}h^{11}}\left( \mutp (9264 + 2789\nu-1305\nu^2) + \delta\,\mutm(8832 - 2730\nu) + 192\sigmatp(22+19\nu)  + 3072 \delta\,\sigmatm\right)\,,\\
\frac{f_{6v}}{(-2\Et)^{3/2}} =& \, \frac{(1+2\Et h^2)^3}{128c^{14}h^{11}}\left( \mutp (156+45\nu-24\nu^2) + \delta\,\mutm(150-45\nu) + 8\sigmatp(7+8\nu)  + 40 \delta\,\sigmatm\right)\,.
\end{align}
\end{subequations}
The quantities entering the phase evolution equation~\eqref{eq:QKphi}
\begin{subequations}
\small
\begin{align}
K =& \, 1 + \frac{3}{c^2h^2} + \frac{3}{4 c^4 h^4} \left[ 35 -10\nu + 2\Et h^2(5-2\nu)\right] + \frac{45\mutp}{8 c^{10}h^{10}}\left[ 21+28\Et h^2+4 \Et^2 h^4\right] \label{eq:Kqkp} \nn \\
& + \frac{1}{c^{12}h^{12}}\Biggl\{ \frac{9\mutp}{32}\left[231(71-10\nu) +420 \Et h^2(68-15\nu) + 980\Et^2 h^4 (11-4\nu) +160 \Et^3 h^6(3-2\nu) \right] \nn \\
& \qquad \qquad + \frac{315\delta\,\mutm}{32} \left(33+60\Et h^2 +20\Et^2 h^4\right)  + \frac{3\sigmatp}{2} \left( 2541 + 5670\Et h^2 +2940\Et^2 h^4 + 200 \Et^3 h^6\right)\Biggr\} \nn \\
& + \frac{1}{c^{14}h^{14}}\Biggl\{ \frac{3\mutp}{64}\left[ 429(6276-2080\nu +105\nu^2) +198 \Et h^2(28608 -12577\nu+ 980\nu^2) + 90\Et^2 h^4(36165-21524\nu+2730\nu^2)\nn \right. \\
& \left. \qquad \qquad \qquad \qquad + 80 \Et^3 h^6 (6099-5093\nu+1155\nu^2) + 120\Et^4 h^8(66-84\nu+41\nu^2) \right] \nn \\
& \qquad \qquad + \frac{27\delta\,\mutm}{128}\left[ 1001(87-10\nu) + 132 \Et h^2(1528-245\nu) + 20\Et^2 h^4(6177 -1400\nu) + 160 \Et^3 h^6 (106-35\nu)  \right] \nn \\
& \qquad \qquad + \frac{\sigmatp}{16} \left[ 5577(439-84\nu) +30492 \Et h^2(199-52\nu) + 22680\Et^2 h^4(185 -69\nu)\nn \right. \\
& \left. \qquad \qquad \qquad \qquad +3920 \Et^3 h^6 (197-114\nu) + 1200\Et^4 h^8(13-14\nu)\right] \nn \\
& \qquad \qquad +\frac{17\delta\,\sigmatm}{8}\left(5577 + 15246 \Et h^2 +11340\Et^2 h^4 +1960 \Et^3 h^6 \right) \nn \\
& \qquad \qquad + \frac{105\zetatp}{16}\left( 429 +990 \Et h^2 +540\Et^2 h^4 +40 \Et^3 h^6\right)\Biggr\}\,,\\
g_{2 v} =& \, \frac{1+2\Et h^2}{8 c^4 h^4}(1+19\nu -3\nu^2) +  \frac{3\mutp}{4 c^{10}h^{10}}\left( 1+ 2\Et h^2 \right)\left( 17 + 4\Et h^2 \right)\nn\\
& + \frac{3}{128 c^{12} h^{12}} \biggl[ \mutp\left( 23911 - 1734\nu + 6 \Et h^2(10271- 1030\nu) +4\Et^2 h^4(8261-1306\nu) + 8 \Et^3 h^6 (279-70\nu) \right)  \nn \\
& + 5 \delta\,\mutm\left(1+2\Et h^2\right)\left(1485 +1348\Et h^2 +116 \Et^2 h^4\right) +16 \sigmatp \left(1+2\Et h^2\right)\left( 1073 +900\Et h^2 + 68\Et^2 h^4\right) \biggr] \nn\\
& + \frac{1}{896 c^{14}h^{14}}\Biggl\{ \mutp\left[ 9(1768600 - 74741\nu -1981\nu^2) + \Et h^2 (47701512 - 2671147\nu - 62790\nu^2)\right. \nn\\
& \left. + 3 \Et^2 h^4 (13241467 - 1011313\nu - 34146\nu^2) + 4 \Et^3 h^6 (2417685 - 303122\nu - 15708\nu^2) + 84 \Et^4 h^8 (4157 - 1263\nu - 50\nu^2) \right] \nn \\
& + 3 \delta\, \mutm\left[2637533 -333900\nu  + 5 \Et h^2 (1762402 -288365 \nu) + \Et^2 h^4 (8552993 - 1878905\nu) + 20 \Et^3 h^6 (124051 - 37086\nu) \right. \nn \\
& \left. +140 \Et^4 h^8 (743-303\nu)\right] + 56 \sigmatp \left[269806 - 38964\nu + 3 \Et h^2(274689 - 47788\nu)  + 2 \Et^2 h^4(337570 - 72387\nu) \right. \nn \\
& \left.  + 4\Et^3 h^6(36331 - 8922\nu) +8 \Et^4 h^8(496-51\nu) \right]  + 56 \delta\,\sigmatm \left( 1+ 2\Et h^2 \right) \left( 50782 + 79987\Et h^2 +27300\Et^2 h^4 + 1132\Et^3 h^6\right) \nn\\
& +210\zetatp\left( 1+ 2\Et h^2 \right)\left( 2063 +1980 \Et h^2 +188\Et^2 h^4 \right)\Biggr\}\,,\\
g_{3 v} =& \, \frac{(1+2\Et h^2)^{3/2}}{32 c^4 h^4}(1-3\nu)\nu +  \frac{3 (1+2\Et h^2)^{3/2}\mutp}{2 c^{10}h^{10}} + \frac{\sqrt{1+2\Et h^2}}{16c^{12} h^{12}} \left[ 3 \mutp\left( 467+17\nu +4 \Et h^2(262+7\nu) +336\Et^2 h^4 \right) \right. \nn \\
& \left. + 15\delta\,\mutm\left( 55+136\Et h^2 + 52\Et^2 h^4\right) +32\sigmatp\left( 19 +44\Et h^2+12 \Et^2 h^4 \right) \right]  \nn \\
& + \frac{(1+2\Et h^2)^{-1/2}}{5376c^{14}h^{14}}\Biggl\{ 3 \mutp\left[ (6710148 + 1186408\nu -129633 \nu^2) + 8 \Et h^2 (3946302 + 693827\nu - 100107\nu^2)\right. \nn\\
& \left. + 24 \Et^2 h^4 (2023121 + 344382\nu - 72226\nu^2) + 64 \Et^3 h^6 (412185 + 61256\nu - 23121\nu^2) + 112 \Et^4 h^8 (34131 + 1970\nu - 3192\nu^2) \right] \nn \\
& + 9 \delta\, \mutm\left( 1+2\Et h^2 \right)\left[5(358236 -52661\nu) + 10 \Et h^2 (517644 -96635 \nu) + 4\Et^2 h^4 (957606 - 237545\nu) + 56 \Et^3 h^6 (11598 - 3785\nu) \right]\nn\\
& + 224 \sigmatp \left( 1+2\Et h^2 \right) \left[49849 - 570\nu + 14 \Et h^2(9025 + 102\nu) + 12 \Et^2 h^4(5583 + 692\nu) + 24\Et^3 h^6(277 + 168\nu)  \right] \nn\\
& + 224 \delta\,\sigmatm\left( 1+2\Et h^2 \right)^2\left( 19753 + 21852\Et h^2 +4116\Et^2 h^4\right) +13440\zetatp\left( 1+2\Et h^2 \right)^2\left( 37+18\Et h^2\right) \Biggr\}\,,\\ 
g_{4 v} =& \, \frac{3 (1+2\Et h^2)^2\mutp}{32 c^{10}h^{10}} + \frac{3 (1+2\Et h^2)}{128 c^{12}h^{12}}\left[ \mutp\left( 473+114\nu +2 \Et h^2(491+120\nu) + 16\Et^2 h^4(9+2\nu) \right) \right. \nn \\ 
& \left. + 5 \delta\,\mutm\left( 1+2\Et h^2\right)\left( 99+16 \Et h^2 \right) -16\sigmatp \left( 1+2\Et h^2\right)\left(5+2\Et h^2 \right)  \right]\nn\\
& + \frac{1}{896 c^{14}h^{14}}\Biggl\{ \mutp\left[ 9(84519 + 40220\nu -4676 \nu^2) + \Et h^2 (3433470 + 1569355\nu - 217812\nu^2)\right. \nn\\
& \left. + \Et^2 h^4 (4808667 + 1992376\nu - 366870\nu^2) + 8 \Et^3 h^6 (253815 + 75617\nu -26187\nu^2) + 420 \Et^4 h^8 (310 +12\nu -49\nu^2) \right] \nn \\
& + 3 \delta\, \mutm\left( 1+2\Et h^2 \right)\left[267368 -47250\nu + \Et h^2 (677822-143885\nu) + 15\Et^2 h^4 (23693-6426\nu) +14 \Et^3 h^6 (1699-570\nu) \right]\nn\\
& + 56\sigmatp\left( 1+2\Et h^2 \right)\left[9(153+160\nu) + 7\Et h^2(449 + 636\nu) + 6 \Et^2 h^4(57+541\nu) -4\Et^3 h^6(2-75\nu)  \right] \nn\\
& + 56\delta\,\sigmatm\left( 1+2\Et h^2 \right)^2\left( 2916+2069\Et h^2 +146\Et^2 h^4\right) +210\zetatp\left( 1+2\Et h^2 \right)^2\left( 61+10\Et h^2\right) \Biggr\}\,,\\
g_{5 v} =& \, \frac{3(1+2\Et h^2)^{5/2}}{80 c^{12}h^{12}}\left[ \mutp(29+15\nu) + 45\delta\,\mutm - 32 \sigmatp \right] \nn \\
& + \frac{(1+2\Et h^2)^{3/2}}{8960 c^{14}h^{14}} \Biggl\{ \mutp\left[ 1756404 + 907930\nu - 198975\nu^2 +4\Et h^2(1012056 + 477989\nu -125790\nu^2) \nn \right.\\
&\left. +84\Et ^2 h^4(15358 + 3928\nu -2355\nu^2) \right]  +3\delta\mutm\left( 656812 - 137585\nu +380\Et h^2(3936-917\nu) + 28\Et^2 h^4(17054-4785\nu)\right)\nn \\
& -672 \sigmatp\left( 359-262\nu + 20\Et h^2(33-34\nu) +4 \Et^2 h^4(61-68\nu) \right) \nn\\
&+ 672\delta\,\sigmatm\left( 393+940\Et h^2 +308\Et^2 h^4 \right) + 13440\left( 1+2\Et h^2 \right)\zetatp\Biggr\}\,,\\
g_{6 v} =& \, \frac{(1+2\Et h^2)^3}{128 c^{12}h^{12}}\left[ \mutp(9+6\nu) + 15\delta\,\mutm - 16 \sigmatp \right] \nn \\
& + \frac{(1+2\Et h^2)^2}{2688 c^{14}h^{14}} \Biggl\{ 3 \mutp\left[34806 +13940\nu -4788\nu^2 +\Et h^2(73476 + 27943\nu -10416\nu^2) + 21 \Et^2 h^4(449+69\nu-74\nu^2) \right] \nn \\
& + 27 \delta\,\mutm\left( 4187-980\nu+\et h^2(8794-2135\nu) + 105\Et^2 h^4(11-3\nu) \right) \nn \\
& - 56 \sigmatp\left(2(151+6\nu) +\Et h^2(571+12\nu) + 6 \Et^2 h^4(16-\nu)\right) \nn \\
& + 56 \delta\,\sigmatm \left( 1+2\Et h^2 \right)\left( 130 + 21 \Et h^2 \right) +210\zetatp \left( 1+2\Et h^2 \right)\Biggr\}\,,\\
g_{7 v} =& \, \frac{(1+2\Et h^2)^{7/2}}{3584 c^{14}h^{14}}\left[ 7 \mutp (2712 +739\nu-399\nu^2) + \delta\,\mutm (20136-4830\nu) - 64 \sigmatp(47+42\nu)+ 64\delta\,\sigmatm\right]\,,\\ 
g_{8 v} =& \, \frac{(1+2\Et h^2)^4}{1024 c^{14}h^{14}}\left[ 6 \mutp(62+12\nu-9\nu^2) + 3 \delta\,\mutm (133-30\nu) - 8  \sigmatp(11+12\nu) - 16\delta\,\sigmatm\right]\,.
\end{align}
\end{subequations}
We do not display the expression of $x$~\eqref{eq:QKx} as a function of $\Et$ and $h$ since it is simply $\tfrac{(G\tmass)^{2/3}}{c^2}(K n)^{2/3}$ where $K$ and $n$ are given above.

\bibliography{RefList_EccentricTides}

\begin{thebibliography}{97}%
\makeatletter
\providecommand \@ifxundefined [1]{%
 \@ifx{#1\undefined}
}%
\providecommand \@ifnum [1]{%
 \ifnum #1\expandafter \@firstoftwo
 \else \expandafter \@secondoftwo
 \fi
}%
\providecommand \@ifx [1]{%
 \ifx #1\expandafter \@firstoftwo
 \else \expandafter \@secondoftwo
 \fi
}%
\providecommand \natexlab [1]{#1}%
\providecommand \enquote  [1]{``#1''}%
\providecommand \bibnamefont  [1]{#1}%
\providecommand \bibfnamefont [1]{#1}%
\providecommand \citenamefont [1]{#1}%
\providecommand \href@noop [0]{\@secondoftwo}%
\providecommand \href [0]{\begingroup \@sanitize@url \@href}%
\providecommand \@href[1]{\@@startlink{#1}\@@href}%
\providecommand \@@href[1]{\endgroup#1\@@endlink}%
\providecommand \@sanitize@url [0]{\catcode `\\12\catcode `\$12\catcode `\&12\catcode `\#12\catcode `\^12\catcode `\_12\catcode `\%12\relax}%
\providecommand \@@startlink[1]{}%
\providecommand \@@endlink[0]{}%
\providecommand \url  [0]{\begingroup\@sanitize@url \@url }%
\providecommand \@url [1]{\endgroup\@href {#1}{\urlprefix }}%
\providecommand \urlprefix  [0]{URL }%
\providecommand \Eprint [0]{\href }%
\providecommand \doibase [0]{https://doi.org/}%
\providecommand \selectlanguage [0]{\@gobble}%
\providecommand \bibinfo  [0]{\@secondoftwo}%
\providecommand \bibfield  [0]{\@secondoftwo}%
\providecommand \translation [1]{[#1]}%
\providecommand \BibitemOpen [0]{}%
\providecommand \bibitemStop [0]{}%
\providecommand \bibitemNoStop [0]{.\EOS\space}%
\providecommand \EOS [0]{\spacefactor3000\relax}%
\providecommand \BibitemShut  [1]{\csname bibitem#1\endcsname}%
\let\auto@bib@innerbib\@empty
\bibitem [{\citenamefont {Abbott}\ \emph {et~al.}(2017{\natexlab{a}})\citenamefont {Abbott} \emph {et~al.}}]{LIGOScientific:2017vwq}%
  \BibitemOpen
  \bibfield  {author} {\bibinfo {author} {\bibfnamefont {B.~P.}\ \bibnamefont {Abbott}} \emph {et~al.} (\bibinfo {collaboration} {LIGO Scientific, Virgo}),\ }\bibfield  {title} {\bibinfo {title} {{GW170817: Observation of Gravitational Waves from a Binary Neutron Star Inspiral}},\ }\href {https://doi.org/10.1103/PhysRevLett.119.161101} {\bibfield  {journal} {\bibinfo  {journal} {Phys. Rev. Lett.}\ }\textbf {\bibinfo {volume} {119}},\ \bibinfo {pages} {161101} (\bibinfo {year} {2017}{\natexlab{a}})},\ \Eprint {https://arxiv.org/abs/1710.05832} {arXiv:1710.05832 [gr-qc]} \BibitemShut {NoStop}%
\bibitem [{\citenamefont {Abbott}\ \emph {et~al.}(2017{\natexlab{b}})\citenamefont {Abbott} \emph {et~al.}}]{LIGOScientific:2017ync}%
  \BibitemOpen
  \bibfield  {author} {\bibinfo {author} {\bibfnamefont {B.~P.}\ \bibnamefont {Abbott}} \emph {et~al.} (\bibinfo {collaboration} {LIGO Scientific, Virgo, Fermi GBM, INTEGRAL, IceCube, AstroSat Cadmium Zinc Telluride Imager Team, IPN, Insight-Hxmt, ANTARES, Swift, AGILE Team, 1M2H Team, Dark Energy Camera GW-EM, DES, DLT40, GRAWITA, Fermi-LAT, ATCA, ASKAP, Las Cumbres Observatory Group, OzGrav, DWF (Deeper Wider Faster Program), AST3, CAASTRO, VINROUGE, MASTER, J-GEM, GROWTH, JAGWAR, CaltechNRAO, TTU-NRAO, NuSTAR, Pan-STARRS, MAXI Team, TZAC Consortium, KU, Nordic Optical Telescope, ePESSTO, GROND, Texas Tech University, SALT Group, TOROS, BOOTES, MWA, CALET, IKI-GW Follow-up, H.E.S.S., LOFAR, LWA, HAWC, Pierre Auger, ALMA, Euro VLBI Team, Pi of Sky, Chandra Team at McGill University, DFN, ATLAS Telescopes, High Time Resolution Universe Survey, RIMAS, RATIR, SKA South Africa/MeerKAT}),\ }\bibfield  {title} {\bibinfo {title} {{Multi-messenger Observations of a Binary Neutron Star Merger}},\ }\href
  {https://doi.org/10.3847/2041-8213/aa91c9} {\bibfield  {journal} {\bibinfo  {journal} {Astrophys. J. Lett.}\ }\textbf {\bibinfo {volume} {848}},\ \bibinfo {pages} {L12} (\bibinfo {year} {2017}{\natexlab{b}})},\ \Eprint {https://arxiv.org/abs/1710.05833} {arXiv:1710.05833 [astro-ph.HE]} \BibitemShut {NoStop}%
\bibitem [{\citenamefont {Abbott}\ \emph {et~al.}(2017{\natexlab{c}})\citenamefont {Abbott} \emph {et~al.}}]{LIGOScientific:2017pwl}%
  \BibitemOpen
  \bibfield  {author} {\bibinfo {author} {\bibfnamefont {B.~P.}\ \bibnamefont {Abbott}} \emph {et~al.} (\bibinfo {collaboration} {LIGO Scientific, Virgo}),\ }\bibfield  {title} {\bibinfo {title} {{Estimating the Contribution of Dynamical Ejecta in the Kilonova Associated with GW170817}},\ }\href {https://doi.org/10.3847/2041-8213/aa9478} {\bibfield  {journal} {\bibinfo  {journal} {Astrophys. J. Lett.}\ }\textbf {\bibinfo {volume} {850}},\ \bibinfo {pages} {L39} (\bibinfo {year} {2017}{\natexlab{c}})},\ \Eprint {https://arxiv.org/abs/1710.05836} {arXiv:1710.05836 [astro-ph.HE]} \BibitemShut {NoStop}%
\bibitem [{\citenamefont {Abbott}\ \emph {et~al.}(2017{\natexlab{d}})\citenamefont {Abbott} \emph {et~al.}}]{LIGOScientific:2017zic}%
  \BibitemOpen
  \bibfield  {author} {\bibinfo {author} {\bibfnamefont {B.~P.}\ \bibnamefont {Abbott}} \emph {et~al.} (\bibinfo {collaboration} {LIGO Scientific, Virgo, Fermi-GBM, INTEGRAL}),\ }\bibfield  {title} {\bibinfo {title} {{Gravitational Waves and Gamma-rays from a Binary Neutron Star Merger: GW170817 and GRB 170817A}},\ }\href {https://doi.org/10.3847/2041-8213/aa920c} {\bibfield  {journal} {\bibinfo  {journal} {Astrophys. J. Lett.}\ }\textbf {\bibinfo {volume} {848}},\ \bibinfo {pages} {L13} (\bibinfo {year} {2017}{\natexlab{d}})},\ \Eprint {https://arxiv.org/abs/1710.05834} {arXiv:1710.05834 [astro-ph.HE]} \BibitemShut {NoStop}%
\bibitem [{\citenamefont {Sotiriou}(2018)}]{Sotiriou:2017obf}%
  \BibitemOpen
  \bibfield  {author} {\bibinfo {author} {\bibfnamefont {T.~P.}\ \bibnamefont {Sotiriou}},\ }\bibfield  {title} {\bibinfo {title} {{Detecting Lorentz Violations with Gravitational Waves from Black Hole Binaries}},\ }\href {https://doi.org/10.1103/PhysRevLett.120.041104} {\bibfield  {journal} {\bibinfo  {journal} {Phys. Rev. Lett.}\ }\textbf {\bibinfo {volume} {120}},\ \bibinfo {pages} {041104} (\bibinfo {year} {2018})},\ \Eprint {https://arxiv.org/abs/1709.00940} {arXiv:1709.00940 [gr-qc]} \BibitemShut {NoStop}%
\bibitem [{\citenamefont {Abbott}\ \emph {et~al.}(2017{\natexlab{e}})\citenamefont {Abbott} \emph {et~al.}}]{LIGOScientific:2017adf}%
  \BibitemOpen
  \bibfield  {author} {\bibinfo {author} {\bibfnamefont {B.~P.}\ \bibnamefont {Abbott}} \emph {et~al.} (\bibinfo {collaboration} {LIGO Scientific, Virgo, 1M2H, Dark Energy Camera GW-E, DES, DLT40, Las Cumbres Observatory, VINROUGE, MASTER}),\ }\bibfield  {title} {\bibinfo {title} {{A gravitational-wave standard siren measurement of the Hubble constant}},\ }\href {https://doi.org/10.1038/nature24471} {\bibfield  {journal} {\bibinfo  {journal} {Nature}\ }\textbf {\bibinfo {volume} {551}},\ \bibinfo {pages} {85} (\bibinfo {year} {2017}{\natexlab{e}})},\ \Eprint {https://arxiv.org/abs/1710.05835} {arXiv:1710.05835 [astro-ph.CO]} \BibitemShut {NoStop}%
\bibitem [{\citenamefont {Abac}\ \emph {et~al.}(2025{\natexlab{a}})\citenamefont {Abac} \emph {et~al.}}]{LIGOScientific:2025pvj}%
  \BibitemOpen
  \bibfield  {author} {\bibinfo {author} {\bibfnamefont {A.~G.}\ \bibnamefont {Abac}} \emph {et~al.} (\bibinfo {collaboration} {LIGO Scientific, VIRGO, KAGRA}),\ }\bibfield  {title} {\bibinfo {title} {{GWTC-4.0: Population Properties of Merging Compact Binaries}},\ }\href@noop {} {\  (\bibinfo {year} {2025}{\natexlab{a}})},\ \Eprint {https://arxiv.org/abs/2508.18083} {arXiv:2508.18083 [astro-ph.HE]} \BibitemShut {NoStop}%
\bibitem [{\citenamefont {Abbott}\ \emph {et~al.}(2021)\citenamefont {Abbott} \emph {et~al.}}]{LIGOScientific:2021qlt}%
  \BibitemOpen
  \bibfield  {author} {\bibinfo {author} {\bibfnamefont {R.}~\bibnamefont {Abbott}} \emph {et~al.} (\bibinfo {collaboration} {LIGO Scientific, KAGRA, VIRGO}),\ }\bibfield  {title} {\bibinfo {title} {{Observation of Gravitational Waves from Two Neutron Star{\textendash}Black Hole Coalescences}},\ }\href {https://doi.org/10.3847/2041-8213/ac082e} {\bibfield  {journal} {\bibinfo  {journal} {Astrophys. J. Lett.}\ }\textbf {\bibinfo {volume} {915}},\ \bibinfo {pages} {L5} (\bibinfo {year} {2021})},\ \Eprint {https://arxiv.org/abs/2106.15163} {arXiv:2106.15163 [astro-ph.HE]} \BibitemShut {NoStop}%
\bibitem [{\citenamefont {Abac}\ \emph {et~al.}(2024{\natexlab{a}})\citenamefont {Abac} \emph {et~al.}}]{LIGOScientific:2024elc}%
  \BibitemOpen
  \bibfield  {author} {\bibinfo {author} {\bibfnamefont {A.~G.}\ \bibnamefont {Abac}} \emph {et~al.} (\bibinfo {collaboration} {LIGO Scientific, KAGRA, VIRGO}),\ }\bibfield  {title} {\bibinfo {title} {{Observation of Gravitational Waves from the Coalescence of a 2.5{\textendash}4.5 $M _{\odot}$ Compact Object and a Neutron Star}},\ }\href {https://doi.org/10.3847/2041-8213/ad5beb} {\bibfield  {journal} {\bibinfo  {journal} {Astrophys. J. Lett.}\ }\textbf {\bibinfo {volume} {970}},\ \bibinfo {pages} {L34} (\bibinfo {year} {2024}{\natexlab{a}})},\ \Eprint {https://arxiv.org/abs/2404.04248} {arXiv:2404.04248 [astro-ph.HE]} \BibitemShut {NoStop}%
\bibitem [{\citenamefont {Fei}\ and\ \citenamefont {Yang}(2024)}]{Fei:2024ruj}%
  \BibitemOpen
  \bibfield  {author} {\bibinfo {author} {\bibfnamefont {Q.}~\bibnamefont {Fei}}\ and\ \bibinfo {author} {\bibfnamefont {Y.}~\bibnamefont {Yang}},\ }\bibfield  {title} {\bibinfo {title} {{Test of the Brans{\textendash}Dicke theory with GW200105 and GW200115}},\ }\href {https://doi.org/10.1088/1572-9494/ad4bbb} {\bibfield  {journal} {\bibinfo  {journal} {Commun. Theor. Phys.}\ }\textbf {\bibinfo {volume} {76}},\ \bibinfo {pages} {075402} (\bibinfo {year} {2024})}\BibitemShut {NoStop}%
\bibitem [{\citenamefont {Morras}\ \emph {et~al.}(2025{\natexlab{a}})\citenamefont {Morras}, \citenamefont {Pratten},\ and\ \citenamefont {Schmidt}}]{Morras:2025xfu}%
  \BibitemOpen
  \bibfield  {author} {\bibinfo {author} {\bibfnamefont {G.}~\bibnamefont {Morras}}, \bibinfo {author} {\bibfnamefont {G.}~\bibnamefont {Pratten}},\ and\ \bibinfo {author} {\bibfnamefont {P.}~\bibnamefont {Schmidt}},\ }\bibfield  {title} {\bibinfo {title} {{Orbital eccentricity in a neutron star - black hole binary}},\ }\href@noop {} {\bibfield  {journal} {\bibinfo  {journal} {arXiv e-prints}\ } (\bibinfo {year} {2025}{\natexlab{a}})},\ \Eprint {https://arxiv.org/abs/2503.15393} {2503.15393 [astro-ph.HE]} \BibitemShut {NoStop}%
\bibitem [{\citenamefont {Planas}\ \emph {et~al.}(2025{\natexlab{a}})\citenamefont {Planas}, \citenamefont {Husa}, \citenamefont {Ramos-Buades},\ and\ \citenamefont {Valencia}}]{Planas:2025plq}%
  \BibitemOpen
  \bibfield  {author} {\bibinfo {author} {\bibfnamefont {M.~d.~L.}\ \bibnamefont {Planas}}, \bibinfo {author} {\bibfnamefont {S.}~\bibnamefont {Husa}}, \bibinfo {author} {\bibfnamefont {A.}~\bibnamefont {Ramos-Buades}},\ and\ \bibinfo {author} {\bibfnamefont {J.}~\bibnamefont {Valencia}},\ }\bibfield  {title} {\bibinfo {title} {{First eccentric inspiral-merger-ringdown analysis of neutron star-black hole mergers}},\ }\href@noop {} {\bibfield  {journal} {\bibinfo  {journal} {arXiv e-prints}\ } (\bibinfo {year} {2025}{\natexlab{a}})},\ \Eprint {https://arxiv.org/abs/2506.01760} {arXiv:2506.01760 [astro-ph.HE]} \BibitemShut {NoStop}%
\bibitem [{\citenamefont {Kacanja}\ \emph {et~al.}(2025)\citenamefont {Kacanja}, \citenamefont {Soni},\ and\ \citenamefont {Nitz}}]{Kacanja:2025kpr}%
  \BibitemOpen
  \bibfield  {author} {\bibinfo {author} {\bibfnamefont {K.}~\bibnamefont {Kacanja}}, \bibinfo {author} {\bibfnamefont {K.}~\bibnamefont {Soni}},\ and\ \bibinfo {author} {\bibfnamefont {A.~H.}\ \bibnamefont {Nitz}},\ }\bibfield  {title} {\bibinfo {title} {{Eccentricity signatures in LIGO-Virgo-KAGRA's BNS and NSBH binaries}},\ }\href@noop {} {\  (\bibinfo {year} {2025})},\ \Eprint {https://arxiv.org/abs/2508.00179} {arXiv:2508.00179 [gr-qc]} \BibitemShut {NoStop}%
\bibitem [{\citenamefont {Jan}\ \emph {et~al.}(2025)\citenamefont {Jan}, \citenamefont {Tsao}, \citenamefont {O'Shaughnessy}, \citenamefont {Shoemaker},\ and\ \citenamefont {Laguna}}]{Jan:2025fps}%
  \BibitemOpen
  \bibfield  {author} {\bibinfo {author} {\bibfnamefont {A.}~\bibnamefont {Jan}}, \bibinfo {author} {\bibfnamefont {B.-J.}\ \bibnamefont {Tsao}}, \bibinfo {author} {\bibfnamefont {R.}~\bibnamefont {O'Shaughnessy}}, \bibinfo {author} {\bibfnamefont {D.}~\bibnamefont {Shoemaker}},\ and\ \bibinfo {author} {\bibfnamefont {P.}~\bibnamefont {Laguna}},\ }\bibfield  {title} {\bibinfo {title} {{GW200105: A detailed study of eccentricity in the neutron star-black hole binary}},\ }\href@noop {} {\  (\bibinfo {year} {2025})},\ \Eprint {https://arxiv.org/abs/2508.12460} {arXiv:2508.12460 [gr-qc]} \BibitemShut {NoStop}%
\bibitem [{\citenamefont {Tiwari}\ \emph {et~al.}(2025)\citenamefont {Tiwari}, \citenamefont {Bhat}, \citenamefont {Shaikh},\ and\ \citenamefont {Kapaida}}]{Tiwari:2025fua}%
  \BibitemOpen
  \bibfield  {author} {\bibinfo {author} {\bibfnamefont {A.}~\bibnamefont {Tiwari}}, \bibinfo {author} {\bibfnamefont {S.~A.}\ \bibnamefont {Bhat}}, \bibinfo {author} {\bibfnamefont {M.~A.}\ \bibnamefont {Shaikh}},\ and\ \bibinfo {author} {\bibfnamefont {S.~J.}\ \bibnamefont {Kapaida}},\ }\bibfield  {title} {\bibinfo {title} {{Testing the nature of GW200105 by probing the frequency evolution of eccentricity}},\ }\href@noop {} {\  (\bibinfo {year} {2025})},\ \Eprint {https://arxiv.org/abs/2509.26152} {arXiv:2509.26152 [astro-ph.HE]} \BibitemShut {NoStop}%
\bibitem [{\citenamefont {Collaboration}(2025)}]{LVK:2025T2400403}%
  \BibitemOpen
  \bibfield  {author} {\bibinfo {author} {\bibfnamefont {L.-V.-K.}\ \bibnamefont {Collaboration}} (\bibinfo {collaboration} {LIGO Scientific, KAGRA, VIRGO}),\ }\bibfield  {title} {\bibinfo {title} {{The LSC-Virgo-KAGRA Observational Science White Paper (2025 Edition)}},\ }\href {https://dcc.ligo.org/LIGO-T2400403/public} {\bibfield  {journal} {\bibinfo  {journal} {LIGO Technical Notes}\ }\textbf {\bibinfo {volume} {LIGO-T2400403}} (\bibinfo {year} {2025})}\BibitemShut {NoStop}%
\bibitem [{\citenamefont {Shah}\ \emph {et~al.}(2024)\citenamefont {Shah}, \citenamefont {Narayan}, \citenamefont {Perkins}, \citenamefont {Foley}, \citenamefont {Chatterjee}, \citenamefont {Cousins},\ and\ \citenamefont {Macias}}]{Shah:2023ozh}%
  \BibitemOpen
  \bibfield  {author} {\bibinfo {author} {\bibfnamefont {V.~G.}\ \bibnamefont {Shah}}, \bibinfo {author} {\bibfnamefont {G.}~\bibnamefont {Narayan}}, \bibinfo {author} {\bibfnamefont {H.~M.~L.}\ \bibnamefont {Perkins}}, \bibinfo {author} {\bibfnamefont {R.~J.}\ \bibnamefont {Foley}}, \bibinfo {author} {\bibfnamefont {D.}~\bibnamefont {Chatterjee}}, \bibinfo {author} {\bibfnamefont {B.}~\bibnamefont {Cousins}},\ and\ \bibinfo {author} {\bibfnamefont {P.}~\bibnamefont {Macias}},\ }\bibfield  {title} {\bibinfo {title} {{Predictions for electromagnetic counterparts to Neutron Star mergers discovered during LIGO-Virgo-KAGRA observing runs 4 and 5}},\ }\href {https://doi.org/10.1093/mnras/stad3711} {\bibfield  {journal} {\bibinfo  {journal} {Mon. Not. Roy. Astron. Soc.}\ }\textbf {\bibinfo {volume} {528}},\ \bibinfo {pages} {1109} (\bibinfo {year} {2024})},\ \Eprint {https://arxiv.org/abs/2310.15240} {arXiv:2310.15240 [astro-ph.HE]} \BibitemShut {NoStop}%
\bibitem [{\citenamefont {Colombo}\ \emph {et~al.}(2024)\citenamefont {Colombo} \emph {et~al.}}]{Colombo:2023une}%
  \BibitemOpen
  \bibfield  {author} {\bibinfo {author} {\bibfnamefont {A.}~\bibnamefont {Colombo}} \emph {et~al.},\ }\bibfield  {title} {\bibinfo {title} {{Multi-messenger prospects for black hole - neutron star mergers in the O4 and O5 runs}},\ }\href {https://doi.org/10.1051/0004-6361/202348384} {\bibfield  {journal} {\bibinfo  {journal} {Astron. Astrophys.}\ }\textbf {\bibinfo {volume} {686}},\ \bibinfo {pages} {A265} (\bibinfo {year} {2024})},\ \Eprint {https://arxiv.org/abs/2310.16894} {arXiv:2310.16894 [astro-ph.HE]} \BibitemShut {NoStop}%
\bibitem [{\citenamefont {Abac}\ \emph {et~al.}(2025{\natexlab{b}})\citenamefont {Abac} \emph {et~al.}}]{ET:2025xjr}%
  \BibitemOpen
  \bibfield  {author} {\bibinfo {author} {\bibfnamefont {A.}~\bibnamefont {Abac}} \emph {et~al.} (\bibinfo {collaboration} {ET}),\ }\bibfield  {title} {\bibinfo {title} {{The Science of the Einstein Telescope}},\ }\href@noop {} {\  (\bibinfo {year} {2025}{\natexlab{b}})},\ \Eprint {https://arxiv.org/abs/2503.12263} {arXiv:2503.12263 [gr-qc]} \BibitemShut {NoStop}%
\bibitem [{\citenamefont {Colpi}\ \emph {et~al.}(2024)\citenamefont {Colpi} \emph {et~al.}}]{LISA:2024hlh}%
  \BibitemOpen
  \bibfield  {author} {\bibinfo {author} {\bibfnamefont {M.}~\bibnamefont {Colpi}} \emph {et~al.} (\bibinfo {collaboration} {LISA}),\ }\bibfield  {title} {\bibinfo {title} {{LISA Definition Study Report}},\ }\href@noop {} {\  (\bibinfo {year} {2024})},\ \Eprint {https://arxiv.org/abs/2402.07571} {arXiv:2402.07571 [astro-ph.CO]} \BibitemShut {NoStop}%
\bibitem [{\citenamefont {Kremer}\ \emph {et~al.}(2018)\citenamefont {Kremer}, \citenamefont {Chatterjee}, \citenamefont {Breivik}, \citenamefont {Rodriguez}, \citenamefont {Larson},\ and\ \citenamefont {Rasio}}]{Kremer:2018tzm}%
  \BibitemOpen
  \bibfield  {author} {\bibinfo {author} {\bibfnamefont {K.}~\bibnamefont {Kremer}}, \bibinfo {author} {\bibfnamefont {S.}~\bibnamefont {Chatterjee}}, \bibinfo {author} {\bibfnamefont {K.}~\bibnamefont {Breivik}}, \bibinfo {author} {\bibfnamefont {C.~L.}\ \bibnamefont {Rodriguez}}, \bibinfo {author} {\bibfnamefont {S.~L.}\ \bibnamefont {Larson}},\ and\ \bibinfo {author} {\bibfnamefont {F.~A.}\ \bibnamefont {Rasio}},\ }\bibfield  {title} {\bibinfo {title} {{LISA Sources in Milky Way Globular Clusters}},\ }\href {https://doi.org/10.1103/PhysRevLett.120.191103} {\bibfield  {journal} {\bibinfo  {journal} {Phys. Rev. Lett.}\ }\textbf {\bibinfo {volume} {120}},\ \bibinfo {pages} {191103} (\bibinfo {year} {2018})},\ \Eprint {https://arxiv.org/abs/1802.05661} {arXiv:1802.05661 [astro-ph.HE]} \BibitemShut {NoStop}%
\bibitem [{\citenamefont {Sedda}(2020)}]{Sedda:2020wzl}%
  \BibitemOpen
  \bibfield  {author} {\bibinfo {author} {\bibfnamefont {M.~A.}\ \bibnamefont {Sedda}},\ }\bibfield  {title} {\bibinfo {title} {{Dissecting the properties of neutron star - black hole mergers originating in dense star clusters}},\ }\href {https://doi.org/10.1038/s42005-020-0310-x} {\bibfield  {journal} {\bibinfo  {journal} {Commun. Phys.}\ }\textbf {\bibinfo {volume} {3}},\ \bibinfo {pages} {43} (\bibinfo {year} {2020})},\ \Eprint {https://arxiv.org/abs/2003.02279} {arXiv:2003.02279 [astro-ph.GA]} \BibitemShut {NoStop}%
\bibitem [{\citenamefont {Andrews}\ \emph {et~al.}(2020)\citenamefont {Andrews}, \citenamefont {Breivik}, \citenamefont {Pankow}, \citenamefont {D'Orazio},\ and\ \citenamefont {Safarzadeh}}]{Andrews:2019plw}%
  \BibitemOpen
  \bibfield  {author} {\bibinfo {author} {\bibfnamefont {J.~J.}\ \bibnamefont {Andrews}}, \bibinfo {author} {\bibfnamefont {K.}~\bibnamefont {Breivik}}, \bibinfo {author} {\bibfnamefont {C.}~\bibnamefont {Pankow}}, \bibinfo {author} {\bibfnamefont {D.~J.}\ \bibnamefont {D'Orazio}},\ and\ \bibinfo {author} {\bibfnamefont {M.}~\bibnamefont {Safarzadeh}},\ }\bibfield  {title} {\bibinfo {title} {{LISA and the Existence of a Fast-Merging Double Neutron Star Formation Channel}},\ }\href {https://doi.org/10.3847/2041-8213/ab5b9a} {\bibfield  {journal} {\bibinfo  {journal} {Astrophys. J. Lett.}\ }\textbf {\bibinfo {volume} {892}},\ \bibinfo {pages} {L9} (\bibinfo {year} {2020})},\ \Eprint {https://arxiv.org/abs/1910.13436} {arXiv:1910.13436 [astro-ph.HE]} \BibitemShut {NoStop}%
\bibitem [{\citenamefont {Abbott}\ \emph {et~al.}(2019)\citenamefont {Abbott} \emph {et~al.}}]{LIGOScientific:2018hze}%
  \BibitemOpen
  \bibfield  {author} {\bibinfo {author} {\bibfnamefont {B.~P.}\ \bibnamefont {Abbott}} \emph {et~al.} (\bibinfo {collaboration} {LIGO Scientific, Virgo}),\ }\bibfield  {title} {\bibinfo {title} {{Properties of the binary neutron star merger GW170817}},\ }\href {https://doi.org/10.1103/PhysRevX.9.011001} {\bibfield  {journal} {\bibinfo  {journal} {Phys. Rev. X}\ }\textbf {\bibinfo {volume} {9}},\ \bibinfo {pages} {011001} (\bibinfo {year} {2019})},\ \Eprint {https://arxiv.org/abs/1805.11579} {arXiv:1805.11579 [gr-qc]} \BibitemShut {NoStop}%
\bibitem [{\citenamefont {{Gamba}}\ \emph {et~al.}(2023)\citenamefont {{Gamba}}, \citenamefont {{Breschi}}, \citenamefont {{Bernuzzi}}, \citenamefont {{Nagar}}, \citenamefont {{Cook}}, \citenamefont {{Doulis}}, \citenamefont {{Fabbri}}, \citenamefont {{Ortiz}}, \citenamefont {{Poudel}}, \citenamefont {{Rashti}}, \citenamefont {{Tichy}},\ and\ \citenamefont {{Ujevic}}}]{Gamba:2023mww}%
  \BibitemOpen
  \bibfield  {author} {\bibinfo {author} {\bibfnamefont {R.}~\bibnamefont {{Gamba}}}, \bibinfo {author} {\bibfnamefont {M.}~\bibnamefont {{Breschi}}}, \bibinfo {author} {\bibfnamefont {S.}~\bibnamefont {{Bernuzzi}}}, \bibinfo {author} {\bibfnamefont {A.}~\bibnamefont {{Nagar}}}, \bibinfo {author} {\bibfnamefont {W.}~\bibnamefont {{Cook}}}, \bibinfo {author} {\bibfnamefont {G.}~\bibnamefont {{Doulis}}}, \bibinfo {author} {\bibfnamefont {F.}~\bibnamefont {{Fabbri}}}, \bibinfo {author} {\bibfnamefont {N.}~\bibnamefont {{Ortiz}}}, \bibinfo {author} {\bibfnamefont {A.}~\bibnamefont {{Poudel}}}, \bibinfo {author} {\bibfnamefont {A.}~\bibnamefont {{Rashti}}}, \bibinfo {author} {\bibfnamefont {W.}~\bibnamefont {{Tichy}}},\ and\ \bibinfo {author} {\bibfnamefont {M.}~\bibnamefont {{Ujevic}}},\ }\bibfield  {title} {\bibinfo {title} {{Analytically improved and numerical-relativity informed effective-one-body model for coalescing binary neutron stars}},\ }\href {https://doi.org/10.48550/arXiv.2307.15125} {\bibfield
  {journal} {\bibinfo  {journal} {arXiv e-prints}\ ,\ \bibinfo {eid} {arXiv:2307.15125}} (\bibinfo {year} {2023})},\ \Eprint {https://arxiv.org/abs/2307.15125} {arXiv:2307.15125 [gr-qc]} \BibitemShut {NoStop}%
\bibitem [{\citenamefont {Albanesi}\ \emph {et~al.}(2025)\citenamefont {Albanesi}, \citenamefont {Gamba}, \citenamefont {Bernuzzi}, \citenamefont {Fontbut{\'e}}, \citenamefont {Gonzalez},\ and\ \citenamefont {Nagar}}]{Albanesi:2025txj}%
  \BibitemOpen
  \bibfield  {author} {\bibinfo {author} {\bibfnamefont {S.}~\bibnamefont {Albanesi}}, \bibinfo {author} {\bibfnamefont {R.}~\bibnamefont {Gamba}}, \bibinfo {author} {\bibfnamefont {S.}~\bibnamefont {Bernuzzi}}, \bibinfo {author} {\bibfnamefont {J.}~\bibnamefont {Fontbut{\'e}}}, \bibinfo {author} {\bibfnamefont {A.}~\bibnamefont {Gonzalez}},\ and\ \bibinfo {author} {\bibfnamefont {A.}~\bibnamefont {Nagar}},\ }\bibfield  {title} {\bibinfo {title} {{Effective-one-body modeling for generic compact binaries with arbitrary orbits}},\ }\href@noop {} {\bibfield  {journal} {\bibinfo  {journal} {arXiv e-prints}\ } (\bibinfo {year} {2025})},\ \Eprint {https://arxiv.org/abs/2503.14580} {arXiv:2503.14580 [gr-qc]} \BibitemShut {NoStop}%
\bibitem [{\citenamefont {Haberland}\ \emph {et~al.}(2025)\citenamefont {Haberland}, \citenamefont {Buonanno},\ and\ \citenamefont {Steinhoff}}]{Haberland:2025luz}%
  \BibitemOpen
  \bibfield  {author} {\bibinfo {author} {\bibfnamefont {M.}~\bibnamefont {Haberland}}, \bibinfo {author} {\bibfnamefont {A.}~\bibnamefont {Buonanno}},\ and\ \bibinfo {author} {\bibfnamefont {J.}~\bibnamefont {Steinhoff}},\ }\bibfield  {title} {\bibinfo {title} {{Modeling matter in seobnrv5thm: Generating fast and accurate effective-one-body waveforms for spin-aligned binary neutron stars}},\ }\href {https://doi.org/10.1103/d3ns-h77x} {\bibfield  {journal} {\bibinfo  {journal} {Phys. Rev. D}\ }\textbf {\bibinfo {volume} {112}},\ \bibinfo {pages} {084024} (\bibinfo {year} {2025})},\ \Eprint {https://arxiv.org/abs/2503.18934} {arXiv:2503.18934 [gr-qc]} \BibitemShut {NoStop}%
\bibitem [{\citenamefont {Colleoni}\ \emph {et~al.}(2025)\citenamefont {Colleoni}, \citenamefont {Ramis~Vidal}, \citenamefont {Johnson-McDaniel}, \citenamefont {Dietrich}, \citenamefont {Haney},\ and\ \citenamefont {Pratten}}]{Colleoni:2023ple}%
  \BibitemOpen
  \bibfield  {author} {\bibinfo {author} {\bibfnamefont {M.}~\bibnamefont {Colleoni}}, \bibinfo {author} {\bibfnamefont {F.~A.}\ \bibnamefont {Ramis~Vidal}}, \bibinfo {author} {\bibfnamefont {N.~K.}\ \bibnamefont {Johnson-McDaniel}}, \bibinfo {author} {\bibfnamefont {T.}~\bibnamefont {Dietrich}}, \bibinfo {author} {\bibfnamefont {M.}~\bibnamefont {Haney}},\ and\ \bibinfo {author} {\bibfnamefont {G.}~\bibnamefont {Pratten}},\ }\bibfield  {title} {\bibinfo {title} {{New gravitational waveform model for precessing binary neutron stars with double-spin effects}},\ }\href {https://doi.org/10.1103/PhysRevD.111.064025} {\bibfield  {journal} {\bibinfo  {journal} {Phys. Rev. D}\ }\textbf {\bibinfo {volume} {111}},\ \bibinfo {pages} {064025} (\bibinfo {year} {2025})},\ \Eprint {https://arxiv.org/abs/2311.15978} {arXiv:2311.15978 [gr-qc]} \BibitemShut {NoStop}%
\bibitem [{\citenamefont {Abac}\ \emph {et~al.}(2024{\natexlab{b}})\citenamefont {Abac}, \citenamefont {Dietrich}, \citenamefont {Buonanno}, \citenamefont {Steinhoff},\ and\ \citenamefont {Ujevic}}]{Abac:2023ujg}%
  \BibitemOpen
  \bibfield  {author} {\bibinfo {author} {\bibfnamefont {A.}~\bibnamefont {Abac}}, \bibinfo {author} {\bibfnamefont {T.}~\bibnamefont {Dietrich}}, \bibinfo {author} {\bibfnamefont {A.}~\bibnamefont {Buonanno}}, \bibinfo {author} {\bibfnamefont {J.}~\bibnamefont {Steinhoff}},\ and\ \bibinfo {author} {\bibfnamefont {M.}~\bibnamefont {Ujevic}},\ }\bibfield  {title} {\bibinfo {title} {{New and robust gravitational-waveform model for high-mass-ratio binary neutron star systems with dynamical tidal effects}},\ }\href {https://doi.org/10.1103/PhysRevD.109.024062} {\bibfield  {journal} {\bibinfo  {journal} {Phys. Rev. D}\ }\textbf {\bibinfo {volume} {109}},\ \bibinfo {pages} {024062} (\bibinfo {year} {2024}{\natexlab{b}})},\ \Eprint {https://arxiv.org/abs/2311.07456} {arXiv:2311.07456 [gr-qc]} \BibitemShut {NoStop}%
\bibitem [{\citenamefont {Ramis~Vidal}\ and\ \citenamefont {et~al}()}]{IMRPhenomXPHMNSBH}%
  \BibitemOpen
  \bibfield  {author} {\bibinfo {author} {\bibfnamefont {F.~A.}\ \bibnamefont {Ramis~Vidal}}\ and\ \bibinfo {author} {\bibnamefont {et~al}},\ }\href@noop {} {}\bibinfo {note} {To be published}\BibitemShut {NoStop}%
\bibitem [{\citenamefont {P\"urrer}(2014)}]{Purrer:2014fza}%
  \BibitemOpen
  \bibfield  {author} {\bibinfo {author} {\bibfnamefont {M.}~\bibnamefont {P\"urrer}},\ }\bibfield  {title} {\bibinfo {title} {{Frequency domain reduced order models for gravitational waves from aligned-spin compact binaries}},\ }\href {https://doi.org/10.1088/0264-9381/31/19/195010} {\bibfield  {journal} {\bibinfo  {journal} {Class. Quant. Grav.}\ }\textbf {\bibinfo {volume} {31}},\ \bibinfo {pages} {195010} (\bibinfo {year} {2014})},\ \Eprint {https://arxiv.org/abs/1402.4146} {arXiv:1402.4146 [gr-qc]} \BibitemShut {NoStop}%
\bibitem [{\citenamefont {Lackey}\ \emph {et~al.}(2017)\citenamefont {Lackey}, \citenamefont {Bernuzzi}, \citenamefont {Galley}, \citenamefont {Meidam},\ and\ \citenamefont {Van Den~Broeck}}]{Lackey:2016krb}%
  \BibitemOpen
  \bibfield  {author} {\bibinfo {author} {\bibfnamefont {B.~D.}\ \bibnamefont {Lackey}}, \bibinfo {author} {\bibfnamefont {S.}~\bibnamefont {Bernuzzi}}, \bibinfo {author} {\bibfnamefont {C.~R.}\ \bibnamefont {Galley}}, \bibinfo {author} {\bibfnamefont {J.}~\bibnamefont {Meidam}},\ and\ \bibinfo {author} {\bibfnamefont {C.}~\bibnamefont {Van Den~Broeck}},\ }\bibfield  {title} {\bibinfo {title} {{Effective-one-body waveforms for binary neutron stars using surrogate models}},\ }\href {https://doi.org/10.1103/PhysRevD.95.104036} {\bibfield  {journal} {\bibinfo  {journal} {Phys. Rev. D}\ }\textbf {\bibinfo {volume} {95}},\ \bibinfo {pages} {104036} (\bibinfo {year} {2017})},\ \Eprint {https://arxiv.org/abs/1610.04742} {arXiv:1610.04742 [gr-qc]} \BibitemShut {NoStop}%
\bibitem [{\citenamefont {Lackey}\ \emph {et~al.}(2019)\citenamefont {Lackey}, \citenamefont {P\"urrer}, \citenamefont {Taracchini},\ and\ \citenamefont {Marsat}}]{Lackey:2018zvw}%
  \BibitemOpen
  \bibfield  {author} {\bibinfo {author} {\bibfnamefont {B.~D.}\ \bibnamefont {Lackey}}, \bibinfo {author} {\bibfnamefont {M.}~\bibnamefont {P\"urrer}}, \bibinfo {author} {\bibfnamefont {A.}~\bibnamefont {Taracchini}},\ and\ \bibinfo {author} {\bibfnamefont {S.}~\bibnamefont {Marsat}},\ }\bibfield  {title} {\bibinfo {title} {{Surrogate model for an aligned-spin effective one body waveform model of binary neutron star inspirals using Gaussian process regression}},\ }\href {https://doi.org/10.1103/PhysRevD.100.024002} {\bibfield  {journal} {\bibinfo  {journal} {Phys. Rev. D}\ }\textbf {\bibinfo {volume} {100}},\ \bibinfo {pages} {024002} (\bibinfo {year} {2019})},\ \Eprint {https://arxiv.org/abs/1812.08643} {arXiv:1812.08643 [gr-qc]} \BibitemShut {NoStop}%
\bibitem [{\citenamefont {Morras}\ \emph {et~al.}(2025{\natexlab{b}})\citenamefont {Morras}, \citenamefont {Pratten},\ and\ \citenamefont {Schmidt}}]{Morras:2025nlp}%
  \BibitemOpen
  \bibfield  {author} {\bibinfo {author} {\bibfnamefont {G.}~\bibnamefont {Morras}}, \bibinfo {author} {\bibfnamefont {G.}~\bibnamefont {Pratten}},\ and\ \bibinfo {author} {\bibfnamefont {P.}~\bibnamefont {Schmidt}},\ }\bibfield  {title} {\bibinfo {title} {{Improved post-Newtonian waveform model for inspiralling precessing-eccentric compact binaries}},\ }\href {https://doi.org/10.1103/PhysRevD.111.084052} {\bibfield  {journal} {\bibinfo  {journal} {Phys. Rev. D}\ }\textbf {\bibinfo {volume} {111}},\ \bibinfo {pages} {084052} (\bibinfo {year} {2025}{\natexlab{b}})},\ \Eprint {https://arxiv.org/abs/2502.03929} {arXiv:2502.03929 [gr-qc]} \BibitemShut {NoStop}%
\bibitem [{\citenamefont {Planas}\ \emph {et~al.}(2025{\natexlab{b}})\citenamefont {Planas}, \citenamefont {Ramos-Buades}, \citenamefont {Garc{\'\i}a-Quir{\'o}s}, \citenamefont {Estell{\'e}s}, \citenamefont {Husa},\ and\ \citenamefont {Haney}}]{Planas:2025feq}%
  \BibitemOpen
  \bibfield  {author} {\bibinfo {author} {\bibfnamefont {M.~d.~L.}\ \bibnamefont {Planas}}, \bibinfo {author} {\bibfnamefont {A.}~\bibnamefont {Ramos-Buades}}, \bibinfo {author} {\bibfnamefont {C.}~\bibnamefont {Garc{\'\i}a-Quir{\'o}s}}, \bibinfo {author} {\bibfnamefont {H.}~\bibnamefont {Estell{\'e}s}}, \bibinfo {author} {\bibfnamefont {S.}~\bibnamefont {Husa}},\ and\ \bibinfo {author} {\bibfnamefont {M.}~\bibnamefont {Haney}},\ }\bibfield  {title} {\bibinfo {title} {{Time-domain phenomenological multipolar waveforms for aligned-spin binary black holes in elliptical orbits}},\ }\href@noop {} {\  (\bibinfo {year} {2025}{\natexlab{b}})},\ \Eprint {https://arxiv.org/abs/2503.13062} {arXiv:2503.13062 [gr-qc]} \BibitemShut {NoStop}%
\bibitem [{\citenamefont {Hulse}\ and\ \citenamefont {Taylor}(1975)}]{Hulse:1974eb}%
  \BibitemOpen
  \bibfield  {author} {\bibinfo {author} {\bibfnamefont {R.~A.}\ \bibnamefont {Hulse}}\ and\ \bibinfo {author} {\bibfnamefont {J.~H.}\ \bibnamefont {Taylor}},\ }\bibfield  {title} {\bibinfo {title} {{Discovery of a pulsar in a binary system}},\ }\href {https://doi.org/10.1086/181708} {\bibfield  {journal} {\bibinfo  {journal} {Astrophys. J. Lett.}\ }\textbf {\bibinfo {volume} {195}},\ \bibinfo {pages} {L51} (\bibinfo {year} {1975})}\BibitemShut {NoStop}%
\bibitem [{\citenamefont {Damour}\ and\ \citenamefont {Deruelle}(1985{\natexlab{a}})}]{DamourDeruelle1}%
  \BibitemOpen
  \bibfield  {author} {\bibinfo {author} {\bibfnamefont {T.}~\bibnamefont {Damour}}\ and\ \bibinfo {author} {\bibfnamefont {N.}~\bibnamefont {Deruelle}},\ }\bibfield  {title} {\bibinfo {title} {General relativistic celestial mechanics of binary systems. i. the post-newtonian motion},\ }\href {https://www.numdam.org/item/AIHPA_1985__43_1_107_0/} {\bibfield  {journal} {\bibinfo  {journal} {Annales de l'I.H.P. Physique th\'eorique}\ }\textbf {\bibinfo {volume} {43}},\ \bibinfo {pages} {107} (\bibinfo {year} {1985}{\natexlab{a}})}\BibitemShut {NoStop}%
\bibitem [{\citenamefont {Damour}\ and\ \citenamefont {Deruelle}(1985{\natexlab{b}})}]{damour1985general}%
  \BibitemOpen
  \bibfield  {author} {\bibinfo {author} {\bibfnamefont {T.}~\bibnamefont {Damour}}\ and\ \bibinfo {author} {\bibfnamefont {N.}~\bibnamefont {Deruelle}},\ }\bibfield  {title} {\bibinfo {title} {General relativistic celestial mechanics of binary systems. i. the post-newtonian motion},\ }in\ \href@noop {} {\emph {\bibinfo {booktitle} {Annales de l'IHP Physique th{\'e}orique}}},\ Vol.~\bibinfo {volume} {43}\ (\bibinfo {year} {1985})\ pp.\ \bibinfo {pages} {107--132}\BibitemShut {NoStop}%
\bibitem [{\citenamefont {Damour}\ and\ \citenamefont {Schaefer}(1988)}]{Damour:1988mr}%
  \BibitemOpen
  \bibfield  {author} {\bibinfo {author} {\bibfnamefont {T.}~\bibnamefont {Damour}}\ and\ \bibinfo {author} {\bibfnamefont {G.}~\bibnamefont {Schaefer}},\ }\bibfield  {title} {\bibinfo {title} {{Higher Order Relativistic Periastron Advances and Binary Pulsars}},\ }\href {https://doi.org/10.1007/BF02828697} {\bibfield  {journal} {\bibinfo  {journal} {Nuovo Cim. B}\ }\textbf {\bibinfo {volume} {101}},\ \bibinfo {pages} {127} (\bibinfo {year} {1988})}\BibitemShut {NoStop}%
\bibitem [{\citenamefont {Sch{\"a}fer}\ and\ \citenamefont {Wex}(1993)}]{Schafer:1993pkg}%
  \BibitemOpen
  \bibfield  {author} {\bibinfo {author} {\bibfnamefont {G.}~\bibnamefont {Sch{\"a}fer}}\ and\ \bibinfo {author} {\bibfnamefont {N.}~\bibnamefont {Wex}},\ }\bibfield  {title} {\bibinfo {title} {{Second post-Newtonian motion of compact binaries}},\ }\href {https://doi.org/10.1016/0375-9601(93)90758-r} {\bibfield  {journal} {\bibinfo  {journal} {Phys. Lett. A}\ }\textbf {\bibinfo {volume} {174}},\ \bibinfo {pages} {196} (\bibinfo {year} {1993})},\ \bibinfo {note} {[Erratum: Phys.Lett.A 177, (1993)]}\BibitemShut {NoStop}%
\bibitem [{\citenamefont {Memmesheimer}\ \emph {et~al.}(2004)\citenamefont {Memmesheimer}, \citenamefont {Gopakumar},\ and\ \citenamefont {Schaefer}}]{Memmesheimer:2004cv}%
  \BibitemOpen
  \bibfield  {author} {\bibinfo {author} {\bibfnamefont {R.-M.}\ \bibnamefont {Memmesheimer}}, \bibinfo {author} {\bibfnamefont {A.}~\bibnamefont {Gopakumar}},\ and\ \bibinfo {author} {\bibfnamefont {G.}~\bibnamefont {Schaefer}},\ }\bibfield  {title} {\bibinfo {title} {{Third post-Newtonian accurate generalized quasi-Keplerian parametrization for compact binaries in eccentric orbits}},\ }\href {https://doi.org/10.1103/PhysRevD.70.104011} {\bibfield  {journal} {\bibinfo  {journal} {Phys. Rev. D}\ }\textbf {\bibinfo {volume} {70}},\ \bibinfo {pages} {104011} (\bibinfo {year} {2004})},\ \Eprint {https://arxiv.org/abs/gr-qc/0407049} {arXiv:gr-qc/0407049} \BibitemShut {NoStop}%
\bibitem [{\citenamefont {Boetzel}\ \emph {et~al.}(2017)\citenamefont {Boetzel}, \citenamefont {Susobhanan}, \citenamefont {Gopakumar}, \citenamefont {Klein},\ and\ \citenamefont {Jetzer}}]{Boetzel:2017zza}%
  \BibitemOpen
  \bibfield  {author} {\bibinfo {author} {\bibfnamefont {Y.}~\bibnamefont {Boetzel}}, \bibinfo {author} {\bibfnamefont {A.}~\bibnamefont {Susobhanan}}, \bibinfo {author} {\bibfnamefont {A.}~\bibnamefont {Gopakumar}}, \bibinfo {author} {\bibfnamefont {A.}~\bibnamefont {Klein}},\ and\ \bibinfo {author} {\bibfnamefont {P.}~\bibnamefont {Jetzer}},\ }\bibfield  {title} {\bibinfo {title} {{Solving post-Newtonian accurate Kepler Equation}},\ }\href {https://doi.org/10.1103/PhysRevD.96.044011} {\bibfield  {journal} {\bibinfo  {journal} {Phys. Rev. D}\ }\textbf {\bibinfo {volume} {96}},\ \bibinfo {pages} {044011} (\bibinfo {year} {2017})},\ \Eprint {https://arxiv.org/abs/1707.02088} {arXiv:1707.02088 [gr-qc]} \BibitemShut {NoStop}%
\bibitem [{\citenamefont {Cho}\ \emph {et~al.}(2022{\natexlab{a}})\citenamefont {Cho}, \citenamefont {Tanay}, \citenamefont {Gopakumar},\ and\ \citenamefont {Lee}}]{Cho:2021oai}%
  \BibitemOpen
  \bibfield  {author} {\bibinfo {author} {\bibfnamefont {G.}~\bibnamefont {Cho}}, \bibinfo {author} {\bibfnamefont {S.}~\bibnamefont {Tanay}}, \bibinfo {author} {\bibfnamefont {A.}~\bibnamefont {Gopakumar}},\ and\ \bibinfo {author} {\bibfnamefont {H.~M.}\ \bibnamefont {Lee}},\ }\bibfield  {title} {\bibinfo {title} {{Generalized quasi-Keplerian solution for eccentric, nonspinning compact binaries at 4PN order and the associated inspiral-merger-ringdown waveform}},\ }\href {https://doi.org/10.1103/PhysRevD.105.064010} {\bibfield  {journal} {\bibinfo  {journal} {Phys. Rev. D}\ }\textbf {\bibinfo {volume} {105}},\ \bibinfo {pages} {064010} (\bibinfo {year} {2022}{\natexlab{a}})},\ \Eprint {https://arxiv.org/abs/2110.09608} {arXiv:2110.09608 [gr-qc]} \BibitemShut {NoStop}%
\bibitem [{\citenamefont {Trestini}(2025)}]{Trestini:2025yyc}%
  \BibitemOpen
  \bibfield  {author} {\bibinfo {author} {\bibfnamefont {D.}~\bibnamefont {Trestini}},\ }\bibfield  {title} {\bibinfo {title} {{Constants of motion and fundamental frequencies for elliptic orbits at fourth post-Newtonian order}},\ }\href@noop {} {\  (\bibinfo {year} {2025})},\ \Eprint {https://arxiv.org/abs/2511.10735} {arXiv:2511.10735 [gr-qc]} \BibitemShut {NoStop}%
\bibitem [{\citenamefont {Tessmer}\ \emph {et~al.}(2010)\citenamefont {Tessmer}, \citenamefont {Hartung},\ and\ \citenamefont {Schafer}}]{Tessmer:2010hp}%
  \BibitemOpen
  \bibfield  {author} {\bibinfo {author} {\bibfnamefont {M.}~\bibnamefont {Tessmer}}, \bibinfo {author} {\bibfnamefont {J.}~\bibnamefont {Hartung}},\ and\ \bibinfo {author} {\bibfnamefont {G.}~\bibnamefont {Schafer}},\ }\bibfield  {title} {\bibinfo {title} {{Motion and gravitational wave forms of eccentric compact binaries with orbital-angular-momentum-aligned spins under next-to-leading order in spin-orbit and leading order in spin(1)-spin(2) and spin-squared couplings}},\ }\href {https://doi.org/10.1088/0264-9381/27/16/165005} {\bibfield  {journal} {\bibinfo  {journal} {Class. Quant. Grav.}\ }\textbf {\bibinfo {volume} {27}},\ \bibinfo {pages} {165005} (\bibinfo {year} {2010})},\ \Eprint {https://arxiv.org/abs/1003.2735} {arXiv:1003.2735 [gr-qc]} \BibitemShut {NoStop}%
\bibitem [{\citenamefont {Klein}\ and\ \citenamefont {Jetzer}(2010)}]{Klein:2010ti}%
  \BibitemOpen
  \bibfield  {author} {\bibinfo {author} {\bibfnamefont {A.}~\bibnamefont {Klein}}\ and\ \bibinfo {author} {\bibfnamefont {P.}~\bibnamefont {Jetzer}},\ }\bibfield  {title} {\bibinfo {title} {{Spin effects in the phasing of gravitational waves from binaries on eccentric orbits}},\ }\href {https://doi.org/10.1103/PhysRevD.81.124001} {\bibfield  {journal} {\bibinfo  {journal} {Phys. Rev. D}\ }\textbf {\bibinfo {volume} {81}},\ \bibinfo {pages} {124001} (\bibinfo {year} {2010})},\ \Eprint {https://arxiv.org/abs/1005.2046} {arXiv:1005.2046 [gr-qc]} \BibitemShut {NoStop}%
\bibitem [{\citenamefont {Tessmer}\ \emph {et~al.}(2013)\citenamefont {Tessmer}, \citenamefont {Hartung},\ and\ \citenamefont {Schafer}}]{Tessmer:2012xr}%
  \BibitemOpen
  \bibfield  {author} {\bibinfo {author} {\bibfnamefont {M.}~\bibnamefont {Tessmer}}, \bibinfo {author} {\bibfnamefont {J.}~\bibnamefont {Hartung}},\ and\ \bibinfo {author} {\bibfnamefont {G.}~\bibnamefont {Schafer}},\ }\bibfield  {title} {\bibinfo {title} {{Aligned Spins: Orbital Elements, Decaying Orbits, and Last Stable Circular Orbit to high post-Newtonian Orders}},\ }\href {https://doi.org/10.1088/0264-9381/30/1/015007} {\bibfield  {journal} {\bibinfo  {journal} {Class. Quant. Grav.}\ }\textbf {\bibinfo {volume} {30}},\ \bibinfo {pages} {015007} (\bibinfo {year} {2013})},\ \Eprint {https://arxiv.org/abs/1207.6961} {arXiv:1207.6961 [gr-qc]} \BibitemShut {NoStop}%
\bibitem [{\citenamefont {Cho}\ \emph {et~al.}(2022{\natexlab{b}})\citenamefont {Cho}, \citenamefont {Porto},\ and\ \citenamefont {Yang}}]{Cho:2022syn}%
  \BibitemOpen
  \bibfield  {author} {\bibinfo {author} {\bibfnamefont {G.}~\bibnamefont {Cho}}, \bibinfo {author} {\bibfnamefont {R.~A.}\ \bibnamefont {Porto}},\ and\ \bibinfo {author} {\bibfnamefont {Z.}~\bibnamefont {Yang}},\ }\bibfield  {title} {\bibinfo {title} {{Gravitational radiation from inspiralling compact objects: Spin effects to the fourth post-Newtonian order}},\ }\href {https://doi.org/10.1103/PhysRevD.106.L101501} {\bibfield  {journal} {\bibinfo  {journal} {Phys. Rev. D}\ }\textbf {\bibinfo {volume} {106}},\ \bibinfo {pages} {L101501} (\bibinfo {year} {2022}{\natexlab{b}})},\ \Eprint {https://arxiv.org/abs/2201.05138} {arXiv:2201.05138 [gr-qc]} \BibitemShut {NoStop}%
\bibitem [{\citenamefont {Henry}\ and\ \citenamefont {Khalil}(2023)}]{Henry:2023tka}%
  \BibitemOpen
  \bibfield  {author} {\bibinfo {author} {\bibfnamefont {Q.}~\bibnamefont {Henry}}\ and\ \bibinfo {author} {\bibfnamefont {M.}~\bibnamefont {Khalil}},\ }\bibfield  {title} {\bibinfo {title} {{Spin effects in gravitational waveforms and fluxes for binaries on eccentric orbits to the third post-Newtonian order}},\ }\href {https://doi.org/10.1103/PhysRevD.108.104016} {\bibfield  {journal} {\bibinfo  {journal} {Phys. Rev. D}\ }\textbf {\bibinfo {volume} {108}},\ \bibinfo {pages} {104016} (\bibinfo {year} {2023})},\ \Eprint {https://arxiv.org/abs/2308.13606} {arXiv:2308.13606 [gr-qc]} \BibitemShut {NoStop}%
\bibitem [{\citenamefont {Damour}\ \emph {et~al.}(2004)\citenamefont {Damour}, \citenamefont {Gopakumar},\ and\ \citenamefont {Iyer}}]{Damour:2004bz}%
  \BibitemOpen
  \bibfield  {author} {\bibinfo {author} {\bibfnamefont {T.}~\bibnamefont {Damour}}, \bibinfo {author} {\bibfnamefont {A.}~\bibnamefont {Gopakumar}},\ and\ \bibinfo {author} {\bibfnamefont {B.~R.}\ \bibnamefont {Iyer}},\ }\bibfield  {title} {\bibinfo {title} {{Phasing of gravitational waves from inspiralling eccentric binaries}},\ }\href {https://doi.org/10.1103/PhysRevD.70.064028} {\bibfield  {journal} {\bibinfo  {journal} {Phys. Rev. D}\ }\textbf {\bibinfo {volume} {70}},\ \bibinfo {pages} {064028} (\bibinfo {year} {2004})},\ \Eprint {https://arxiv.org/abs/gr-qc/0404128} {arXiv:gr-qc/0404128} \BibitemShut {NoStop}%
\bibitem [{\citenamefont {Konigsdorffer}\ and\ \citenamefont {Gopakumar}(2006)}]{Konigsdorffer:2006zt}%
  \BibitemOpen
  \bibfield  {author} {\bibinfo {author} {\bibfnamefont {C.}~\bibnamefont {Konigsdorffer}}\ and\ \bibinfo {author} {\bibfnamefont {A.}~\bibnamefont {Gopakumar}},\ }\bibfield  {title} {\bibinfo {title} {{Phasing of gravitational waves from inspiralling eccentric binaries at the third-and-a-half post-Newtonian order}},\ }\href {https://doi.org/10.1103/PhysRevD.73.124012} {\bibfield  {journal} {\bibinfo  {journal} {Phys. Rev. D}\ }\textbf {\bibinfo {volume} {73}},\ \bibinfo {pages} {124012} (\bibinfo {year} {2006})},\ \Eprint {https://arxiv.org/abs/gr-qc/0603056} {arXiv:gr-qc/0603056} \BibitemShut {NoStop}%
\bibitem [{\citenamefont {Boetzel}\ \emph {et~al.}(2019)\citenamefont {Boetzel}, \citenamefont {Mishra}, \citenamefont {Faye}, \citenamefont {Gopakumar},\ and\ \citenamefont {Iyer}}]{Boetzel:2019nfw}%
  \BibitemOpen
  \bibfield  {author} {\bibinfo {author} {\bibfnamefont {Y.}~\bibnamefont {Boetzel}}, \bibinfo {author} {\bibfnamefont {C.~K.}\ \bibnamefont {Mishra}}, \bibinfo {author} {\bibfnamefont {G.}~\bibnamefont {Faye}}, \bibinfo {author} {\bibfnamefont {A.}~\bibnamefont {Gopakumar}},\ and\ \bibinfo {author} {\bibfnamefont {B.~R.}\ \bibnamefont {Iyer}},\ }\bibfield  {title} {\bibinfo {title} {{Gravitational-wave amplitudes for compact binaries in eccentric orbits at the third post-Newtonian order: Tail contributions and postadiabatic corrections}},\ }\href {https://doi.org/10.1103/PhysRevD.100.044018} {\bibfield  {journal} {\bibinfo  {journal} {Phys. Rev. D}\ }\textbf {\bibinfo {volume} {100}},\ \bibinfo {pages} {044018} (\bibinfo {year} {2019})},\ \Eprint {https://arxiv.org/abs/1904.11814} {arXiv:1904.11814 [gr-qc]} \BibitemShut {NoStop}%
\bibitem [{\citenamefont {Arun}\ \emph {et~al.}(2008{\natexlab{a}})\citenamefont {Arun}, \citenamefont {Blanchet}, \citenamefont {Iyer},\ and\ \citenamefont {Qusailah}}]{Arun:2007sg}%
  \BibitemOpen
  \bibfield  {author} {\bibinfo {author} {\bibfnamefont {K.~G.}\ \bibnamefont {Arun}}, \bibinfo {author} {\bibfnamefont {L.}~\bibnamefont {Blanchet}}, \bibinfo {author} {\bibfnamefont {B.~R.}\ \bibnamefont {Iyer}},\ and\ \bibinfo {author} {\bibfnamefont {M.~S.~S.}\ \bibnamefont {Qusailah}},\ }\bibfield  {title} {\bibinfo {title} {{Inspiralling compact binaries in quasi-elliptical orbits: The Complete 3PN energy flux}},\ }\href {https://doi.org/10.1103/PhysRevD.77.064035} {\bibfield  {journal} {\bibinfo  {journal} {Phys. Rev. D}\ }\textbf {\bibinfo {volume} {77}},\ \bibinfo {pages} {064035} (\bibinfo {year} {2008}{\natexlab{a}})},\ \Eprint {https://arxiv.org/abs/0711.0302} {arXiv:0711.0302 [gr-qc]} \BibitemShut {NoStop}%
\bibitem [{\citenamefont {Arun}\ \emph {et~al.}(2009)\citenamefont {Arun}, \citenamefont {Blanchet}, \citenamefont {Iyer},\ and\ \citenamefont {Sinha}}]{Arun:2009mc}%
  \BibitemOpen
  \bibfield  {author} {\bibinfo {author} {\bibfnamefont {K.~G.}\ \bibnamefont {Arun}}, \bibinfo {author} {\bibfnamefont {L.}~\bibnamefont {Blanchet}}, \bibinfo {author} {\bibfnamefont {B.~R.}\ \bibnamefont {Iyer}},\ and\ \bibinfo {author} {\bibfnamefont {S.}~\bibnamefont {Sinha}},\ }\bibfield  {title} {\bibinfo {title} {{Third post-Newtonian angular momentum flux and the secular evolution of orbital elements for inspiralling compact binaries in quasi-elliptical orbits}},\ }\href {https://doi.org/10.1103/PhysRevD.80.124018} {\bibfield  {journal} {\bibinfo  {journal} {Phys. Rev. D}\ }\textbf {\bibinfo {volume} {80}},\ \bibinfo {pages} {124018} (\bibinfo {year} {2009})},\ \Eprint {https://arxiv.org/abs/0908.3854} {arXiv:0908.3854 [gr-qc]} \BibitemShut {NoStop}%
\bibitem [{\citenamefont {Arun}\ \emph {et~al.}(2008{\natexlab{b}})\citenamefont {Arun}, \citenamefont {Blanchet}, \citenamefont {Iyer},\ and\ \citenamefont {Qusailah}}]{Arun:2007rg}%
  \BibitemOpen
  \bibfield  {author} {\bibinfo {author} {\bibfnamefont {K.~G.}\ \bibnamefont {Arun}}, \bibinfo {author} {\bibfnamefont {L.}~\bibnamefont {Blanchet}}, \bibinfo {author} {\bibfnamefont {B.~R.}\ \bibnamefont {Iyer}},\ and\ \bibinfo {author} {\bibfnamefont {M.~S.~S.}\ \bibnamefont {Qusailah}},\ }\bibfield  {title} {\bibinfo {title} {{Tail effects in the 3PN gravitational wave energy flux of compact binaries in quasi-elliptical orbits}},\ }\href {https://doi.org/10.1103/PhysRevD.77.064034} {\bibfield  {journal} {\bibinfo  {journal} {Phys. Rev. D}\ }\textbf {\bibinfo {volume} {77}},\ \bibinfo {pages} {064034} (\bibinfo {year} {2008}{\natexlab{b}})},\ \Eprint {https://arxiv.org/abs/0711.0250} {arXiv:0711.0250 [gr-qc]} \BibitemShut {NoStop}%
\bibitem [{\citenamefont {Vines}\ and\ \citenamefont {Flanagan}(2013)}]{Vines:2010ca}%
  \BibitemOpen
  \bibfield  {author} {\bibinfo {author} {\bibfnamefont {J.~E.}\ \bibnamefont {Vines}}\ and\ \bibinfo {author} {\bibfnamefont {E.~E.}\ \bibnamefont {Flanagan}},\ }\bibfield  {title} {\bibinfo {title} {{Post-1-Newtonian quadrupole tidal interactions in binary systems}},\ }\href {https://doi.org/10.1103/PhysRevD.88.024046} {\bibfield  {journal} {\bibinfo  {journal} {Phys. Rev. D}\ }\textbf {\bibinfo {volume} {88}},\ \bibinfo {pages} {024046} (\bibinfo {year} {2013})},\ \Eprint {https://arxiv.org/abs/1009.4919} {arXiv:1009.4919 [gr-qc]} \BibitemShut {NoStop}%
\bibitem [{\citenamefont {Damour}\ \emph {et~al.}(2012)\citenamefont {Damour}, \citenamefont {Nagar},\ and\ \citenamefont {Villain}}]{Damour:2012yf}%
  \BibitemOpen
  \bibfield  {author} {\bibinfo {author} {\bibfnamefont {T.}~\bibnamefont {Damour}}, \bibinfo {author} {\bibfnamefont {A.}~\bibnamefont {Nagar}},\ and\ \bibinfo {author} {\bibfnamefont {L.}~\bibnamefont {Villain}},\ }\bibfield  {title} {\bibinfo {title} {{Measurability of the tidal polarizability of neutron stars in late-inspiral gravitational-wave signals}},\ }\href {https://doi.org/10.1103/PhysRevD.85.123007} {\bibfield  {journal} {\bibinfo  {journal} {Phys. Rev. D}\ }\textbf {\bibinfo {volume} {85}},\ \bibinfo {pages} {123007} (\bibinfo {year} {2012})},\ \Eprint {https://arxiv.org/abs/1203.4352} {arXiv:1203.4352 [gr-qc]} \BibitemShut {NoStop}%
\bibitem [{\citenamefont {Steinhoff}\ \emph {et~al.}(2016)\citenamefont {Steinhoff}, \citenamefont {Hinderer}, \citenamefont {Buonanno},\ and\ \citenamefont {Taracchini}}]{Steinhoff:2016rfi}%
  \BibitemOpen
  \bibfield  {author} {\bibinfo {author} {\bibfnamefont {J.}~\bibnamefont {Steinhoff}}, \bibinfo {author} {\bibfnamefont {T.}~\bibnamefont {Hinderer}}, \bibinfo {author} {\bibfnamefont {A.}~\bibnamefont {Buonanno}},\ and\ \bibinfo {author} {\bibfnamefont {A.}~\bibnamefont {Taracchini}},\ }\bibfield  {title} {\bibinfo {title} {{Dynamical Tides in General Relativity: Effective Action and Effective-One-Body Hamiltonian}},\ }\href {https://doi.org/10.1103/PhysRevD.94.104028} {\bibfield  {journal} {\bibinfo  {journal} {Phys. Rev. D}\ }\textbf {\bibinfo {volume} {94}},\ \bibinfo {pages} {104028} (\bibinfo {year} {2016})},\ \Eprint {https://arxiv.org/abs/1608.01907} {arXiv:1608.01907 [gr-qc]} \BibitemShut {NoStop}%
\bibitem [{\citenamefont {Banihashemi}\ and\ \citenamefont {Vines}(2020)}]{Banihashemi:2018xfb}%
  \BibitemOpen
  \bibfield  {author} {\bibinfo {author} {\bibfnamefont {B.}~\bibnamefont {Banihashemi}}\ and\ \bibinfo {author} {\bibfnamefont {J.}~\bibnamefont {Vines}},\ }\bibfield  {title} {\bibinfo {title} {{Gravitomagnetic tidal effects in gravitational waves from neutron star binaries}},\ }\href {https://doi.org/10.1103/PhysRevD.101.064003} {\bibfield  {journal} {\bibinfo  {journal} {Phys. Rev. D}\ }\textbf {\bibinfo {volume} {101}},\ \bibinfo {pages} {064003} (\bibinfo {year} {2020})},\ \Eprint {https://arxiv.org/abs/1805.07266} {arXiv:1805.07266 [gr-qc]} \BibitemShut {NoStop}%
\bibitem [{\citenamefont {Abdelsalhin}\ \emph {et~al.}(2018)\citenamefont {Abdelsalhin}, \citenamefont {Gualtieri},\ and\ \citenamefont {Pani}}]{Abdelsalhin:2018reg}%
  \BibitemOpen
  \bibfield  {author} {\bibinfo {author} {\bibfnamefont {T.}~\bibnamefont {Abdelsalhin}}, \bibinfo {author} {\bibfnamefont {L.}~\bibnamefont {Gualtieri}},\ and\ \bibinfo {author} {\bibfnamefont {P.}~\bibnamefont {Pani}},\ }\bibfield  {title} {\bibinfo {title} {{Post-Newtonian spin-tidal couplings for compact binaries}},\ }\href {https://doi.org/10.1103/PhysRevD.98.104046} {\bibfield  {journal} {\bibinfo  {journal} {Phys. Rev. D}\ }\textbf {\bibinfo {volume} {98}},\ \bibinfo {pages} {104046} (\bibinfo {year} {2018})},\ \Eprint {https://arxiv.org/abs/1805.01487} {arXiv:1805.01487 [gr-qc]} \BibitemShut {NoStop}%
\bibitem [{\citenamefont {Henry}\ \emph {et~al.}(2020{\natexlab{a}})\citenamefont {Henry}, \citenamefont {Faye},\ and\ \citenamefont {Blanchet}}]{HFB19}%
  \BibitemOpen
  \bibfield  {author} {\bibinfo {author} {\bibfnamefont {Q.}~\bibnamefont {Henry}}, \bibinfo {author} {\bibfnamefont {G.}~\bibnamefont {Faye}},\ and\ \bibinfo {author} {\bibfnamefont {L.}~\bibnamefont {Blanchet}},\ }\bibfield  {title} {\bibinfo {title} {{Tidal effects in the equations of motion of compact binary systems to next-to-next-to-leading post-Newtonian order}},\ }\href {https://doi.org/10.1103/PhysRevD.101.064047} {\bibfield  {journal} {\bibinfo  {journal} {Phys. Rev. D}\ }\textbf {\bibinfo {volume} {101}},\ \bibinfo {pages} {064047} (\bibinfo {year} {2020}{\natexlab{a}})},\ \Eprint {https://arxiv.org/abs/1912.01920} {arXiv:1912.01920 [gr-qc]} \BibitemShut {NoStop}%
\bibitem [{\citenamefont {Henry}\ \emph {et~al.}(2020{\natexlab{b}})\citenamefont {Henry}, \citenamefont {Faye},\ and\ \citenamefont {Blanchet}}]{HFB20a}%
  \BibitemOpen
  \bibfield  {author} {\bibinfo {author} {\bibfnamefont {Q.}~\bibnamefont {Henry}}, \bibinfo {author} {\bibfnamefont {G.}~\bibnamefont {Faye}},\ and\ \bibinfo {author} {\bibfnamefont {L.}~\bibnamefont {Blanchet}},\ }\bibfield  {title} {\bibinfo {title} {{Tidal effects in the gravitational-wave phase evolution of compact binary systems to next-to-next-to-leading post-Newtonian order}},\ }\href {https://doi.org/10.1103/PhysRevD.102.044033} {\bibfield  {journal} {\bibinfo  {journal} {Phys. Rev. D}\ }\textbf {\bibinfo {volume} {102}},\ \bibinfo {pages} {044033} (\bibinfo {year} {2020}{\natexlab{b}})},\ \bibinfo {note} {[Erratum: Phys.Rev.D 108, 089901 (2023)]},\ \Eprint {https://arxiv.org/abs/2005.13367} {arXiv:2005.13367 [gr-qc]} \BibitemShut {NoStop}%
\bibitem [{\citenamefont {Henry}\ \emph {et~al.}(2020{\natexlab{c}})\citenamefont {Henry}, \citenamefont {Faye},\ and\ \citenamefont {Blanchet}}]{HFB20b}%
  \BibitemOpen
  \bibfield  {author} {\bibinfo {author} {\bibfnamefont {Q.}~\bibnamefont {Henry}}, \bibinfo {author} {\bibfnamefont {G.}~\bibnamefont {Faye}},\ and\ \bibinfo {author} {\bibfnamefont {L.}~\bibnamefont {Blanchet}},\ }\bibfield  {title} {\bibinfo {title} {{Hamiltonian for tidal interactions in compact binary systems to next-to-next-to-leading post-Newtonian order}},\ }\href {https://doi.org/10.1103/PhysRevD.102.124074} {\bibfield  {journal} {\bibinfo  {journal} {Phys. Rev. D}\ }\textbf {\bibinfo {volume} {102}},\ \bibinfo {pages} {124074} (\bibinfo {year} {2020}{\natexlab{c}})},\ \Eprint {https://arxiv.org/abs/2009.12332} {arXiv:2009.12332 [gr-qc]} \BibitemShut {NoStop}%
\bibitem [{\citenamefont {Cheung}\ and\ \citenamefont {Solon}(2020)}]{Cheung:2020sdj}%
  \BibitemOpen
  \bibfield  {author} {\bibinfo {author} {\bibfnamefont {C.}~\bibnamefont {Cheung}}\ and\ \bibinfo {author} {\bibfnamefont {M.~P.}\ \bibnamefont {Solon}},\ }\bibfield  {title} {\bibinfo {title} {{Tidal Effects in the Post-Minkowskian Expansion}},\ }\href {https://doi.org/10.1103/PhysRevLett.125.191601} {\bibfield  {journal} {\bibinfo  {journal} {Phys. Rev. Lett.}\ }\textbf {\bibinfo {volume} {125}},\ \bibinfo {pages} {191601} (\bibinfo {year} {2020})},\ \Eprint {https://arxiv.org/abs/2006.06665} {arXiv:2006.06665 [hep-th]} \BibitemShut {NoStop}%
\bibitem [{\citenamefont {K\"alin}\ \emph {et~al.}(2020)\citenamefont {K\"alin}, \citenamefont {Liu},\ and\ \citenamefont {Porto}}]{Kalin:2020lmz}%
  \BibitemOpen
  \bibfield  {author} {\bibinfo {author} {\bibfnamefont {G.}~\bibnamefont {K\"alin}}, \bibinfo {author} {\bibfnamefont {Z.}~\bibnamefont {Liu}},\ and\ \bibinfo {author} {\bibfnamefont {R.~A.}\ \bibnamefont {Porto}},\ }\bibfield  {title} {\bibinfo {title} {{Conservative Tidal Effects in Compact Binary Systems to Next-to-Leading Post-Minkowskian Order}},\ }\href {https://doi.org/10.1103/PhysRevD.102.124025} {\bibfield  {journal} {\bibinfo  {journal} {Phys. Rev. D}\ }\textbf {\bibinfo {volume} {102}},\ \bibinfo {pages} {124025} (\bibinfo {year} {2020})},\ \Eprint {https://arxiv.org/abs/2008.06047} {arXiv:2008.06047 [hep-th]} \BibitemShut {NoStop}%
\bibitem [{\citenamefont {Mougiakakos}\ \emph {et~al.}(2022)\citenamefont {Mougiakakos}, \citenamefont {Riva},\ and\ \citenamefont {Vernizzi}}]{Mougiakakos:2022sic}%
  \BibitemOpen
  \bibfield  {author} {\bibinfo {author} {\bibfnamefont {S.}~\bibnamefont {Mougiakakos}}, \bibinfo {author} {\bibfnamefont {M.~M.}\ \bibnamefont {Riva}},\ and\ \bibinfo {author} {\bibfnamefont {F.}~\bibnamefont {Vernizzi}},\ }\bibfield  {title} {\bibinfo {title} {{Gravitational Bremsstrahlung with Tidal Effects in the Post-Minkowskian Expansion}},\ }\href {https://doi.org/10.1103/PhysRevLett.129.121101} {\bibfield  {journal} {\bibinfo  {journal} {Phys. Rev. Lett.}\ }\textbf {\bibinfo {volume} {129}},\ \bibinfo {pages} {121101} (\bibinfo {year} {2022})},\ \Eprint {https://arxiv.org/abs/2204.06556} {arXiv:2204.06556 [hep-th]} \BibitemShut {NoStop}%
\bibitem [{\citenamefont {Mandal}\ \emph {et~al.}(2024{\natexlab{a}})\citenamefont {Mandal}, \citenamefont {Mastrolia}, \citenamefont {Silva}, \citenamefont {Patil},\ and\ \citenamefont {Steinhoff}}]{Mandal:2023hqa}%
  \BibitemOpen
  \bibfield  {author} {\bibinfo {author} {\bibfnamefont {M.~K.}\ \bibnamefont {Mandal}}, \bibinfo {author} {\bibfnamefont {P.}~\bibnamefont {Mastrolia}}, \bibinfo {author} {\bibfnamefont {H.~O.}\ \bibnamefont {Silva}}, \bibinfo {author} {\bibfnamefont {R.}~\bibnamefont {Patil}},\ and\ \bibinfo {author} {\bibfnamefont {J.}~\bibnamefont {Steinhoff}},\ }\bibfield  {title} {\bibinfo {title} {{Renormalizing Love: tidal effects at the third post-Newtonian order}},\ }\href {https://doi.org/10.1007/JHEP02(2024)188} {\bibfield  {journal} {\bibinfo  {journal} {JHEP}\ }\textbf {\bibinfo {volume} {02}},\ \bibinfo {pages} {188}},\ \Eprint {https://arxiv.org/abs/2308.01865} {arXiv:2308.01865 [hep-th]} \BibitemShut {NoStop}%
\bibitem [{\citenamefont {Mandal}\ \emph {et~al.}(2024{\natexlab{b}})\citenamefont {Mandal}, \citenamefont {Mastrolia}, \citenamefont {Patil},\ and\ \citenamefont {Steinhoff}}]{Patil2024}%
  \BibitemOpen
  \bibfield  {author} {\bibinfo {author} {\bibfnamefont {M.~K.}\ \bibnamefont {Mandal}}, \bibinfo {author} {\bibfnamefont {P.}~\bibnamefont {Mastrolia}}, \bibinfo {author} {\bibfnamefont {R.}~\bibnamefont {Patil}},\ and\ \bibinfo {author} {\bibfnamefont {J.}~\bibnamefont {Steinhoff}},\ }\bibfield  {title} {\bibinfo {title} {{Radiating Love: adiabatic tidal fluxes and modes up to next-to-next-to-leading post-Newtonian order}},\ }\href@noop {} {\  (\bibinfo {year} {2024}{\natexlab{b}})},\ \Eprint {https://arxiv.org/abs/2412.01706} {arXiv:2412.01706 [gr-qc]} \BibitemShut {NoStop}%
\bibitem [{\citenamefont {Bernard}\ \emph {et~al.}(2024)\citenamefont {Bernard}, \citenamefont {Dones},\ and\ \citenamefont {Mougiakakos}}]{Bernard:2023eul}%
  \BibitemOpen
  \bibfield  {author} {\bibinfo {author} {\bibfnamefont {L.}~\bibnamefont {Bernard}}, \bibinfo {author} {\bibfnamefont {E.}~\bibnamefont {Dones}},\ and\ \bibinfo {author} {\bibfnamefont {S.}~\bibnamefont {Mougiakakos}},\ }\bibfield  {title} {\bibinfo {title} {{Tidal effects up to next-to-next-to-leading post-Newtonian order in massless scalar-tensor theories}},\ }\href {https://doi.org/10.1103/PhysRevD.109.044006} {\bibfield  {journal} {\bibinfo  {journal} {Phys. Rev. D}\ }\textbf {\bibinfo {volume} {109}},\ \bibinfo {pages} {044006} (\bibinfo {year} {2024})},\ \Eprint {https://arxiv.org/abs/2310.19679} {arXiv:2310.19679 [gr-qc]} \BibitemShut {NoStop}%
\bibitem [{\citenamefont {Dones}\ \emph {et~al.}(2025)\citenamefont {Dones}, \citenamefont {Henry},\ and\ \citenamefont {Bernard}}]{Dones:2024odv}%
  \BibitemOpen
  \bibfield  {author} {\bibinfo {author} {\bibfnamefont {E.}~\bibnamefont {Dones}}, \bibinfo {author} {\bibfnamefont {Q.}~\bibnamefont {Henry}},\ and\ \bibinfo {author} {\bibfnamefont {L.}~\bibnamefont {Bernard}},\ }\bibfield  {title} {\bibinfo {title} {{Tidal contributions to the full gravitational waveform to the second-and-a-half post-Newtonian order}},\ }\href {https://doi.org/10.1103/PhysRevD.111.084043} {\bibfield  {journal} {\bibinfo  {journal} {Phys. Rev. D}\ }\textbf {\bibinfo {volume} {111}},\ \bibinfo {pages} {084043} (\bibinfo {year} {2025})},\ \Eprint {https://arxiv.org/abs/2412.14249} {arXiv:2412.14249 [gr-qc]} \BibitemShut {NoStop}%
\bibitem [{\citenamefont {Dones}\ and\ \citenamefont {Bernard}(2025)}]{Dones:2025zbs}%
  \BibitemOpen
  \bibfield  {author} {\bibinfo {author} {\bibfnamefont {E.}~\bibnamefont {Dones}}\ and\ \bibinfo {author} {\bibfnamefont {L.}~\bibnamefont {Bernard}},\ }\bibfield  {title} {\bibinfo {title} {{Tidal effects in gravitational and scalar waveforms and fluxes to one post-Newtonian order in massless scalar-tensor theories}},\ }\href@noop {} {\  (\bibinfo {year} {2025})},\ \Eprint {https://arxiv.org/abs/2507.07676} {arXiv:2507.07676 [gr-qc]} \BibitemShut {NoStop}%
\bibitem [{\citenamefont {Henry}(2026)}]{paperII}%
  \BibitemOpen
  \bibfield  {author} {\bibinfo {author} {\bibfnamefont {Q.}~\bibnamefont {Henry}},\ }\bibfield  {title} {\bibinfo {title} {{Adiabatic tides in compact binaries on quasi-elliptic orbits: Radiation at the second-and-a-half relative post-Newtonian order}},\ }\href@noop {} {\  (\bibinfo {year} {2026})},\ \Eprint {https://arxiv.org/abs/2601.01794} {arXiv:2601.01794 [gr-qc]} \BibitemShut {NoStop}%
\bibitem [{\citenamefont {Damour}\ \emph {et~al.}(1991)\citenamefont {Damour}, \citenamefont {Soffel},\ and\ \citenamefont {Xu}}]{Damour:1990pi}%
  \BibitemOpen
  \bibfield  {author} {\bibinfo {author} {\bibfnamefont {T.}~\bibnamefont {Damour}}, \bibinfo {author} {\bibfnamefont {M.}~\bibnamefont {Soffel}},\ and\ \bibinfo {author} {\bibfnamefont {C.-m.}\ \bibnamefont {Xu}},\ }\bibfield  {title} {\bibinfo {title} {{General relativistic celestial mechanics. 1. Method and definition of reference systems}},\ }\href {https://doi.org/10.1103/PhysRevD.43.3273} {\bibfield  {journal} {\bibinfo  {journal} {Phys. Rev. D}\ }\textbf {\bibinfo {volume} {43}},\ \bibinfo {pages} {3273} (\bibinfo {year} {1991})}\BibitemShut {NoStop}%
\bibitem [{\citenamefont {Damour}\ \emph {et~al.}(1992)\citenamefont {Damour}, \citenamefont {Soffel},\ and\ \citenamefont {Xu}}]{Damour:1991yw}%
  \BibitemOpen
  \bibfield  {author} {\bibinfo {author} {\bibfnamefont {T.}~\bibnamefont {Damour}}, \bibinfo {author} {\bibfnamefont {M.}~\bibnamefont {Soffel}},\ and\ \bibinfo {author} {\bibfnamefont {C.-m.}\ \bibnamefont {Xu}},\ }\bibfield  {title} {\bibinfo {title} {{General relativistic celestial mechanics. 2. Translational equations of motion}},\ }\href {https://doi.org/10.1103/PhysRevD.45.1017} {\bibfield  {journal} {\bibinfo  {journal} {Phys. Rev. D}\ }\textbf {\bibinfo {volume} {45}},\ \bibinfo {pages} {1017} (\bibinfo {year} {1992})}\BibitemShut {NoStop}%
\bibitem [{\citenamefont {Bini}\ \emph {et~al.}(2012)\citenamefont {Bini}, \citenamefont {Damour},\ and\ \citenamefont {Faye}}]{Bini:2012gu}%
  \BibitemOpen
  \bibfield  {author} {\bibinfo {author} {\bibfnamefont {D.}~\bibnamefont {Bini}}, \bibinfo {author} {\bibfnamefont {T.}~\bibnamefont {Damour}},\ and\ \bibinfo {author} {\bibfnamefont {G.}~\bibnamefont {Faye}},\ }\bibfield  {title} {\bibinfo {title} {{Effective action approach to higher-order relativistic tidal interactions in binary systems and their effective one body description}},\ }\href {https://doi.org/10.1103/PhysRevD.85.124034} {\bibfield  {journal} {\bibinfo  {journal} {Phys. Rev. D}\ }\textbf {\bibinfo {volume} {85}},\ \bibinfo {pages} {124034} (\bibinfo {year} {2012})},\ \Eprint {https://arxiv.org/abs/1202.3565} {arXiv:1202.3565 [gr-qc]} \BibitemShut {NoStop}%
\bibitem [{\citenamefont {Love}(1911)}]{love1911_geodynamics}%
  \BibitemOpen
  \bibfield  {author} {\bibinfo {author} {\bibfnamefont {A.~E.~H.}\ \bibnamefont {Love}},\ }\href {https://archive.org/details/cu31924060184367/page/n9/mode/2up} {\emph {\bibinfo {title} {Some Problems of Geodynamics: Being an Essay to Which the Adams Prize in the University of Cambridge Was Adjudged in 1911}}}\ (\bibinfo  {publisher} {Cambridge University Press},\ \bibinfo {year} {1911})\BibitemShut {NoStop}%
\bibitem [{\citenamefont {Damour}\ and\ \citenamefont {Nagar}(2009)}]{Damour:2009vw}%
  \BibitemOpen
  \bibfield  {author} {\bibinfo {author} {\bibfnamefont {T.}~\bibnamefont {Damour}}\ and\ \bibinfo {author} {\bibfnamefont {A.}~\bibnamefont {Nagar}},\ }\bibfield  {title} {\bibinfo {title} {{Relativistic tidal properties of neutron stars}},\ }\href {https://doi.org/10.1103/PhysRevD.80.084035} {\bibfield  {journal} {\bibinfo  {journal} {Phys. Rev. D}\ }\textbf {\bibinfo {volume} {80}},\ \bibinfo {pages} {084035} (\bibinfo {year} {2009})},\ \Eprint {https://arxiv.org/abs/0906.0096} {arXiv:0906.0096 [gr-qc]} \BibitemShut {NoStop}%
\bibitem [{\citenamefont {Binnington}\ and\ \citenamefont {Poisson}(2009)}]{Binnington:2009bb}%
  \BibitemOpen
  \bibfield  {author} {\bibinfo {author} {\bibfnamefont {T.}~\bibnamefont {Binnington}}\ and\ \bibinfo {author} {\bibfnamefont {E.}~\bibnamefont {Poisson}},\ }\bibfield  {title} {\bibinfo {title} {{Relativistic theory of tidal Love numbers}},\ }\href {https://doi.org/10.1103/PhysRevD.80.084018} {\bibfield  {journal} {\bibinfo  {journal} {Phys. Rev. D}\ }\textbf {\bibinfo {volume} {80}},\ \bibinfo {pages} {084018} (\bibinfo {year} {2009})},\ \Eprint {https://arxiv.org/abs/0906.1366} {arXiv:0906.1366 [gr-qc]} \BibitemShut {NoStop}%
\bibitem [{\citenamefont {Landry}\ and\ \citenamefont {Poisson}(2015)}]{Landry:2015zfa}%
  \BibitemOpen
  \bibfield  {author} {\bibinfo {author} {\bibfnamefont {P.}~\bibnamefont {Landry}}\ and\ \bibinfo {author} {\bibfnamefont {E.}~\bibnamefont {Poisson}},\ }\bibfield  {title} {\bibinfo {title} {{Tidal deformation of a slowly rotating material body. External metric}},\ }\href {https://doi.org/10.1103/PhysRevD.91.104018} {\bibfield  {journal} {\bibinfo  {journal} {Phys. Rev. D}\ }\textbf {\bibinfo {volume} {91}},\ \bibinfo {pages} {104018} (\bibinfo {year} {2015})},\ \Eprint {https://arxiv.org/abs/1503.07366} {arXiv:1503.07366 [gr-qc]} \BibitemShut {NoStop}%
\bibitem [{\citenamefont {Le~Tiec}\ \emph {et~al.}(2021)\citenamefont {Le~Tiec}, \citenamefont {Casals},\ and\ \citenamefont {Franzin}}]{LeTiec:2020bos}%
  \BibitemOpen
  \bibfield  {author} {\bibinfo {author} {\bibfnamefont {A.}~\bibnamefont {Le~Tiec}}, \bibinfo {author} {\bibfnamefont {M.}~\bibnamefont {Casals}},\ and\ \bibinfo {author} {\bibfnamefont {E.}~\bibnamefont {Franzin}},\ }\bibfield  {title} {\bibinfo {title} {{Tidal Love Numbers of Kerr Black Holes}},\ }\href {https://doi.org/10.1103/PhysRevD.103.084021} {\bibfield  {journal} {\bibinfo  {journal} {Phys. Rev. D}\ }\textbf {\bibinfo {volume} {103}},\ \bibinfo {pages} {084021} (\bibinfo {year} {2021})},\ \Eprint {https://arxiv.org/abs/2010.15795} {arXiv:2010.15795 [gr-qc]} \BibitemShut {NoStop}%
\bibitem [{Sup()}]{SuppMaterial1}%
  \BibitemOpen
  \href@noop {} {\bibinfo {title} {The ancillary file \texttt{Conservative\_Radiative\_dynamics\_adiabatic\_tides\_2.5PN.m} contains the conservative and raidative dynamics with tidal contributions at relative 2.5pn.}}\BibitemShut {Stop}%
\bibitem [{\citenamefont {Duffing}(1918)}]{duffing1918erzwungene}%
  \BibitemOpen
  \bibfield  {author} {\bibinfo {author} {\bibfnamefont {G.}~\bibnamefont {Duffing}},\ }\href@noop {} {\emph {\bibinfo {title} {Erzwungene Schwingungen bei ver{\"a}nderlicher Eigenfrequenz und ihre technische Bedeutung}}},\ \bibinfo {number} {41-42}\ (\bibinfo  {publisher} {Vieweg},\ \bibinfo {year} {1918})\BibitemShut {NoStop}%
\bibitem [{\citenamefont {Mickens}(1996)}]{mickens1996oscillations}%
  \BibitemOpen
  \bibfield  {author} {\bibinfo {author} {\bibfnamefont {R.~E.}\ \bibnamefont {Mickens}},\ }\href {https://unina2.on-line.it/sebina/repository/catalogazione/documenti/Mickens%20-%20Oscillations%20in%20planar%20dynamic%20systems.pdf} {\emph {\bibinfo {title} {Oscillations in planar dynamic systems}}},\ Vol.~\bibinfo {volume} {37}\ (\bibinfo  {publisher} {World Scientific},\ \bibinfo {year} {1996})\BibitemShut {NoStop}%
\bibitem [{\citenamefont {Poincar{\'e}}(1893)}]{Poincare1893MethodesNouvelles2}%
  \BibitemOpen
  \bibfield  {author} {\bibinfo {author} {\bibfnamefont {H.}~\bibnamefont {Poincar{\'e}}},\ }\href {https://archive.org/details/in.ernet.dli.2015.153619/page/n9/mode/2up} {\emph {\bibinfo {title} {Les {m{\'e}thodes} nouvelles de la m{\'e}canique c{\'e}leste: {T}ome {II}}}}\ (\bibinfo  {publisher} {Gauthier‐Villars et Fils},\ \bibinfo {address} {Paris},\ \bibinfo {year} {1893})\BibitemShut {NoStop}%
\bibitem [{\citenamefont {Loutrel}\ \emph {et~al.}(2019)\citenamefont {Loutrel}, \citenamefont {Liebersbach}, \citenamefont {Yunes},\ and\ \citenamefont {Cornish}}]{Loutrel:2018ydu}%
  \BibitemOpen
  \bibfield  {author} {\bibinfo {author} {\bibfnamefont {N.}~\bibnamefont {Loutrel}}, \bibinfo {author} {\bibfnamefont {S.}~\bibnamefont {Liebersbach}}, \bibinfo {author} {\bibfnamefont {N.}~\bibnamefont {Yunes}},\ and\ \bibinfo {author} {\bibfnamefont {N.}~\bibnamefont {Cornish}},\ }\bibfield  {title} {\bibinfo {title} {{The eccentric behavior of inspiralling compact binaries}},\ }\href {https://doi.org/10.1088/1361-6382/aaf2a9} {\bibfield  {journal} {\bibinfo  {journal} {Class. Quant. Grav.}\ }\textbf {\bibinfo {volume} {36}},\ \bibinfo {pages} {025004} (\bibinfo {year} {2019})},\ \Eprint {https://arxiv.org/abs/1810.03521} {arXiv:1810.03521 [gr-qc]} \BibitemShut {NoStop}%
\bibitem [{\citenamefont {Flanagan}\ and\ \citenamefont {Hinderer}(2008)}]{Flanagan:2007ix}%
  \BibitemOpen
  \bibfield  {author} {\bibinfo {author} {\bibfnamefont {E.~E.}\ \bibnamefont {Flanagan}}\ and\ \bibinfo {author} {\bibfnamefont {T.}~\bibnamefont {Hinderer}},\ }\bibfield  {title} {\bibinfo {title} {{Constraining neutron star tidal Love numbers with gravitational wave detectors}},\ }\href {https://doi.org/10.1103/PhysRevD.77.021502} {\bibfield  {journal} {\bibinfo  {journal} {Phys. Rev. D}\ }\textbf {\bibinfo {volume} {77}},\ \bibinfo {pages} {021502} (\bibinfo {year} {2008})},\ \Eprint {https://arxiv.org/abs/0709.1915} {arXiv:0709.1915 [astro-ph]} \BibitemShut {NoStop}%
\bibitem [{\citenamefont {Rezzolla}\ and\ \citenamefont {Ecker}(2025)}]{Rezzolla:2025pft}%
  \BibitemOpen
  \bibfield  {author} {\bibinfo {author} {\bibfnamefont {L.}~\bibnamefont {Rezzolla}}\ and\ \bibinfo {author} {\bibfnamefont {C.}~\bibnamefont {Ecker}},\ }\bibfield  {title} {\bibinfo {title} {{On the maximum compactness of neutron stars}},\ }\href@noop {} {\  (\bibinfo {year} {2025})},\ \Eprint {https://arxiv.org/abs/2510.12870} {arXiv:2510.12870 [gr-qc]} \BibitemShut {NoStop}%
\bibitem [{\citenamefont {Blanchet}\ and\ \citenamefont {Iyer}(2003)}]{Blanchet:2002mb}%
  \BibitemOpen
  \bibfield  {author} {\bibinfo {author} {\bibfnamefont {L.}~\bibnamefont {Blanchet}}\ and\ \bibinfo {author} {\bibfnamefont {B.~R.}\ \bibnamefont {Iyer}},\ }\bibfield  {title} {\bibinfo {title} {{Third postNewtonian dynamics of compact binaries: Equations of motion in the center-of-mass frame}},\ }\href {https://doi.org/10.1088/0264-9381/20/4/309} {\bibfield  {journal} {\bibinfo  {journal} {Class. Quant. Grav.}\ }\textbf {\bibinfo {volume} {20}},\ \bibinfo {pages} {755} (\bibinfo {year} {2003})},\ \Eprint {https://arxiv.org/abs/gr-qc/0209089} {arXiv:gr-qc/0209089} \BibitemShut {NoStop}%
\bibitem [{\citenamefont {Mandal}\ \emph {et~al.}(2024{\natexlab{c}})\citenamefont {Mandal}, \citenamefont {Mastrolia}, \citenamefont {Patil},\ and\ \citenamefont {Steinhoff}}]{Mandal:2024iug}%
  \BibitemOpen
  \bibfield  {author} {\bibinfo {author} {\bibfnamefont {M.~K.}\ \bibnamefont {Mandal}}, \bibinfo {author} {\bibfnamefont {P.}~\bibnamefont {Mastrolia}}, \bibinfo {author} {\bibfnamefont {R.}~\bibnamefont {Patil}},\ and\ \bibinfo {author} {\bibfnamefont {J.}~\bibnamefont {Steinhoff}},\ }\bibfield  {title} {\bibinfo {title} {{Radiating Love: adiabatic tidal fluxes and modes up to next-to-next-to-leading post-Newtonian order}},\ }\href@noop {} {\bibfield  {journal} {\bibinfo  {journal} {Phys. Rev. D}\ } (\bibinfo {year} {2024}{\natexlab{c}})},\ \Eprint {https://arxiv.org/abs/2412.01706} {arXiv:2412.01706 [gr-qc]} \BibitemShut {NoStop}%
\bibitem [{\citenamefont {Henry}\ \emph {et~al.}(2023)\citenamefont {Henry}, \citenamefont {Larrouturou},\ and\ \citenamefont {Le~Poncin-Lafitte}}]{Henry:2023guc}%
  \BibitemOpen
  \bibfield  {author} {\bibinfo {author} {\bibfnamefont {Q.}~\bibnamefont {Henry}}, \bibinfo {author} {\bibfnamefont {F.}~\bibnamefont {Larrouturou}},\ and\ \bibinfo {author} {\bibfnamefont {C.}~\bibnamefont {Le~Poncin-Lafitte}},\ }\bibfield  {title} {\bibinfo {title} {{Electromagnetic fields in compact binaries: A post-Newtonian approach}},\ }\href {https://doi.org/10.1103/PhysRevD.108.024020} {\bibfield  {journal} {\bibinfo  {journal} {Phys. Rev. D}\ }\textbf {\bibinfo {volume} {108}},\ \bibinfo {pages} {024020} (\bibinfo {year} {2023})},\ \Eprint {https://arxiv.org/abs/2303.17536} {arXiv:2303.17536 [gr-qc]} \BibitemShut {NoStop}%
\bibitem [{\citenamefont {Henry}\ \emph {et~al.}(2024)\citenamefont {Henry}, \citenamefont {Larrouturou},\ and\ \citenamefont {Le~Poncin-Lafitte}}]{Henry:2023len}%
  \BibitemOpen
  \bibfield  {author} {\bibinfo {author} {\bibfnamefont {Q.}~\bibnamefont {Henry}}, \bibinfo {author} {\bibfnamefont {F.}~\bibnamefont {Larrouturou}},\ and\ \bibinfo {author} {\bibfnamefont {C.}~\bibnamefont {Le~Poncin-Lafitte}},\ }\bibfield  {title} {\bibinfo {title} {{Electromagnetic fields in compact binaries: Post-Newtonian wave generation and application to double white dwarfs systems}},\ }\href {https://doi.org/10.1103/PhysRevD.109.084048} {\bibfield  {journal} {\bibinfo  {journal} {Phys. Rev. D}\ }\textbf {\bibinfo {volume} {109}},\ \bibinfo {pages} {084048} (\bibinfo {year} {2024})},\ \Eprint {https://arxiv.org/abs/2310.03785} {arXiv:2310.03785 [gr-qc]} \BibitemShut {NoStop}%
\bibitem [{\citenamefont {Lagrange}(1770)}]{lagrange1770}%
  \BibitemOpen
  \bibfield  {author} {\bibinfo {author} {\bibfnamefont {J.-L.}\ \bibnamefont {Lagrange}},\ }\bibfield  {title} {\bibinfo {title} {Nouvelle m{\'e}thode pour r{\'e}soudre les {\'e}quations litt{\'e}rales par le moyen des s{\'e}ries},\ }\href {https://archive.ph/20120630182344/http://gdz.sub.uni-goettingen.de/no_cache/dms/load/img/?IDDOC=41070} {\bibfield  {journal} {\bibinfo  {journal} {M{\'e}moires de l'Acad{\'e}mie Royale des Sciences et Belles-Lettres de Berlin}\ }\textbf {\bibinfo {volume} {24}},\ \bibinfo {pages} {251} (\bibinfo {year} {1770})},\ \bibinfo {note} {lu devant l'Acad\'emie les 18 janvier et 5 avril 1770.}\BibitemShut {Stop}%
\bibitem [{\citenamefont {Lagrange}\ and\ \citenamefont {Legendre}(1799)}]{lagrange_legendre1799}%
  \BibitemOpen
  \bibfield  {author} {\bibinfo {author} {\bibfnamefont {J.-L.}\ \bibnamefont {Lagrange}}\ and\ \bibinfo {author} {\bibfnamefont {A.-M.}\ \bibnamefont {Legendre}},\ }\bibfield  {title} {\bibinfo {title} {Rapport sur deux mémoires d'analyse du professeur burmann},\ }\href {https://gallica.bnf.fr/ark:/12148/bpt6k3217h} {\bibfield  {journal} {\bibinfo  {journal} {M{\'e}moires de l'Institut national des sciences et arts. Sciences math{\'e}matiques et physiques}\ }\textbf {\bibinfo {volume} {2}},\ \bibinfo {pages} {13} (\bibinfo {year} {1799})}\BibitemShut {NoStop}%
\bibitem [{\citenamefont {Laplace}(1825)}]{Laplace1825}%
  \BibitemOpen
  \bibfield  {author} {\bibinfo {author} {\bibfnamefont {P.-S.}\ \bibnamefont {Laplace}},\ }\href {https://archive.org/details/dli.ernet.493093} {\emph {\bibinfo {title} {Supplément au cinquième volume du Traité de mécanique céleste}}}\ (\bibinfo  {publisher} {Bachelier},\ \bibinfo {year} {1825})\BibitemShut {NoStop}%
\bibitem [{\citenamefont {Tisserand}(1889)}]{tisserand1889traité}%
  \BibitemOpen
  \bibfield  {author} {\bibinfo {author} {\bibfnamefont {F.}~\bibnamefont {Tisserand}},\ }\href {https://archive.org/details/traitdemcani04tissuoft/page/n5/mode/2up} {\emph {\bibinfo {title} {Trait{\'e} de m{\'e}canique c{\'e}leste}}}\ (\bibinfo  {publisher} {Gauthier-Villars et fils},\ \bibinfo {year} {1889})\BibitemShut {NoStop}%
\bibitem [{\citenamefont {Sridhar}\ \emph {et~al.}(2025)\citenamefont {Sridhar}, \citenamefont {Bhattacharyya}, \citenamefont {Paul},\ and\ \citenamefont {Mishra}}]{Sridhar:2024zms}%
  \BibitemOpen
  \bibfield  {author} {\bibinfo {author} {\bibfnamefont {O.}~\bibnamefont {Sridhar}}, \bibinfo {author} {\bibfnamefont {S.}~\bibnamefont {Bhattacharyya}}, \bibinfo {author} {\bibfnamefont {K.}~\bibnamefont {Paul}},\ and\ \bibinfo {author} {\bibfnamefont {C.~K.}\ \bibnamefont {Mishra}},\ }\bibfield  {title} {\bibinfo {title} {{Spin effects in the phasing formula of eccentric compact binary inspirals up to the third post-Newtonian order}},\ }\href {https://doi.org/10.1103/7xw4-flsn} {\bibfield  {journal} {\bibinfo  {journal} {Phys. Rev. D}\ }\textbf {\bibinfo {volume} {112}},\ \bibinfo {pages} {024026} (\bibinfo {year} {2025})},\ \Eprint {https://arxiv.org/abs/2412.10909} {arXiv:2412.10909 [gr-qc]} \BibitemShut {NoStop}%
\bibitem [{\citenamefont {Gradshteyn}\ and\ \citenamefont {Ryzhik}(2007)}]{gradshteyn2007}%
  \BibitemOpen
  \bibfield  {author} {\bibinfo {author} {\bibfnamefont {I.~S.}\ \bibnamefont {Gradshteyn}}\ and\ \bibinfo {author} {\bibfnamefont {I.~M.}\ \bibnamefont {Ryzhik}},\ }\href@noop {} {\emph {\bibinfo {title} {Table of integrals, series, and products}}},\ \bibinfo {edition} {seventh}\ ed.\ (\bibinfo {year} {2007})\BibitemShut {NoStop}%
\end{thebibliography}%

\end{document}